\documentclass[longauth]{aa}  

\usepackage[mathlines,switch]{lineno}
\usepackage{graphicx}
\usepackage{txfonts}
\usepackage{listings}
\usepackage{amsmath}
\usepackage{booktabs}
\usepackage{array}
\usepackage{natbib,twoopt}
\usepackage{upgreek}
\usepackage{color}
\usepackage{comment}
\usepackage[nolist,printonlyused,nohyperlinks]{acronym}
\bibpunct{(}{)}{;}{a}{}{,}

\newcommandtwoopt{\citeads}[3][][]{\href{http://adsabs.harvard.edu/abs/#3}%
{\citealp[#1][#2]{#3}}}
\newcommandtwoopt{\citepads}[3][][]{\href{http://adsabs.harvard.edu/abs/#3}%
{\citep[#1][#2]{#3}}}
\newcommandtwoopt{\citetads}[3][][]{\href{http://adsabs.harvard.edu/abs/#3}%
{\citet[#1][#2]{#3}}}
\newcommandtwoopt{\citeyearads}[3][][]%
{\href{http://adsabs.harvard.edu/abs/#3}{\citeyear[#1][#2]{#3}}} 

\newcommand{\HI}{\ion{H}{i}}
\newcommand{\mJybeam}{mJy beam$^{-1}$}
\newcommand{\kms}{ km s$^{-1}$}

\acrodef{ADC}{Analogue-to-Digital Converter}
\acrodef{ADU}{Analogue to Digital Unit}
\acrodef{AGN}{Active Galactic Nucleus}
\acrodefplural{AGN}[AGNs]{Active Galactic Nuclei}
\acrodef{ALERT}{Apertif-LOFAR Exploration of the Radio Transient sky.}
\acrodef{ALTA}{Apertif Long-Term Archive}
\acrodef{AMBER}[{\sc AMBER}]{Apertif Monitor for Bursts Encountered in Real-Time}
\acrodef{Apertif}{APERture Tile In Focus}
\acrodef{Apertif-X}{Apertif correlator}
\acrodef{FEBF}{Apertif Front-End Beamformer}
\acrodef{Artamis}[ARTAMIS]{the All-Round Telescope Array Monitoring and Information System}
\acrodef{ARTS}{Apertif Radio Transient System}
\acrodef{ARTS0}{ARTS Pulsar timing machine}
\acrodef{ATDB}{Apertif Task DataBase}
\acrodef{beam}{Group of beamlets or beamlet-channels that point in the same direction}
\acrodef{beamlet}{Beam formed subband by the Apertif FE BF, a small beam spanning one subband}
\acrodef{BER}{Bit Error Rate}
\acrodef{BF}{Beamformer}
\acrodef{BG}{Block Generator}
\acrodef{BN}{Back Node}
\acrodef{bps}{bits per second}
\acrodef{Bps}{bytes per second}
\acrodef{BRAM}{Block RAM} 
\acrodef{BSN}{Block Sequence Number}
\acrodef{BW}{BandWidth}
\acrodef{CB}{Compound Beam}
\acrodef{CCU}{Central Control Unit}
\acrodef{CDR}{Critical Design Review}
\acrodef{CHAN}{Channel, a frequency channel within a beamlet}
\acrodef{channel}{Unit frequency band within a beamlet}
\acrodef{cint}{Complex integer (type cast)}
\acrodef{Corner-turn}{Transposes data in time (in series)}
\acrodef{COTS}{Commercial Of The Shelf}
\acrodef{CPU}{Central Processing Unit}
\acrodef{CR}{Change Request}
\acrodef{CRC}{Cyclic Redundancy Check}
\acrodef{Cross-connect}{Transposes data in space (in parallel)}
\acrodef{CW}{Carrier Wave (single frequency signal)}
\acrodef{DARC}[{\sc DARC}]{Data Analysis of Real-time Candidates}
\acrodef{DB}{Data Buffer}
\acrodef{DM}{dispersion measure}
\acrodef{DNS}{double neutron star}
\acrodef{DP}{Data Path (streaming interface)}
\acrodef{DSP}{Digital Signal Processing}
\acrodef{DT}{Delay Tracking}
\acrodef{eop}{End of Packet (or frame, or block)}
\acrodef{ESD}{Electro Static Discharge}
\acrodef{FB}{FilterBank}
\acrodef{Fchan_a}{Channel filterbank in Arts BF for SC3 and SC4}
\acrodef{Fchan_x}{Channel filterbank in Apertif X}
\acrodef{FF}{Flip flop (1 bit storage in FPGA)}
\acrodef{FFT}{Fast Fourier Transform}
\acrodef{FIFO}{First In First Out (buffer in FPGA)}
\acrodef{FIR}{Finite Impulse Response (digital filter)}
\acrodef{FN}{Front Node}
\acrodef{FoV}{Field of View}
\acrodef{FPA}{Focal Plane Array (= PAF)}
\acrodef{FPGA}{Field-Programmable Gate Array}
\acrodef{FR}{Functional Requirement}
\acrodef{FRB}{Fast Radio Burst} 
\acrodef{FS}{Fringe Stopping}
\acrodef{Fsub}{Subband filterbank in Apertif FE BF}
\acrodef{FW}{Firmware}
\acrodef{GbE}{Gigabit Ethernet}
\acrodef{GPU}{Graphics processing unit}
\acrodef{HDL}{Hardware Description Language}
\acrodef{HF}{High Frequency}
\acrodef{HEM}{HMC Extension Board (provides UniBoard2 with HMC and extra optical IO)}
\acrodef{HMC}{Hybrid Memory Cube}
\acrodef{HW}{Hardware}
\acrodef{IAB}{Incoherent-Array Beam}
\acrodef{iablet}{A incoherently array beamformed beamlet using both polarizations}
\acrodef{IBF}{Incoherent BeamFormer}
\acrodef{iblet}{Fixed index of a data slot that carries a selected subband or a selected beamlet}
\acrodef{IGM}{InterGalactic Medium}
\acrodef{Im}{Imaginary}
\acrodef{int}{Signed integer}
\acrodef{I/O}{Input/Output}
\acrodef{IO}[I/O]{Input/Output}
\acrodef{IP}{Internet Protocol}
\acrodef{ISM}{InterStellar Medium}
\acrodef{LCU}{Local Control Unit}
\acrodef{LNA}{Low-Noise Amplifier}
\acrodef{LOFAR}{LOw Frequency ARray}
\acrodef{LSbit}{Least Significant bit}
\acrodef{LUT}{Look Up Table (boolean logic in FPGA)}
\acrodef{M20K}{Block RAM in Arria10 FPGA, 1024 x 20b so about 2 kByte}
\acrodef{M9K}{Block RAM in Stratix IV FPGA, 1024 x 9b so about 1 kByte}
\acrodef{ML}{Machine Learning}
\acrodef{MAC}{Monitoring And Control}
\acrodef{MFFE}{Multi Frequency Front End}
\acrodef{MISO}{Master In Slave Out (for memory mapped interface) [14]}
\acrodef{MM}{Memory Mapped}
\acrodef{MOSI}{Master Out Slave In (for memory mapped interface) [14]}
\acrodef{MSbit}{Most Significant bit}
\acrodef{node}{Processing node (PN), typically one FPGA chip}
\acrodef{Nof}{Number of}
\acrodef{OEB}{Optical-Electrical Board}
\acrodef{PAF}{phased array feed}
\acrodef{PBS}{Product Breakdown Structure}
\acrodef{PDR}{Preliminary Design Review}
\acrodef{PFB}{Polyphase Filter Bank}
\acrodef{PL}{Pipeline processing}
\acrodef{PN}{Processing Node}
\acrodef{PN2}{Processing Node on UniBoard2)}
\acrodef{power beam}{Full Stokes power values: I, Q, U, V}
\acrodef{PPS}{Pulse Per Second}
\acrodef{PSR}{Pulsar}
\acrodef{PT}{Phase Tracking}
\acrodef{Pulsar}{Pulsating radio star}
\acrodef{Rband}{Reorder TAB output for Arts BF - PL interface}
\acrodef{Rbeam}{Reorder and select beamlets for KCB = 40 compound beam directions Aperitif BF}
\acrodef{Re}{Real}
\acrodef{RF}{Radio Frequency}
\acrodef{RFI}{Radio Frequency Interference}
\acrodef{RO}{Radio Observatory}
\acrodef{RoHS}{Restriction of Hazardous Substances}
\acrodef{Rsub}{Reorder and select subbands for CBBW = 300 MHz in Aperitif BF}
\acrodef{RT}{Radio Telescope}
\acrodef{Rx}{Receive}
\acrodef{SB}{Synthesized Beam}
\acrodef{SC}{Science Case}
\acrodef{SEFD}{System-Equivalent Flux Density}
\acrodef{SFR}{Star Formation Rate}
\acrodef{SISO}{Source In Sink Out (for streaming interface) [14]}
\acrodef{SKA}{Square Kilometre Array}
\acrodef{SST}{Subband statistics}
\acrodef{S/N}{Signal-to-Noise ratio}
\acrodef{SNR}{Signal to Noise Ratio}
\acrodef{SODIMM}{Small Outline Dual In-line Memory Module}
\acrodef{sop}{Start of Packet (or frame, or block)}
\acrodef{SOSI}{Source Out Sink In (for streaming interface) [14]}
\acrodef{SP}{Signal Path, 1 CB consists of Npol = 2 SP}
\acrodef{sps}{Samples per second}
\acrodef{SR}{Science Requirement}
\acrodef{ST}{Streaming, statistics}
\acrodef{stream}{One stream of data describe by a SOSI signal}
\acrodef{subband}{Coarse channel frequency band, unit output of the Apertif FE BF filterbank}
\acrodef{SW}{Software}
\acrodef{TAB}{Tied-Array Beam}
\acrodef{TABF}{ARTS Tied-Array Beamformer}
\acrodef{tablet}{A coherently array beamformed beamlet, a grating or pencil beam within the CB }
\acrodef{Tant}{Transpose: split over subbands and group S = 64 (>= Nant) antenna elements in the PAF }
\acrodef{Tarray}{Transpose: split over beamlets and group Ntp = Npol * Ndish paths in the array (= Tdish+ Tpol)}
\acrodef{TB}{Tracking Beam}
\acrodef{Tband}{Transpose: split over tablets and group Nband = 16 bands for the full CBBW}
\acrodef{TBB}{Transient Buffer Board}
\acrodef{TCP}{Transmission Control Protocol}
\acrodef{Tdish}{Transpose: split over beamlets and group Ndish = 12 dishes}
\acrodef{Telescope}{The whole WSRT array}
\acrodef{Tint}{Transpose: split over beamlets and group Nint_x = 800000 beamlet time samples}
\acrodef{Tint_x}{Transpose: split over beamlets and group Nint_x = 800000 beamlet time samples}
\acrodef{Tinv}{Inverse transpose of Tint_x at Arts BF output }
\acrodef{TP}{Telescope Path, 1 TP contains all NCB = 37 SP for one single polarization from one dish}
\acrodef{Tpol}{Transpose: split over beamlets and group Npol = 2 polarizations}
\acrodef{Transpose}{Swaps indices in an signal array [15]}
\acrodef{Tx}{Transmit}
\acrodef{UCP}{UniBoard Control Protocol}
\acrodef{UDP}{User Datagram Protocol}
\acrodef{uint}{Unsigned integer}
\acrodef{UNB}{Uniboard}
\acrodef{UNB2}{Uniboard$^{2}$}
\acrodef{UniBoard}{Digital processing board with 8 Altera Stratix IV FPGAs}
\acrodef{UniBoard2}{Digital processing board with 4 Altera Arria 10 FPGAs}
\acrodef{UPE}{UniBoard Python Environment}
\acrodef{UTC}{Coordinated Universal Time}
\acrodef{VLBI}{Very Large Baseline Interferometry}
\acrodef{VO}{Virtual Observatory}
\acrodef{voltage beam}{Dual polarization sample values with phase information:  Xre, Xim, Yre, Yim}
\acrodef{WSRT}{Westerbork Synthesis Radio Telescope}
\acrodef{X}{Correlator}

\begin{document} 

\title{Apertif, Phased~Array~Feeds for the Westerbork Synthesis Radio~Telescope}
\subtitle{System Overview and Performance Characteristics}

\author{
     W.~A.~van~Cappellen  \inst{\ref{astron}}
\and T.~A.~Oosterloo      \inst{\ref{astron} \and \ref{kapteyn}}
\and M.~A.~W.~Verheijen   \inst{\ref{kapteyn}}
\and E.~A.~K.~Adams       \inst{\ref{astron} \and \ref{kapteyn}}
\and B.~Adebahr           \inst{\ref{airub}}
\and R.~Braun             \inst{\ref{skao}}
\and K.~M.~Hess           \inst{\ref{astron} \and \ref{kapteyn}}
\and H.~Holties           \inst{\ref{astron}}
\and J.~M.~van~der~Hulst  \inst{\ref{kapteyn}}
\and B.~Hut               \inst{\ref{astron}}
\and E.~Kooistra          \inst{\ref{astron}}
\and J.~van~Leeuwen    \inst{\ref{astron} \and \ref{uva}}
\and G.~M.~Loose      \inst{\ref{astron}}
\and R.~Morganti   \inst{\ref{astron} \and \ref{kapteyn}}
\and V.~A.~Moss           \inst{\ref{csiro} \and \ref{sydney} \and \ref{astron}}
\and E.~Orr\'u          \inst{\ref{astron}}
\and M.~Ruiter            \inst{\ref{astron}}
\and A.~P.~Schoenmakers    \inst{\ref{astron}}
\and N.~J.~Vermaas         \inst{\ref{astron}}
\and S.~J.~Wijnholds       \inst{\ref{astron}}
\and A.~S.~van~Amesfoort  \inst{\ref{astron}}
\and M.~J.~Arts            \inst{\ref{astron}}
\and J.~J.~Attema          \inst{\ref{escience}}
\and L.~Bakker
\and C.~G.~Bassa           \inst{\ref{astron}}
\and J.~E.~Bast           \inst{\ref{astron}}
\and P.~Benthem       \inst{\ref{astron}}
\and R.~Beukema           \inst{\ref{astron}}
\and R.~Blaauw            \inst{\ref{astron}}
\and W.~J.~G.~de~Blok      \inst{\ref{astron} \and \ref{uct} \and \ref{kapteyn}}
\and M.~Bouwhuis          \inst{\ref{nikhef}}
\and R.~H.~van~den~Brink~  \inst{\ref{tricas}}
\and L.~Connor           \inst{\ref{uva} \and \ref{caltech}}
\and A.~H.~W.~M.~Coolen      \inst{\ref{astron}}
\and S.~Damstra           \inst{\ref{astron}}
\and G.~N.~J.~van~Diepen    \inst{\ref{astron}}
\and R.~de~Goei           \inst{\ref{astron}}
\and H.~Dénes            \inst{\ref{astron}}
\and M.~Drost             \inst{\ref{astron}}
\and N.~Ebbendorf         \inst{\ref{astron}}
\and B.~S.~Frank         \inst{\ref{sarao} \and \ref{idia} \and \ref{uct}}
\and D.~W.~Gardenier       \inst{\ref{astron} \and \ref{uva}}
\and M.~Gerbers            \inst{\ref{astron}}
\and Y.~G.~Grange          \inst{\ref{astron}}
\and T.~Grit               \inst{\ref{astron}}
\and A.~W.~Gunst           \inst{\ref{astron}}
\and N.~Gupta             \inst{\ref{iucaa} \and \ref{astron}}
\and M.~V.~~Ivashina       \inst{\ref{chalmers}}
\and G.~I.~G.~J\'ozsa     \inst{\ref{sarao} \and \ref{dperhodes} \and \ref{argelander}}
\and G.~H.~Janssen         \inst{\ref{astron} \and \ref{radboud}}
\and A.~Koster          \inst{\ref{astron}}
\and G.~H.~Kruithof        \inst{\ref{astron}}
\and S.~J.~Kuindersma      \inst{\ref{astron}}
\and A.~Kutkin            \inst{\ref{astron} \and \ref{lebedev}}
\and D.~M.~Lucero         \inst{\ref{kapteyn} \and \ref{virginiatech}}
\and Y.~Maan          \inst{\ref{astron}}
\and F.~M.~Maccagni        \inst{\ref{inaf}}
\and J.~van~der~Marel     \inst{\ref{astron}}
\and A.~Mika              \inst{\ref{astron}}
\and J.~Morawietz         \inst{\ref{astron}}
\and H.~Mulder            \inst{\ref{astron}}
\and E.~Mulder            \inst{\ref{astron}}
\and M.~J.~Norden          \inst{\ref{astron}}
\and A.~R.~Offringa        \inst{\ref{astron} \and \ref{kapteyn}}
\and L.~C.~Oostrum        \inst{\ref{astron} \and \ref{uva} \and \ref{escience}}
\and R.~E.~Overeem         \inst{\ref{astron}}
\and Z.~Paragi            \inst{\ref{jive}}
\and H.~J.~Pepping         \inst{\ref{astron}}
\and E.~Petroff           \inst{\ref{uva} \and \ref{veni} \and \ref{mcgill}}
\and D.~J.~Pisano          \inst{\ref{wvu}}
\and A.~G.~Polatidis       \inst{\ref{astron}}
\and P.~Prasad             \inst{\ref{asml}}
\and J.~P.~R.~de~Reijer     \inst{\ref{astron}}
\and J.~W.~Romein          \inst{\ref{astron}}
\and J.~Schaap            \inst{\ref{astron}}
\and G.~W.~Schoonderbeek   \inst{\ref{astron}}
\and R.~Schulz            \inst{\ref{astron}}
\and D.~van~der~Schuur    \inst{\ref{astron}}
\and A.~Sclocco           \inst{\ref{escience}}
\and J.~J.~Sluman          \inst{\ref{astron}}
\and R.~Smits             \inst{\ref{astron}}
\and B.~W.~Stappers      \inst{\ref{uom}}
\and S.~M.~Straal        \inst{\ref{nyuad} \and \ref{nyuadcfa}}
\and K.~J.~C.~Stuurwold      \inst{\ref{astron}}
\and J.~Verstappen     \inst{\ref{kapteyn} \and \ref{astron}}
\and D.~Vohl              \inst{\ref{astron}}
\and K.~J.~Wierenga        
\and E.~E.~M.~Woestenburg   \inst{\ref{astron}}
\and A.~W.~Zanting         
\and J.~Ziemke            \inst{\ref{astron} \and \ref{oslo}}
} 

%% ----------------- Institutes --------------- 
\institute{ASTRON, the Netherlands Institute for Radio Astronomy, Oude Hoogeveensedijk 4,7991 PD Dwingeloo, The Netherlands\label{astron}
  \and
Kapteyn Astronomical Institute, PO Box 800, 9700 AV Groningen, The Netherlands\label{kapteyn}
  \and
Astronomisches Institut der Ruhr-Universit{\"a}t Bochum (AIRUB), Universit{\"a}tsstrasse 150, 44780 Bochum, Germany\label{airub}
  \and
SKA Organization, Jodrell Bank, Lower Withington, Macclesfield, Cheshire, SK11 9DL, United Kingdom\label{skao}
  \and
Anton Pannekoek Institute, University of Amsterdam, Postbus 94249, 1090 GE Amsterdam, The Netherlands\label{uva}
  \and
CSIRO Astronomy and Space Science, Australia Telescope National Facility, PO Box 76, Epping NSW 1710, Australia\label{csiro}
  \and
Sydney Institute for Astronomy, School of Physics, University of Sydney, Sydney, New South Wales 2006, Australia\label{sydney}
  \and
South African Radio Astronomy Observatory (SARAO), 2 Fir Street, Observatory, 7925, South Africa\label{sarao}
  \and
Department of Physics and Electronics, Rhodes University, PO Box 94, Makhanda, 6140, South Africa\label{dperhodes}
  \and
Argelander-Institut f\"ur Astronomie, Auf dem H\"ugel 71, D-53121 Bonn, Germany\label{argelander}
  \and
Dept.\ of Electrical Engineering, Chalmers University of Technology, Gothenburg, Sweden\label{chalmers}
  \and
The Inter-University Institute for Data Intensive Astronomy (IDIA), and University of Cape Town, Private Bag X3, Rondebosch, 7701, South Africa\label{idia}
  \and
Department of Astronomy, University of Cape Town, Private Bag X3, Rondebosch 7701, South Africa\label{uct}
  \and
Veni Fellow\label{veni}
  \and
Department of Physics, McGill University, 3600 rue University, Montreal, QC H3A 2T8, Canada\label{mcgill}
  \and
Tricas Industrial Design \& Engineering, Zwolle, The Netherlands\label{tricas}
  \and
Netherlands eScience Center, Science Park 140, 1098 XG, Amsterdam, The Netherlands\label{escience}
  \and
Astro Space Center of Lebedev Physical Institute, Profsoyuznaya Str. 84/32, 117997 Moscow, Russia\label{lebedev}
  \and
ASML Netherlands B.V., De Run 6501, 5504 DR Veldhoven, The Netherlands\label{asml}
  \and
Cahill Center for Astronomy, California Institute of Technology, Pasadena, CA, USA\label{caltech}
  \and
Jodrell Bank Centre for Astrophysics, Department of Physics and Astronomy, University of Manchester, Manchester M13 9PL, UK\label{uom}
  \and
Joint Institute for VLBI ERIC, Oude Hoogeveensedijk 4, 7991 PD Dwingeloo, The Netherlands\label{jive}
  \and
INAF, Osservatorio Astronomico di Cagliari, via della Scienza 5, 09047 Selargius, CA, Italy\label{inaf}
  \and
Inter-University Centre for Astronomy and Astrophysics, Post Bag 4, Ganeshkhind, Pune 411 007, India\label{iucaa}
  \and
NYU Abu Dhabi, PO Box 129188, Abu Dhabi, United Arab Emirates\label{nyuad}
  \and
Center for Astro, Particle, and Planetary Physics (CAP$^3$), NYU Abu Dhabi, PO Box 129188, Abu Dhabi, United Arab Emirates\label{nyuadcfa}
  \and
NIKHEF, National Institute for Subatomic Physics, NL-1098 XG Amsterdam, the Netherlands\label{nikhef}
  \and
Department of Astrophysics/IMAPP, Radboud University, PO Box 9010, 6500 GL Nijmegen, The Netherlands\label{radboud}
  \and
West Virginia University, White Hall, Box 6315, Morgantown, WV 26506\label{wvu}
  \and
Department of Physics, Virginia Polytechnic Institute and State University, 50 West Campus Drive, Blacksburg, VA 24061, USA\label{virginiatech}
  \and
University of Oslo Center for Information Technology, P.O. Box 1059, 0316 Oslo, Norway\label{oslo}}

 \abstract {We describe the APERture Tile In Focus (Apertif) system, a phased array feed (PAF) upgrade of the Westerbork Synthesis Radio Telescope which has transformed this telescope into a  high-sensitivity, wide field-of-view L-band imaging and transient survey instrument. Using novel PAF technology, up to 40 partially overlapping beams can be formed on the sky simultaneously, significantly increasing the survey speed of the telescope. With this upgraded instrument, an imaging survey covering an area of 2300 deg$^2$ is being performed which will deliver both continuum and spectral line data sets, of which the first data has been publicly released. In addition, a time domain transient and pulsar survey covering 15\,000~deg$^2$ is in progress. An overview of the Apertif science drivers, hardware and software of the upgraded telescope is presented, along with its key performance characteristics.}

\titlerunning{Apertif, Phased Array Feeds for the Westerbork Telescope}
\authorrunning{W.A. van Cappellen et al.}
\maketitle
%%-------------------------------------------------------------------
\section{Introduction}\label{sec:introduction}
Since the start of radio astronomy, the study of the sky at radio wavelengths has provided a unique view of a diverse range of physical phenomena occurring  in our own Milky Way, in external galaxies, and even out to the edge of the observable Universe.  This has been done by tracing both the radio continuum emission as well as the  emission and absorption from radio spectral lines,  in particular the 21-cm line of  atomic neutral hydrogen (hereafter \HI).
This exploration of the radio sky could not have been done without the availability of large surveys. The legacy value of these surveys for the entire astronomical community is, therefore, huge (see e.g.\ \citealt{Norris17} for an overview on continuum surveys and \citealt{Giovanelli15} for \HI\ surveys). 

A number of blind large-area high-resolution continuum surveys  are currently available. The main ones, covering large fractions of sky, include  the  NRAO VLA Sky Survey (NVSS, \citealt{1998AJ....115.1693C}), the Faint Images of the Radio Sky at Twenty cm (FIRST, \citealt{1994ASPC...61..165B}), the Sydney University/Molonglo Sky Survey (SUMSS, \citealt{1999AJ....117.1578B}), the  TIFR GMRT Sky Survey (TGSS, \citealt{2016arXiv160304368I}),  the Westerbork Northern Sky Survey (WENSS,  \citealt{1997A&AS..124..259R}) and, now on-going, the LOFAR \citep{lofar2013} Two-metre Sky Survey (LoTSS, \citealt{Shimwell19}), the LOFAR LBA Sky Survey (LoLSS, \citealt{deGasperin2021}), the Evolutionary Map of the Universe survey (EMU, \citealt{EMU2011}), the  Galactic and Extra-Galactic All-sky MWA survey (GLEAM, \citealt{Gleam2015}), and the VLA Sky Survey (VLASS, \citealt{vlass2020}). The list of topics they have enabled to be addressed is far too long to be described in detail, but they include star formation and nuclear activity in galaxies, magnetism across cosmic time, the life-cycle of radio AGN and their impact on the evolution of galaxies, and detailed studies of individual galaxies and clusters of galaxies (see e.g.\ \citealt{Norris17,Prandoni15} and references therein). 

Large-area \HI\ surveys have been more time consuming to perform due to the low survey speed for spectral line work of existing interferometers. Therefore the main surveys have been done with single-dish telescopes and have low spatial resolution. The largest \HI\ surveys are the \HI\ Parkes All Sky Survey (HIPASS, \citealt{Barnes01}) and the Arecibo Legacy Fast ALFA Survey (ALFALFA, \citealt{Haynes18}). The main results from these surveys focus on the role of neutral hydrogen in galaxy evolution and the relation between gas content and star-formation (e.g.\ \citealt{Catinella18}, see \citealt{Giovanelli15, Blyth15} for overviews). These surveys have been performed with single-dish telescopes, i.e.\ at relatively low spatial resolution. This has limited the science questions  they can address.

Although the above surveys have led to many important results, one has to maintain the impetus of discovery by continuously improving the performance of telescopes so that new surveys with better characteristics can be performed. A number of recent technological developments have opened up opportunities for major steps forward by increasing the survey speed of radio interferometers  enabling more powerful large radio surveys at high spatial resolution at frequencies around 1 GHz. 
Particularly relevant is the development  of  phased-array feeds (PAF) with low cross talk and low noise. In addition, the increased processing power of modern computing hardware (e.g.\ for  beam forming), together with the capability of transporting and handling very large data streams, have  made it possible to build low-noise PAF systems which combine broadband coverage at high frequency resolution and good sensitivity, with an increased field of view.  This combination offers a significant improvement in survey speed,  for both spectral line and continuum observations, giving unprecedented possibilities for carrying out a new generation of radio surveys. 
Such a step is also clearly a top priority on the way to the Square Kilometre Array (SKA) telescope, as can be seen from the SKA science case \citep{Braun15}.  

Taking advantage of these recent technological developments, the \ac{Apertif} PAF system was developed for  the Westerbork Synthesis Radio Telescope (WSRT) and completed in 2019.
Apertif aims at expanding the capabilities of the telescope, with the main characteristic allowing it to image large areas of the sky a factor 39 faster than before with an increase in survey speed of a factor 15 (see Sect.\ \ref{sec:performance}).
 
In this paper, we provide an overview of the Apertif system, including key elements of its design, and its performance characteristics. Section~\ref{sec:wsrt} provides a brief introduction of the WSRT. Section~\ref{sec:earlypafs} provides a historical perspective on phased array feed technology and describes how the idea of Apertif was born. The science drivers and the system architecture of the upgraded WSRT are provided in Secs.~\ref{sec:sciencedrivers} and \ref{sec:systemoverview}. In Sec.~\ref{sec:rfi} the Radio Frequency Interference (RFI) environment is described, followed by a description of the phased array feed in Sec.~\ref{sec:paf}, which includes the antenna array, receivers and digital beam forming. The upgraded central processing (correlator and data writer) is presented in Sec.~\ref{sec:centralprocessing}, and the monitoring and control systems in Sec.~\ref{sec:mac}. Sections~\ref{sec:alta} and \ref{sec:operations} present the long-term archive and the operational model, followed by the measured system performance in Sec.~\ref{sec:performance}.

\begin{figure*}
	\centering
	\includegraphics[width=17cm]{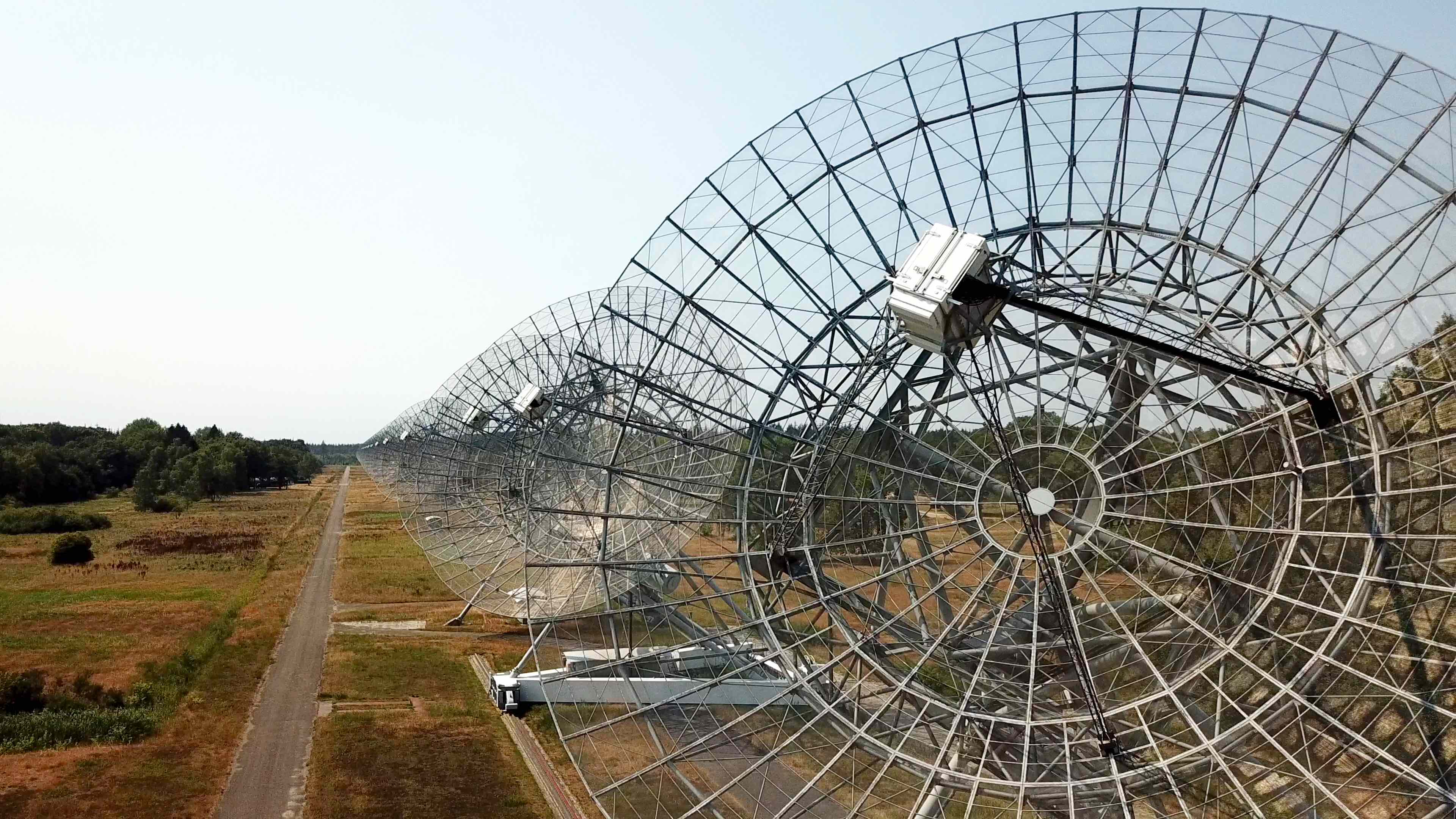}
	\caption{Dishes of the Westerbork Synthesis Radio Telescope.}
	\label{fig:WSRTphoto}
\end{figure*}

\section{The Westerbork Synthesis Radio Telescope}\label{sec:wsrt}
The Westerbork Synthesis Radio Telescope (WSRT) was designed and built in the 1960s and was officially inaugurated in 1970. It was one of the first few telescopes that explored the use of dish-based interferometry. An overview of the rich history of discoveries and technical innovations of this radio telescope over the past 50 years can be found in \citet{WSRT50book}. 

The WSRT consists of a linear east-west array of 14 identical steerable 25-m dishes on an equatorial mount, with a focal ratio f/D = 0.35 (Fig.~\ref{fig:WSRTphoto}). Instead of a solid surface, the reflectors use a metal mesh surface to reduce weight and wind load. The feed is located in a cabin at the primary focus of each reflector. Given that the WSRT is a linear array, every observation requires a full 12-hour period in order to exploit the rotation of the Earth to image an area on the sky.  
The Apertif system  uses 12 of the 14 dishes (see Sect.\ \ref{sec:uvcoverage}), with the remaining two  dedicated to other projects, for example as part of the European VLBI Network (EVN) .  The location of ten dishes is fixed. The remaining four are on a rail track and can be re-positioned to adjust the uv coverage of the interferometer, although this possibility is not used currently.  The shortest baseline of the array is 36 m while the longest is 2412 m.  At the observing frequency of Apertif this gives a spatial resolution of about 12 $\times$ $12/\sin\delta$ arcsec with $\delta$ the declination of the observation.  The field of view of a single WSRT dish with a single receiver has a full width at half maximum of about 35 arcminutes at 1.4~GHz.  
The installation of the Apertif system was combined with a major refurbishing (cleaning and painting) of the metal structures that support the dishes, and the electronics that drive the dish pointing motors and the angular encoders have been replaced.

\section{Early phased array feed experiments on the WSRT}\label{sec:earlypafs}
For many years, reflector antennas with multiple feeds used electrically large feed antennas (such as horns), providing widely separated beams on the sky. Research by Ron Ekers and Rick Fisher in the late 1970s and early 1980s provided the notion that the electromagnetic fields in the focal plane of a reflector antenna carry the information to correct for distortions of the reflector surface, and/or observe a contiguous field of view that is much larger than the single beam provided by traditional feed antennas. However, the antenna technology to sample the complete electro-magnetic field in the focal plane and the processing algorithms to form multiple closely spaced, or overlapping, beams in relatively deep reflectors was not available then. In the mid-1990s, ASTRON initiated a research program towards the Square Kilometre Array, which triggered research into dense aperture arrays, such as phased arrays with an electrical spacing smaller than half a wavelength. Reflector antennas with array feeds were not actively pursued in the ASTRON program, but were identified as ''hybrid solutions''. Amongst others, Rick Fisher continued investigations into wide field of view reflector feeds, and how the output of these feeds should be processed. In 1998 ASTRON started a  collaboration with Dan Schaubert (University of Massachusetts) on arrays of  tapered-slot or ''Vivaldi'' antennas. Originating from defense applications, such arrays demonstrated a constant effective area over a wide frequency range \citep{schaubert2000,schaubert2003}. Besides their application in aperture arrays, it was realised that the dense sampling enabled by the Vivaldi array would be a game-changer when placed as antenna array in the focal plane of a reflector as well (van Ardenne, 2020, priv. comm.). This application, combined with the research on the noise behaviour of such arrays \citep{Weem2001}, formed the technological basis of the European research program ''FARADAY'' (2001--2004). The FARADAY project experimentally demonstrated that a densely packed phased array feed could form multiple overlapping high-efficiency beams on the sky, using a Vivaldi array and an RF beam former at 2--5 GHz \citep{crimico02,doi:10.1117/12.554895}.

Triggered by the success of the FARADAY project, the PHAROS project \citep{4584487} continued the WSRT PAF research with the objective of demonstrating low-noise PAF operation. For this purpose the first cryogenically cooled PAF was designed, operating in the 4--8 GHz band and combining 24 active elements into 4 high-efficiency RF beams (13 elements per beam). 

The results of these early PAF experiments confirmed the potential of PAFs for the WSRT. A full-scale demonstration of PAF capabilities would require an advanced high-speed digital backend, which led to the definition of a series of prototypes known as DIGESTIF \citep{PAST2010}. The \mbox{DIGESTIF} system provided the ultimate demonstration of L-band PAF technology for Apertif on the WSRT.

%--------------------------------------------------------------------
\section{Science drivers}\label{sec:sciencedrivers}
Two major imaging surveys are performed with Apertif, a relatively shallow wide-area imaging survey (using a single 12-hour observation per pointing) covering an area up to 2300 deg$^2$, and  a  medium-deep survey (using $10 \times 12$ hour per pointing) covering up to 150 deg$^2$. Both these surveys will deliver continuum, polarisation, and spectral line data sets. The first data of these surveys have been publicly released\footnote{http://hdl.handle.net/21.12136/B014022C-978B-40F6-96C6-1A3B1F4A3DB0} (see also Adams et al.\ in prep). In addition, a time domain transient and pulsar survey covering 15\,000 deg$^2$ is performed. 

The areas covered by the Apertif surveys have primarily been chosen such that  there is good overlap with regions that have important publicly available ancillary spectroscopic and imaging data, such as the data from  the Sloan Digital Sky Survey  (SDSS)\footnote{www.sdss.org},  the Panoramic Survey Telescope and Rapid Response System survey (Pan-STARRS)\footnote{www.ifa.hawaii.edu/pswww},  the  Hobby-Eberly Telescope Dark Energy Experiment (HetDex)\footnote{www.hetdex.org},  the Herschel-Atlas project\footnote{www.h-atlas.org}, and  the Mapping Nearby Galaxies at APO (MaNGA)\footnote{www.sdss.org/surveys/manga} and the WHT Enhanced Area Velocity Explorer (WEAVE)\footnote{www.ing.iac.es/weave/about.html} spectroscopic surveys. However, because  the imaging surveys are focused on extra-galactic science,   regions of low Galactic latitude are avoided. This, combined with the fact that in order to obtain complete $uv$ coverage with Apertif,  observations over the full hour angle range from --6 h to + 6 h are needed, efficient scheduling of the survey observations puts strong constraints on the location of the survey areas. As a result, for part of the area covered by the imaging surveys  only limited additional data from other wavebands are available.
In addition, survey observations are scheduled to be 11.5 h in duration instead of the standard 12 h, to allow for calibration and telescope movement in between observations.

Below we briefly describe the main goals of the imaging surveys. More details can be found in Hess et al.\ (in preparation), which also describes the observation scheduling, and the associated constraints.

\subsection{H\,{\small I} and OH surveys}
\label{HIsurveys}
 
The two recent extragalactic \HI\ emission surveys, HIPASS and ALFALFA, have detected over 25\,000 galaxies and are a major resource for statistical studies of the \HI\ properties of galaxies. However, these surveys were carried out with single-dish telescopes and, although sensitive, they have very limited angular resolution and therefore the large majority of objects detected are spatially unresolved.  The Apertif \HI\ surveys aim to make the next step by providing the spatially resolved \HI\ properties for a large number of galaxies for studies of galaxy structure and evolution.

Detailed spatially resolved studies of the \HI\ emission in galaxies has been one of the key science areas of the WSRT since the 1970s and these have greatly contributed to improving our understanding of galaxies. However, although these studies have been very successful,  they have been limited to targeted observations of relatively small samples of galaxies in the nearby Universe, with the most recent examples being     the WHISP \citep{Hulst01}, the Atlas$^{\rm 3D}$ \citep{Serra12}, and the HALOGAS \citep{Heald11} surveys.

The goal of Apertif is to expand such imaging studies to much larger, blindly selected samples of galaxies in the nearby Universe. This is possible thanks to the improved survey speed, the  high spatial resolution and the high sensitivity of Apertif. The  5-$\sigma$ column density detection limit reached in the data cubes of the wide-area survey at full resolution for a profile width of 20 \kms\ is $\sim$1--2 $\times 10^{20}~\rm cm ^{-2}$ (depending on declination) and is just below  $10^{20}~\rm cm ^{-2}$ for the medium-deep survey.  Because of the specific array layout of the WSRT, at somewhat lower resolution ($30''$)  the column density sensitivities are about a factor 3 better. The sensitivity and resolution of the Apertif wide-area \HI\ survey are similar to that of the wide-area \HI\ survey planned with the Australia Square Kilometre Array Pathfinder (ASKAP,  \citealt{Wallaby2020}).

\begin{figure}
    \centering
    \includegraphics[angle=0,width=9cm]{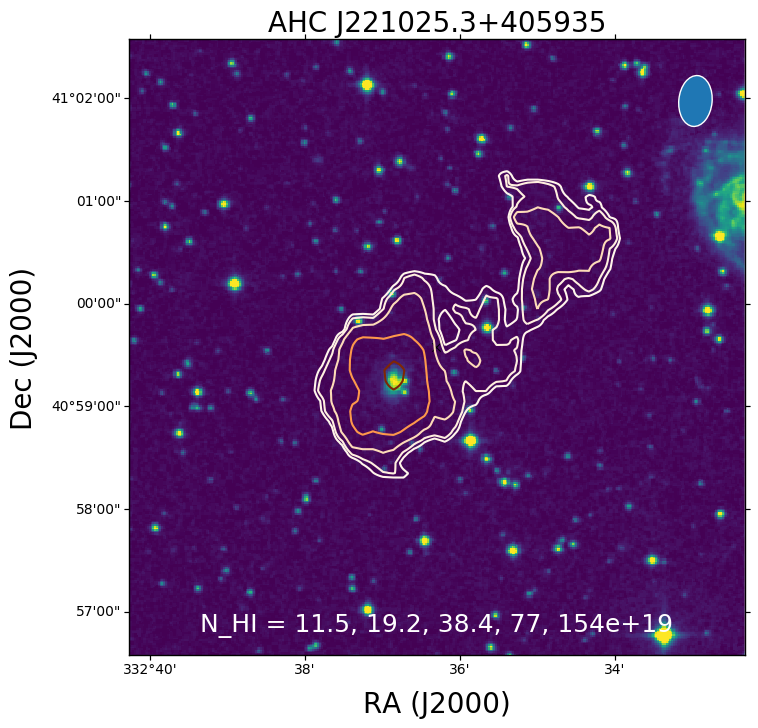}
    \caption{Example of \HI\ emission (contours) detected in a small galaxy group (colour scale) in one of the Apertif pointings of the \HI\ wide-area survey (Hess et al.\ in prep). The contour levels correspond to 3, 5, 10, 20, 40  times the noise level.}
    \label{fig:HIgroup}
\end{figure}

One of the main aims of the \HI\ emission survey is to better understand the processes that govern  the removal and accretion of gas for galaxies of all types and sizes and the role this plays in the evolution of galaxies (see e.g.\ \citealt{2008A&ARv..15..189S} for a review). 
This requires  spatially resolving the \HI\ properties of galaxies and the gas around them in different environments, including isolated areas such as voids,  the numerous groups of galaxies that exist in the nearby Universe, and  the most extreme galaxy densities in clusters. The observed properties of galaxies strongly depend on the cosmic environment in which they are located, implying that the  environment   plays an important role during a galaxy's formation  as well as during the subsequent evolution.  A variety of physical mechanisms affect the growth and shaping of galaxies such as accretion, merging, tidal interactions, and ram-pressure stripping.  
Resolved \HI\ emission studies are particularly sensitive to trace  these processes and  an example of such a case found  with Apertif is shown in Fig.\ \ref{fig:HIgroup}. The availability of such information for a large number of galaxies in different environments  will provide important new information on how galaxies evolve.

Another main topic for the Apertif \HI\ survey is the search for the smallest gas-rich galaxies in the local Universe. Such objects are sensitive probes of the limits of galaxy formation. Very few of such objects are known (e.g.\ Leo T, \citealt{2008MNRAS.384..535R}, and Leo P, \citealt{Giovanelli13}). The visible mass of such systems is completely dominated by \HI\ and only a small number of stars is present. This is possibly the result of such galaxies being almost too small to compress and cool the gas to form stars and suggests that there is a lower limit to the size of a galaxy that can form stars.   How such objects formed and evolve, and whether indeed lower mass limit for star bearing galaxies exists, critically depends on the detailed conditions of normal and dark matter  in the early Universe, and the predicted number of such small gas-rich galaxies in the local Universe differs greatly  between different theories of galaxy formation.     It is therefore important to improve our statistical knowledge on the occurrence of such systems.
The large area covered by the  \HI\ surveys is particularly suited for making an inventory of these objects in the local Universe which will provide important constraints of theories of galaxy formation.

The imaging surveys will also provide information on \HI\ observed in absorption against relatively strong continuum sources. Given the frequency (and, therefore, redshift), Apertif is particularly relevant tracing \HI\ in absorption in the circumnuclear regions of active galactic nuclei (AGN; see \citealt{Morganti18} for a review). The presence and kinematics of the \HI\ in these objects can trace circumnuclear disks, gas accreting and fuelling the central super massive black hole, as well as the presence of gas outflows that can be driven by radio jets (see e.g.\ \citealt{Morganti05}). 
The latter can provide information on how the AGN is impacting galaxy evolution. 
The WSRT has already given  important contributions to such studies (e.g.\  \citealt{Morganti05,2015A&A...575A..44G}) and has also provided one of the largest available surveys of  \HI\ absorption in the nearby Universe (\citealt{Maccagni17}). However, these studies are based on pointed observations  of radio sources for which the redshift is known. Such a pre-selection can introduce biases in the results (see e.g.\ \citealt{Chowdhury20}).
Hence, the main step forward by Apertif will be to instead provide a blind inventory of the presence of \HI\ in absorption without this a priori target selection. Thus, it will be possible to  inspect  continuum radio sources detected by Apertif for absorption regardless of other source properties.

The sensitivity for H\,{\small I} absorption  depends on the strength of the background radio continuum and the optical depth of the H\,{\small I} line, which in turn depends on the spin temperature and the column density of the H\,{\small I} gas. 
The shallow \HI\ survey will reach an optical  depth of  $\tau \sim 0.03$ (for 30 km s$^{-1}$ velocity resolution) against continuum sources stronger than $\sim$100 \mJybeam, well suited for tracing atomic neutral hydrogen in active galactic nuclei (see e.g.\ \citealt{Morganti18}). For brighter sources, Apertif will also allow to look for broad shallow absorption, such as those indicating the presence of fast outflows (\citealt{Morganti05,2015A&A...575A..44G}).
   
The main limitation will come from the relatively hostile environment with regard to radio frequency interference at the lower frequencies covered by Apertif (i.e.\ below 1290 MHz, see Sec.\ \ref{sec:rfi}).
The search for \HI\ absorption out to $z=0.25$ is expected to deliver a few hundred new detections of gas associated with radio galaxies, greatly expanding the parts of parameter space of radio AGN where the presence of \HI\ absorption can be studied.  A comparison with other \HI\ absorption surveys from SKA pathfinders and precursors is discussed in \cite{Maccagni17} and \cite{Morganti18}.

Finally,  Apertif also allows to do interesting large-area surveys for OH megamasers, and in particular of their redshift evolution. The  main spectral lines of OH megamasers are emitted at 1665 and 1667 MHz, with two satellite lines at 1612 and 1720 MHz. Therefore, given the observing band of Apertif, the main lines can be detected out to redshift $z\sim 0.3$.
OH megamasers are found in luminous- and ultra-luminous infrared galaxies  and are associated with galaxy mergers and their intense nuclear star formation activity. This, and the fact that the redshift of the OH megamaser comes automatically from a detection, make them a useful tracer of the merger history in the Universe. In addition, with the broad observing band of Apertif it is also possible to measure the line rations between the main and the satellite lines for a subset of the sources, which gives useful information about the  physical conditions in the star forming regions. The results on the first OH megamaser found with Apertif have  been published \citep{Hess2021}.

\subsection{Continuum surveys}

%------------------ Comparison continuum surveys 
\begin{figure}
    \centering
    \includegraphics[angle=0,width=9cm]{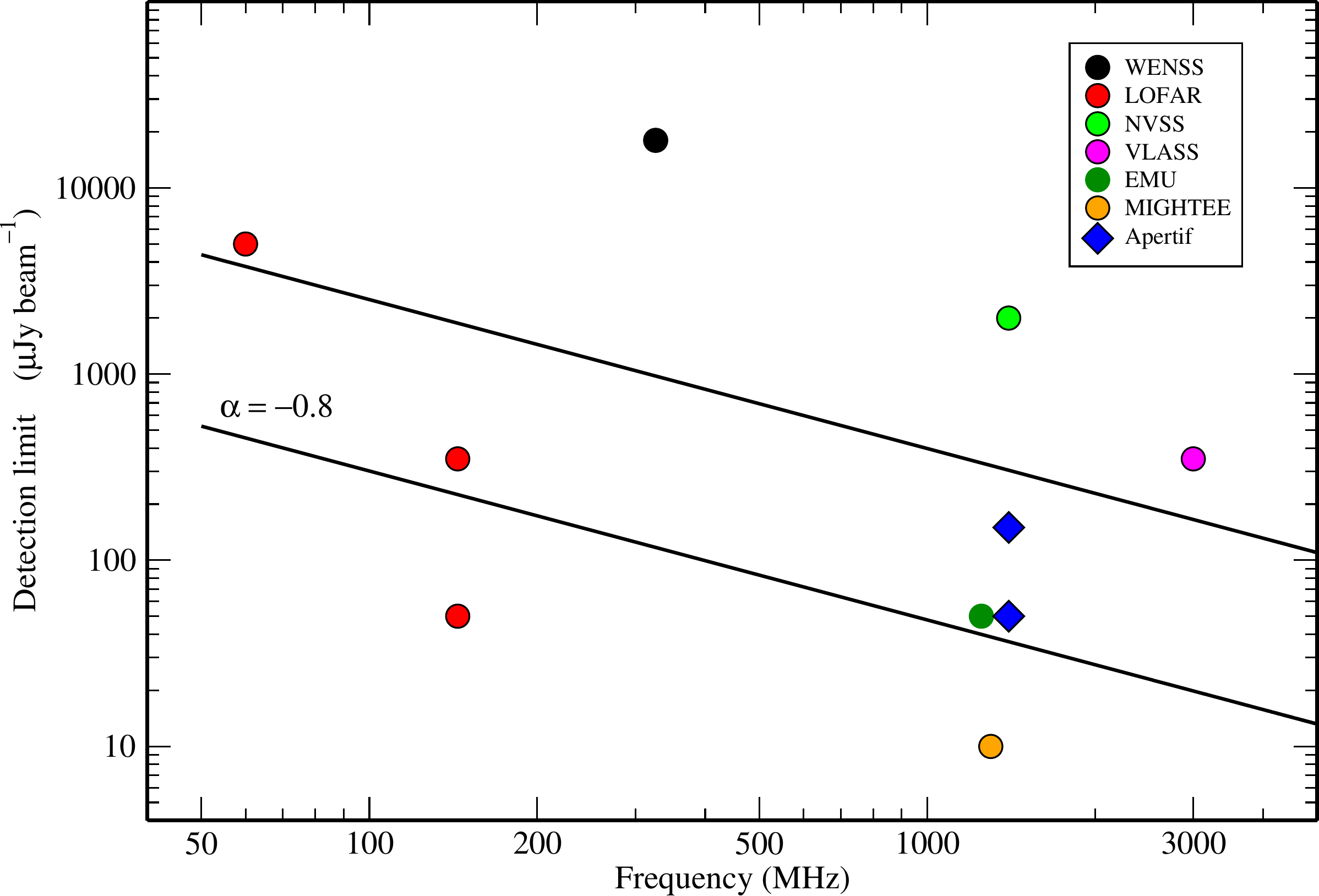}
    \caption{Comparison of the 5-$\sigma$ detection limits of a number of radio continuum surveys and the Apertif surveys. The lines indicate a spectral index $\alpha=-0.8$ which is typical of extragalactic radio sources. The noise for the EMU survey was taken from \citet{Norris17}. The two LOFAR High-Band Antenna (HBA) points refer to LoTSS \citep{Shimwell19} and to the LOFAR deep fields \citep{Tasse2021}. The latter cover limited areas (of a few tens of square degrees). The MIGHTEE survey, although very deep,  only covers a limited number of square degrees.}
    \label{fig:Surveys}
\end{figure}
%--------------------------------------

%------------------ Continuum Image one pointing 
\begin{figure*}
    \centering
    \includegraphics[angle=-90,width=20cm]{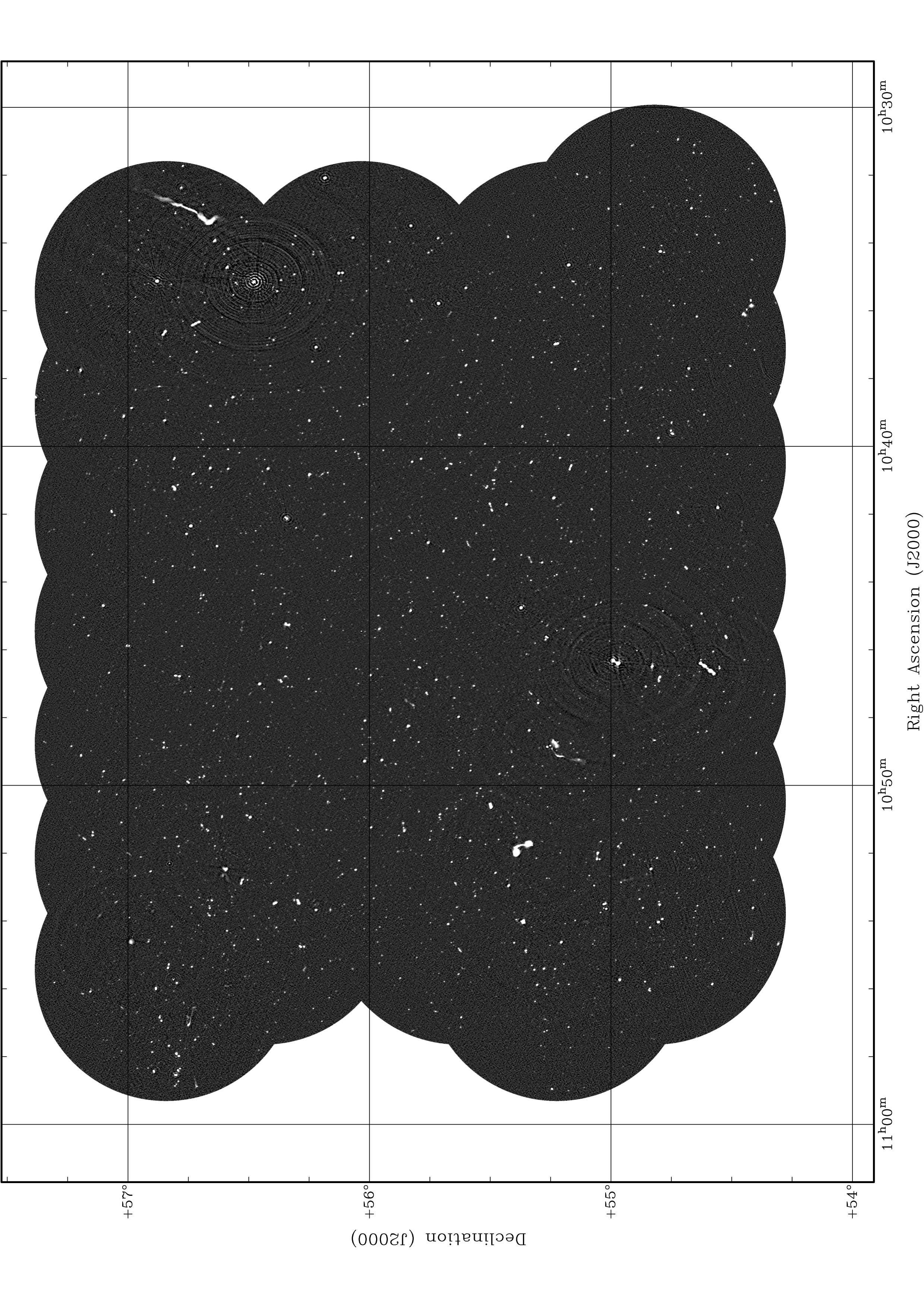}
    \caption{Continuum mosaic obtained from one 11.5-h Apertif observation, centred on the Lockman Hole region. A number of extended sources are visible  (including a giant radio galaxy on the top right). The noise level is about 35 $\mu$
    Jy beam$^{-1}$, except close to a few very bright sources where calibration artefacts are still visible. See  Sec.\ \ref{sec:performance} for more details. }
    \label{fig:ContinuumPointing}
\end{figure*}
%--------------------------------------
The images and  source catalogues resulting from large-area radio continuum surveys like those listed in the introduction, are among the most important databases for many branches  of astronomy. 
The Apertif continuum surveys bring an important addition to this list. A comparison between various surveys is shown in Fig.\ \ref{fig:Surveys}. 
The continuum surveys with Apertif will provide images and catalogues at 1.4~GHz with a resolution of $12^{\prime\prime}/\sin\delta$  and reaching, in the case of the wide-area survey,  a noise level about 20 times lower than NVSS and with 3 times higher angular resolution (see Sec.\  \ref{sec:sensitivity}). An example of the image obtained from a single Apertif pointing is shown in Fig.\ \ref{fig:ContinuumPointing}. 

The data from the Apertif continuum surveys can be used to address a wide range of topics such as galaxy evolution, source counts, the life-cycle of radio galaxies and the polarised properties of radio sources. Furthermore, for star-forming galaxies, the combination of the radio continuum and the \HI\ emission, which are simultaneously obtained, will allow to connect star formation and gas properties. 

A unique aspect is the synergy with the LOFAR surveys, both the  LoTTS at 150~MHz \citep{Shimwell19} as well as the LOFAR  60~MHz survey \citep{deGasperin2021}, because both LOFAR and Apertif observe in the northern sky and there is large overlap between their survey areas. A fortunate circumstance is that  the relative sensitivities of LOFAR and Apertif match the spectrum of  typical radio sources and the spatial resolutions are similar (see  Fig.\ \ref{fig:Surveys}). The combination Apertif-LOFAR will be particularly useful for identifying and studying radio sources in extreme phases of their evolution, such as remnants and restarted radio sources, and diffuse relic and halo emission in cluster of galaxies. For  resolved sources this will instantly yield spectral index and spectral curvature maps which are a very rich source of information to constrain many physical parameters and processes and are powerful tools to study radio sources with extreme radio spectra.

A proof of concept study of work  based on combining Apertif data with LOFAR images is presented by \cite{Morganti20}.
A significant  number of  radio sources in the remnant and restarted phases were found. This has provided new  constraints on the timescales of the duty-cycles of AGN activity and has confirmed  results from numerical simulations.  Given that this was based on only one Apertif pointing, in the Lockman Hole region, it already shows the potential of combining radio images at 150 and 1400 MHz. 

The continuum images from the medium-deep survey will reach the confusion limit and this is particularly interesting, among other topics, for investigating  radio source counts. 
Similar studies have been carried out using the WSRT (e.g.\ \citealt{Prandoni18,Bonato20}), but only for limited areas. 
Radio source  counts are tightly related to the evolutionary properties of the sources  \citep{2016A&ARv..24...13P,Prandoni18,Norris17} and are mostly tied to the faint end of the source population. Up to now, deep observations investigating this population were limited to small fields where local density variations may influence the results. The large survey area of the Apertif surveys will allow mitigation of these local effects. 

Polarised radio continuum emission contains the information about the magnetic field within the observed sources and along the line-of-sight between the observer and the sources themselves. The two components of magnetic field (perpendicular to the line-of-sight and parallel to it) can be recovered using the observed polarisation angle and the rotation measure respectively. These two quantities allow a reconstruction of the three dimensional morphology of the magnetic field. The accuracy of the reconstructed magnetic field properties is dependent on the number of detected polarised sources per area on the sky. The most recent reconstruction of the Milky Way magnetic field structure by \citet{2012A&A...542A..93O} is based on the NVSS and therefore limited by the detected number density of sources ($\sim$1/deg$^2$; \citealt{2009ApJ...702.1230T}). First analysis of pointings from the Apertif wide-area survey shows number densities of 30-40 sources/deg$^2$ allowing an improved reconstruction of the Milky Way magnetic field and of a number of large extragalactic nearby objects.

Additional information on the magnetic field along the line-of-sight and in the polarised sources themselves is their polarised spectrum. It is known that most sources become severely depolarised below 1 GHz. Most models and analytical calculations suggest a $\lambda^2$- to $\lambda^4$-dependence of the degree of polarisation on wavelength $\lambda$ \citep{1998MNRAS.299..189S}, but accurate spectra have only been measured for a few sources \citep{2015MNRAS.450.3579S}. The synergy of the LOTSS survey with the Apertif wide-area survey allows a first statistical analysis of the polarised spectrum of radio sources. Due to the strong depolarisation at LOFAR wavelengths, all polarised sources detected in LOTSS will be detected by Apertif. The combination of this information will for the first time allow an investigation of the  depolarisation mechanisms and the magnetic field properties causing it.

Information about the evolution of the magnetic field over cosmic timescales is  contained in  polarised source counts. Currently such analyses are limited to the wide and shallow data of the NVSS \citep{2002A&A...396..463M,2004MNRAS.349.1267T,2014ApJ...787...99S} or to small fields of a few square degrees with sensitivities comparable to those obtainable with Apertif  \citep{2007ApJ...666..201T,2010ApJ...714.1689G,2010MNRAS.402.2792S,2014MNRAS.441.2555H}. All these latter studies show a rise in the degree of polarisation towards fainter sources, but the scatter is of the order of a few in fractional polarisation and number of detected sources. The measurements are most likely influenced by the small number of detected sources  and by local variations due to the small observing area. The Apertif surveys will allow us to trace a sufficient number of polarised sources ($\sim10^5$) over several thousand square degrees down to flux densities of several tens of $\upmu$Jy. This allows for the first time an accurate characterisation of the polarised sky. To study the evolution of the magnetic field over cosmic timescales, information on the redshifts of sources is needed. The synergy of the WEAVE survey with the Apertif and LOFAR radio surveys, in combination with the number of sources Apertif will detect, allows us to investigate redshift dependent polarised source counts. This is a huge step in determining the evolution of the magnetic field up to redshifts of $z\approx2$.

\subsection{Time domain surveys}

Many transient and time variable phenomena are observable in the radio sky, 
including radio afterglows of supernovae, gamma-ray bursts and gravitational-wave events, 
on $\sim$month timescales. % \citep{2015A&ARv..23....3D,2007RSPTA.365.1241V}, 
Its wide field of view and high sensitivity make Apertif a powerful instrument for finding such imaging transients \citep{2021A&A...650A.131B}.
Faster transients include solar radio bursts lasting seconds to minutes, % \citep{1984SoPh...90..111B,2018A&A...615A..89Z}, 
and radio pulses from Galactic pulsars, some as short as microseconds. % \citep{lk05}.  
However, the most recent and tantalising addition to this time-domain transient list was the discovery of bright ($\sim$Jy), short ($\sim$ms) radio pulses of extragalactic origin by \citet{Lorimer2007}, sources now known as \acfp{FRB}. 

Since their discovery in 2007, the study of FRBs has become a well-developed field in its own right, with more than 100 sources now reported \citep[see][for a review]{phl19}. The origins of these radio bursts are not yet known, but theories invoking young, highly magnetised neutron stars, black holes, and the explosions involved in their birth have gained traction \citep[for a theory review, see][]{pww+19}. A subset of the published FRB sources has been observed to repeat, supporting stable or non-cataclysmic progenitor theories \citep{spitler2016,chime19-8repeaters,2020arXiv200103595F}.

The first FRBs were serendipitously discovered in all-sky surveys for radio pulsars. The systems necessary to find FRBs
require high time resolution ($\lesssim 1$ms) and at least modest fractional bandwidth ($\Delta \nu/\nu \gtrsim 0.25$). The Berkeley Parkes Swinburne Recorder (BPSR) employing single pulse search software on \acp{GPU} at the Parkes radio telescope proved particularly effective in early FRB searches \citep{kjv+10,Thornton2013,Champion2016}.

Increasingly, dedicated search hardware and software have been developed with the primary goal of discovering (and in some cases precisely localising) large numbers of new FRB sources; many with the stated goal of discovering new FRB pulses in real time. The majority of these systems employ \acp{GPU}, \acp{FPGA} and other specialised hardware on dedicated compute clusters to perform the most computationally expensive tasks such as dedispersion 
\citep{2012ApJ...744..109S,2016MNRAS.458..718C,chime2018overview,2018ApJS..236....8L}.

Despite the rapid advance of the field in recent years, many outstanding challenges remain in understanding the FRB
phenomenon. Despite the high all-sky rate of FRBs of several thousand per day \citep{Lawrence2017}, many early
instruments (such as Parkes) reported a new event only every few months due to the limited 
\ac{FoV} inherent to single dishes or single receivers. Some of the most basic properties of the FRB population remain unknown; the underlying distributions of the population in pulse duration, dispersion measure, distance, energy, and spectral structure have all been difficult to pin down. This is partly due to a small sample size, but also in part due to insufficient instrumental resolution in the case of the pulse duration, dispersion measure, and spectral structure distributions.

Recent research was able to preserve data which capture the Stokes parameters of new FRBs and interpret the polarisation
properties of individual bursts for a growing fraction of the population. Many of these bursts have been seen to be
highly linearly polarised \citep{msh+18,bannister19,2020arXiv200103595F}, and analysis of the linear polarisation in
some cases has revealed the presence of a strong magnetic medium local to the FRB source \citep{msh+18}. Polarisation
may be critically important for understanding the emission and local environment of FRBs. However, preserving
polarisation information requires a survey to either keep full-polarisation data for all survey observations, to be
analysed post-detection; or to search data in real-time while the full-polarisation data remains in a memory buffer, to
be excised and saved for a burst. Each choice comes with its own set of challenges for data storage and real-time data processing power, respectively. 

A subset of the FRB population has been observed to repeat, producing multiple bursts at the same (or very similar) dispersion measure, with some sources now observed to repeat over many years
\citep{spitler2016,2020A&A...635A..61O}. Sustained follow up of an FRB source is needed to eventually detect repeats. However, at present only 20 repeating sources have been published and the overall fraction of repeating sources in the FRB population is unknown. Whether all FRBs eventually repeat is an open question, one that is being tackled currently both with observational efforts \citep{2020arXiv200103595F} 
and modelling \citep{2019A&A...632A.125G,2021A&A...647A..30G}.

In addition to the challenges of understanding the underlying population(s) of FRBs, their physical properties, and
their progenitors, there are also technical challenges involved in their discovery at scale. With next generation
telescopes such as the Square Kilometre Array (SKA) it will no longer be possible to preserve the raw survey data for
offline searches \citep{2015aska.confE..55M}. Instead, new sources will need to be identified in real time to capture
the telescope data for later analysis. New automated FRB search techniques and pipelines taking advantage of
classification and machine learning tools are being developed to prepare for this future reality \citep{cl18}. 

To address all these challenges, new FRB search efforts are increasingly employing interferometers to survey the sky \citep{caleb2017,bannister2017,2017arXiv170906104M,2018ApJS..236....8L,chime2018overview}. Interferometers, coherently or incoherently combining signals from many smaller elements or dishes, have the advantage of a large instantaneous \ac{FoV}. Recent technological advances have resulted in new receivers such as \acp{PAF}, which place many dipoles at the focus of each dish of a telescope array.

One of the largest challenges of interferometric radio astronomy has always been computation. Beam forming within the telescope \ac{FoV} requires a powerful correlator to combine the signals from all elements in phase. This is more difficult still when combining the multi-element \ac{PAF} systems to form beams on the sky. Forming coherent beams and searching the time stream of each for impulsive radio signals such as FRBs provides an added technical challenge. 

Faster and more agile processing units available in recent years have made it possible to form more beams and search
them quickly, in some cases in real time. These searches still require large compute clusters to deal with the massive
amounts of data streaming from the telescope and distribute it over many processing nodes. Many FRB search efforts,
including the searches described here with \ac{Apertif}, now house dedicated computing clusters on-site to search the data in real time for pulses.

By combining different elements of the array and the feed, the larger \ac{FoV} of an interferometric array can be
sampled by many smaller beams, enabling a much more precise localisation of any new source.
The instantaneous localisation ability of an interferometer depends on the length of the longest baseline, 
 which in the case of WSRT is long in the E-W direction but very short in the N-S direction,
but even more precise localisation is possible for
brighter signals that appear in several beams, as demonstrated in, e.g., \citet{2020MNRAS.499.4716C}.
% or where the raw voltage streams from the telescope are still available for offline correlation and beam forming. 

Real-time searches are also critical in searching for prompt emission from the FRB source in other wavelength regimes, but at other radio frequencies as well. Previous multi-wavelength searches for related emission following real-time FRB detections were unsuccessful. However, these early efforts triggered follow-up several hours after the initial FRB \citep{FRB140514}.

It is still unknown how broadband FRB pulses can be, and down to what radio frequencies they are detectable. 
In simultaneous FRB observations between Apertif and LOFAR, 
FRB emission has now been seen down to the bottom of the LOFAR HBA band \citep{2020arXiv201208348P}.
Triggered searches at such low frequencies, even for one-off FRBs, are made possible by
the delay that the interstellar dispersion introduces between the 1.4\,GHz and 150\,MHz arrival times of \emph{the same pulse}.
Exploiting this delay, which is of order minutes \citep{2017arXiv170906104M}, 
allows for studies of e.g. the FRB emission mechanism, and the plasma environment local to the host.
Such coincident searches do, however, require real-time classification and alerting. 

An ideal observing setup to tackle all of these challenges would combine large \ac{FoV} with high spatial resolution on the sky, for a high rate of localised FRBs. 
To address as many of the population unknowns as possible, such a system should also be able to resolve FRBs in time and frequency and capture polarisation information. 
The \ac{ARTS} is designed with these considerations in mind. 
\ac{Apertif} provides the large instantaneous \ac{FoV}, while the  WSRT interferometer allows for precise spatial resolution.
ARTS processes a bandwidth of 300\,MHz centred on 1370\,MHz.

ARTS also addresses some of the challenges above by combining a range of innovations:
first, delivering high time and frequency resolution over a frequency range of 1220\,MHz to 1520\,MHz;
second,  the ability to capture full-Stokes polarisation data with a new FRB search pipeline
and third, a machine learning classifier to better identify and trigger on FRB candidates.

\subsection{Serendipitous discovery}

Apart from the science planned for Apertif, its large field of view implies that the potential for detecting rare objects or making serendipitous discoveries is high. This is illustrated by the fact that in the survey data taken so far, 16 intra-hour variable  radio sources were  found
serendipitously  \citep{Oosterloo20}.  Such variable sources are among the rarest objects on the sky. They are extremely small sources which are made to `twinkle' because their radiation passes through turbulent foreground plasma located in the solar neighbourhood  and they give a unique view of the structure and conditions of the nearby ionised interstellar medium. They can change  a factor 5 in flux on time scales of minutes and are easily recognised in Apertif images. Before Apertif, searches with various radio telescopes had revealed only a handful of such sources   over the entire sky and  Apertif has quadrupled the number of such rare objects known in less than a year, illustrating the power of a large field of view.

\begin{figure*}
\centering
	\includegraphics[width=17cm]{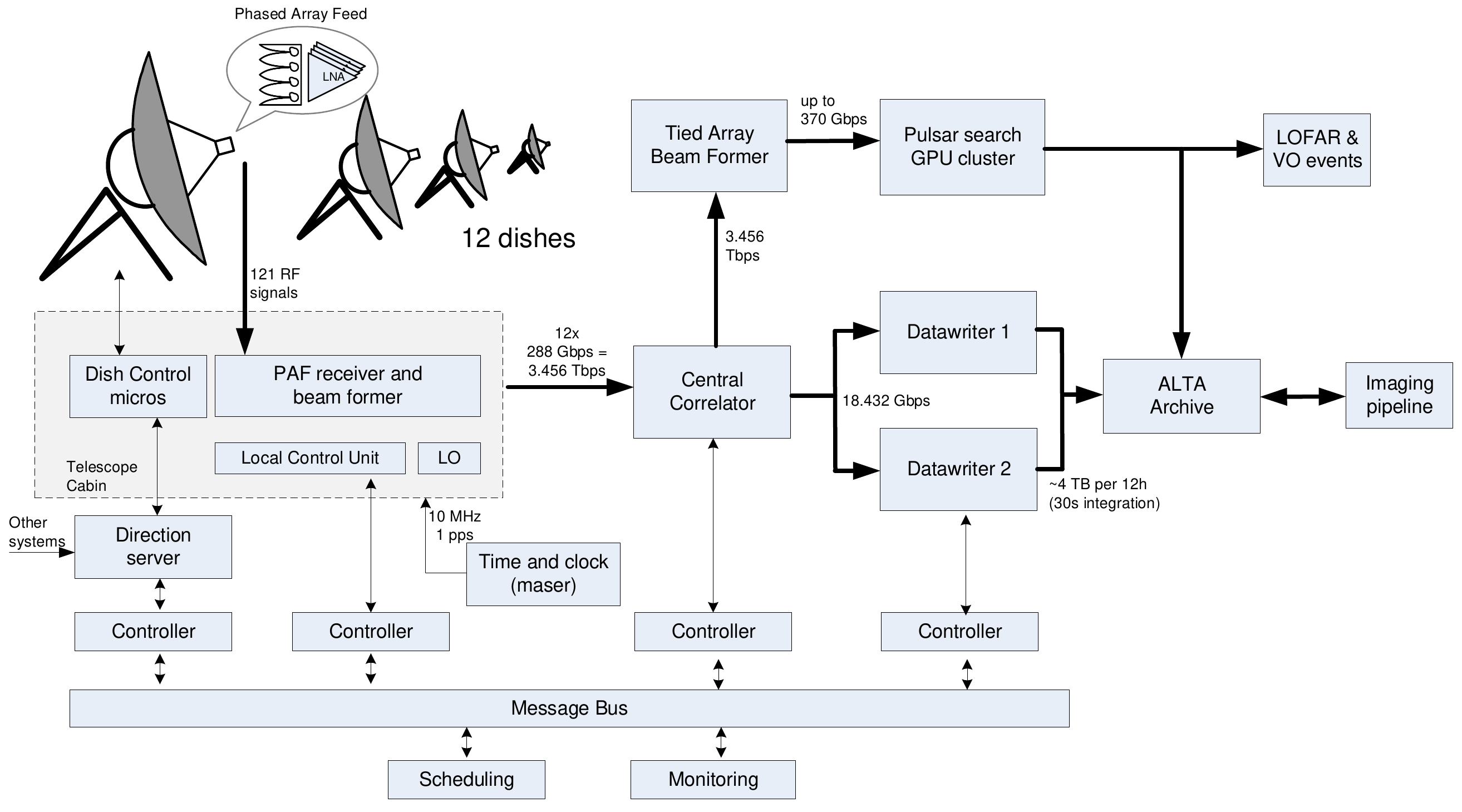}
	\caption{Overview of the Apertif -- ARTS system.}
	\label{fig:SystemOverview}
\end{figure*}

\section{Apertif System Overview} \label{sec:systemoverview}
\begin{figure}
	\centering
	\includegraphics[width=7cm]{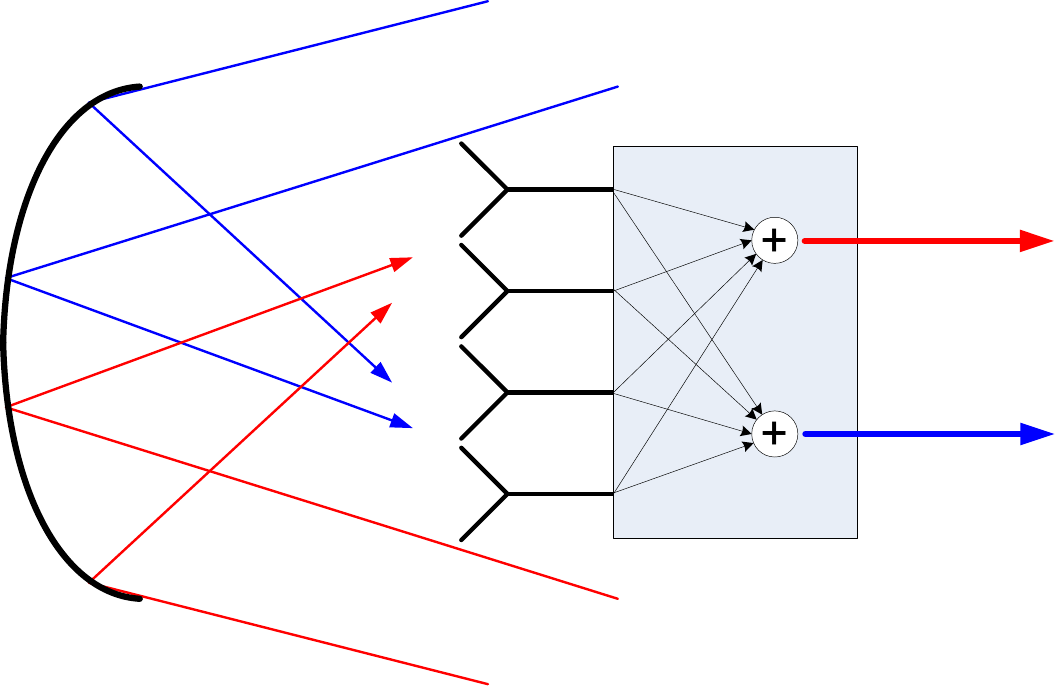}
	\caption{Concept of operation of a dense phased array feed.}
	\label{fig:PAFprinciple}
\end{figure}

The Apertif upgrade replaces the well-known Multi-Frequency Front End (MFFE) \citep{1991ASPC...19...42T}, the IF system, the frequency and clock distribution system, the correlator, most of the cabling, and all software to monitor, control and operate the system.
Figure \ref{fig:SystemOverview} shows the top-level block diagram of the WSRT-Apertif system. Apertif samples the EM field in the focal plane of the reflectors with a densely packed array of antenna elements. The spacing between the elements is 10~cm, corresponding to $0.38\lambda$ at the lowest and $0.58\lambda$ at the highest operating frequency. After filtering and amplification, the RF signals are transported over coaxial RF cables to a Telescope Cabin located under every dish. In the cabin, the RF signals are converted to baseband, digitised and combined into 40 dual-polarised compound beams. To form one compound beam (CB) on the sky, the signals received by antenna array elements are weighted and added in a beam former. Apertif uses a bi-scalar beam former: The signals of \emph{all} array elements of the same polarisation are used for \emph{every} compound beam in that polarisation. Because the beam former is implemented in the digital domain, multiple copies of the input signals are easily made. Using many parallel digital beam formers that each apply different weights, many compound beams (each pointing in a slightly different direction) are formed simultaneously. This principle is illustrated in Fig.~\ref{fig:PAFprinciple}. In Apertif, every dish delivers 40 compound beams. The maximum electronic scan angle is determined by the size of the PAF, the focal length and the diameter of the main reflector. The dimensions of the Apertif PAF are the largest possible in the prime focus cabin of the WSRT reflectors, and provide a good balance between the scan range ($\pm 1.5$ deg in all directions), and the capital costs. The PAF weighs about 6 times less than the pre-Apertif frontend. The instantaneous bandwidth of Apertif, 300 MHz, is defined by the filters in the analog electronics. The number of simultaneous beams is defined by the digital processing capacity.

The partitioning of the system was driven by the desire to build a feasible, low risk, and easy to maintain system. Considering the constraints on physical size, weight and power consumption, only the RF part of the feed is located in the feed cabin at the primary focus of the reflector. Consequently, the weight of the systems in the feed cabin is reduced from about 300~kg before the Apertif upgrade, to only 50~kg. The Telescope Cabin under the dish is relatively large. This lowers the required level of integration of the RF and digital back-end, reducing complexity and development risks. It also allows easy access to the equipment for maintenance and repair. By using a Faraday cage inside the Telescope Cabin, the interference generated by the receivers and back-ends can be very effectively shielded from the antennas. Inside the Faraday cage, all RF parts are shielded to avoid self generated RFI of the digital back-end, and its power supply, leaking into the system. A point of attention in this architecture are the phase variations of the coaxial RF cables that connect the feed antenna to the Telescope Cabin.

 Time and frequency references are provided by a maser and a GPS receiver. The central correlator, with an aggregated input datastream of 3.456 Tbps, correlates every beam of all dishes into visibilities. The visibilities are integrated and stored in the datawriter. This is the end of the real-time system. The resulting datasets are sent to ASTRON's headquarter in Dwingeloo where an automated calibration and imaging pipeline produces calibrated image cubes. Both the calibrated cubes and the raw visibilities are stored into the Apertif Long-Term Archive (ALTA), which is the main interface to the users. Alternatively, the signals from the dishes are tapped-off in the correlator and processed by an independent GPU-based back-end for tied-array observations. \citep{2014htu..conf...79V,ARTS2}.

The beam pattern of the Apertif feed offers a great level of control, enabling aperture efficiencies of 75\% and higher, and control over sidelobes and cross-polarisation levels. As described by \citet{5782591}, accurate knowledge of the beam patterns on the sky is crucial for synthesis radio telescopes to produce high dynamic range, noise limited, images. Receiver gain variations of the PAF can lead to variations of the compound beam and consequently a calibration scheme is required. An online calibration subsystem has been designed to measure the analogue gain variations during an observation and to compensate the variations in the (digital) beam former to stabilise the beam on the sky. At the time of writing, the online calibration system has not been fully implemented yet because the PAF system is inherently stable enough between two calibrations, i.e. a period of 2 -- 3 weeks. 

Table \ref{table:1} lists the main performance characteristics of the WSRT-Apertif system. 

\begin{table}
	\caption{Key performance characteristics of the WSRT-Apertif} % title of Table
	\label{table:1} % is used to refer this table in the text
	\centering % used for centering table
	\begin{tabular}{l r l} % centered columns (4 columns)
		\hline\hline % inserts double horizontal lines
		Characteristic & Value & Unit \\
		\hline
		Frequency range & 1130--1750 & MHz \\ % table heading
		Processed bandwidth & 300 & MHz \\ % table heading
		Number of beams & 40 \\ % table heading
		Field of view single observation $^a$ & 10.5 & deg$^2$ \\ % table heading
		Effective field of view for surveys $^a$ & 5.25& deg$^2$\\
		Sensitivity $^a$& 63 & m$^2$K$^{-1}$ \\
		SEFD $^a$ & 43.7 & Jy \\
		Number of dishes & 12 & \\		
		Baseline lengths & 36--2700 & m \\
		Spectral resolution & 12.2 & kHz \\
		\hline %inserts single line
		$^a$ At 1400 MHz.
	\end{tabular}
\end{table}
\section{Impact of RFI}\label{sec:rfi}
\begin{figure*}
	\centering
	\includegraphics[width=17cm]{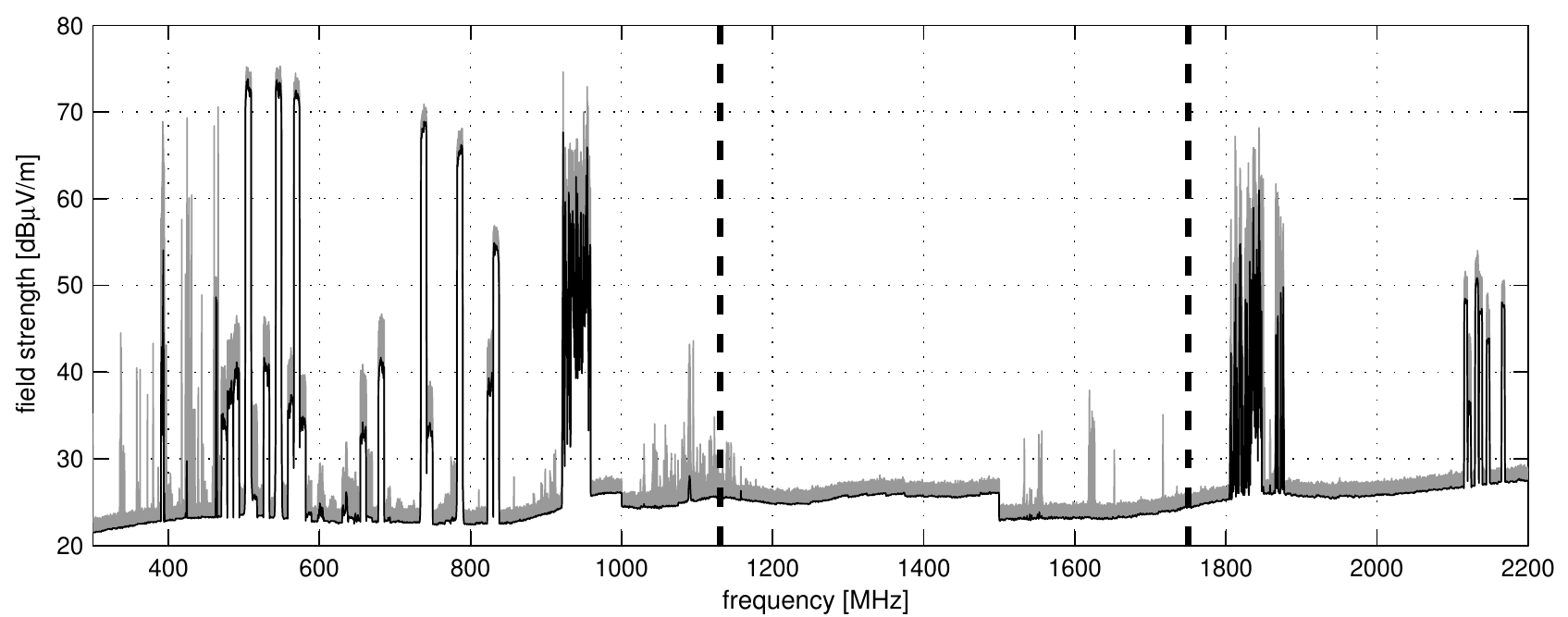}
	\caption{Spectrum measured by the monitoring station at the WSRT site towards the horizon. WSRT-Apertif operates between the two vertical dashed lines. Small jumps in the noise floor are caused by varying antennas and receiver settings for each frequency band. The grey areas indicate time variability of the RFI.}
	\label{fig:wsrtspectrum}
\end{figure*}

Fig.~\ref{fig:wsrtspectrum} shows a measured spectrum at the WSRT site with a low-gain antenna towards the horizon. The main RFI sources are the GSM bands around 850~MHz and 1850~MHz, as well as five terrestrial TV bands below 800 MHz of the TV tower at Smilde, located 12.6~km west of the WSRT. In particular these TV bands proved to be troublesome. The TV stations use a modulation schema using a large number of subcarriers spread over an 8~MHz band.  From measurements in a WSRT dish, the expected worst case power from these signals at the input of the first Low Noise Amplifier (LNA) is approximately -50~dBm. Although these interferers are outside the Apertif frequency band of operation, they necessitate filters throughout the entire RF system to reject the creation of second and third order intermodulation products. These filters lead to an increase of the noise figure. Without these filters, the noise floor in the system would not be determined by thermal noise, but by a noise floor created by these intermodulation products. At the WSRT site, emission from high-power ADS-B aircraft secondary surveillance radar operating at 1090~MHz is also present. Its peak power is much higher than shown in Fig.~\ref{fig:wsrtspectrum} because it is pulsed. 

The frequency range between 1300 and 1500~MHz  is relatively RFI free. At frequencies below 1300~MHz significant parts of the band are badly affected by RFI, in particular for the shorter baselines. The RFI  is dominated by Galileo bands E5 and E6, GPS L2, and GLONASS, making much of this frequency range unusable.  Only a few small windows are relatively free of RFI, for example the frequency range 1180 -- 1195~MHz. In the frequency range 1452 --~1492 MHz RFI is present due to 5G transmitters in the area.

\section{Phased array feed}\label{sec:paf}
\begin{figure}
	\resizebox{\hsize}{!}{\includegraphics{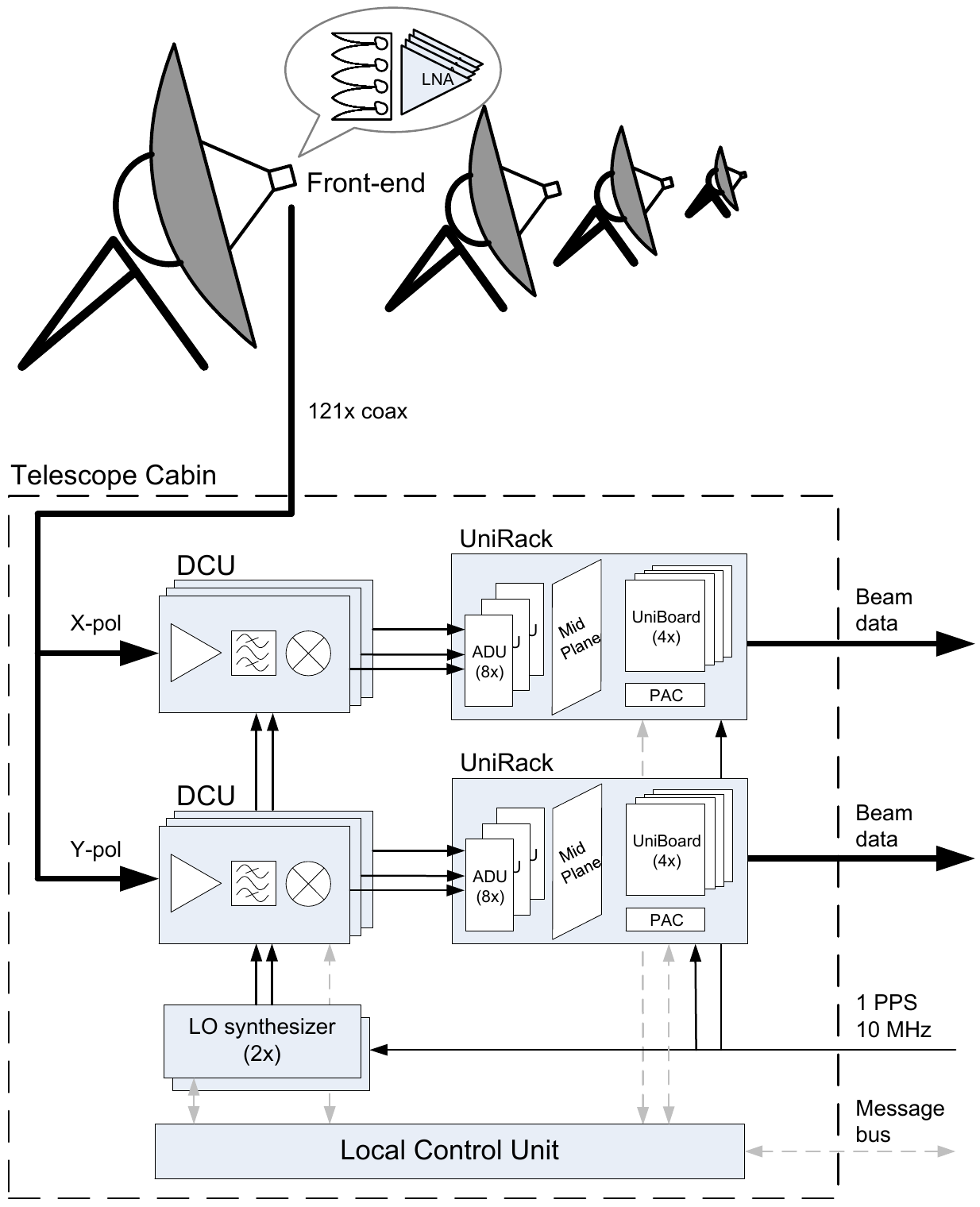}}
	\caption{Overview of the PAF system at every dish.}
	\label{fig:dishoverview}
\end{figure}

This section describes the PAF analogue and digital hardware installed at every dish, delivering 40 dual polarised beams to the central systems. An overview of the PAF system at every dish is show in Fig.~\ref{fig:dishoverview}. The major design driver for the antenna and RF system is the wide frequency band of operation and the required noise figure. 

\subsection{Antenna array}

\begin{figure*}
	\centering
	\includegraphics[width=17cm]{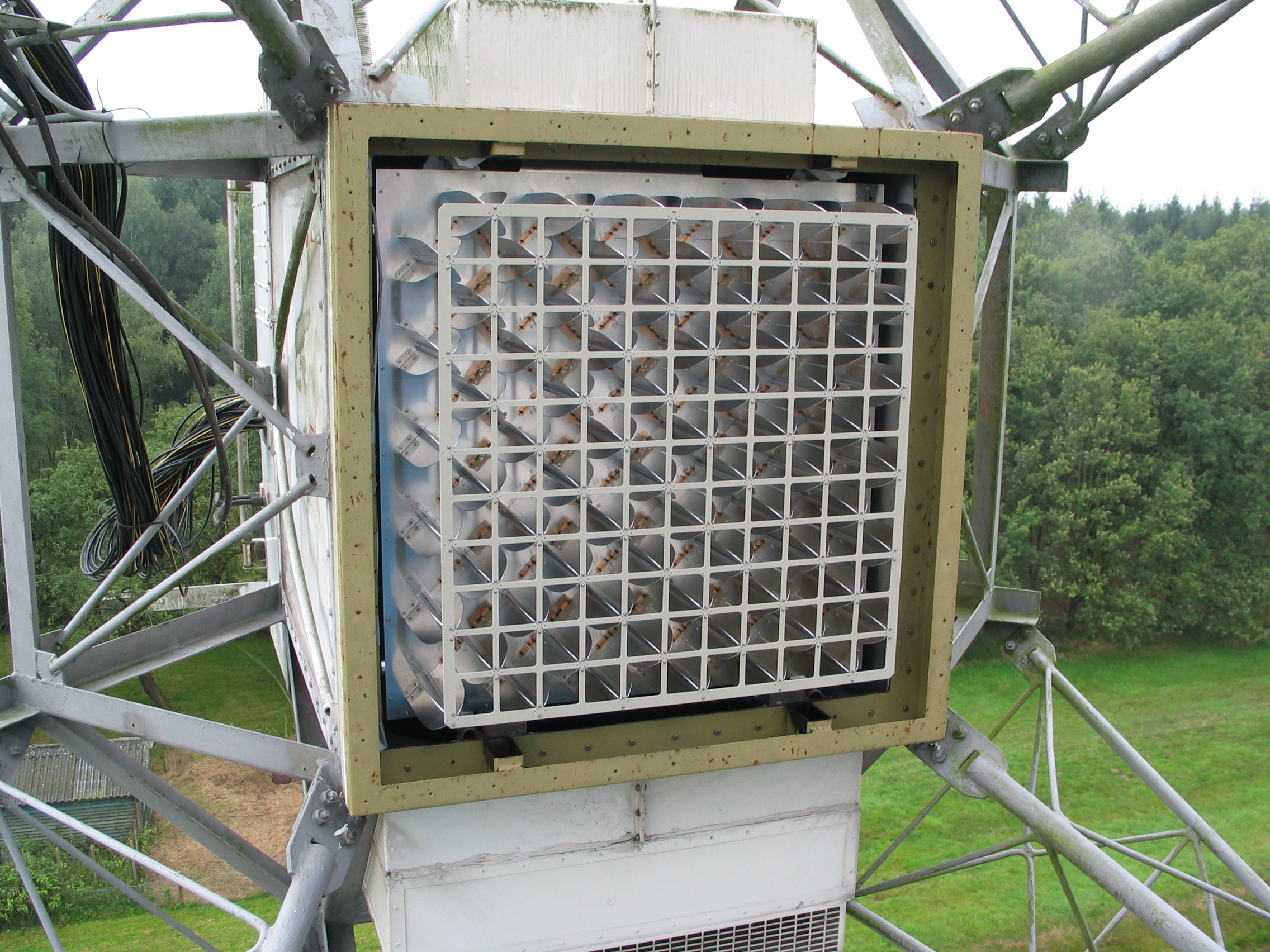}
	\caption{Photo of the Apertif Phased Array Feed. For the photograph, the radome has been removed.}
	\label{fig:photofeed}
\end{figure*}
\begin{figure}
	\centering
	\resizebox{5cm}{!}{\includegraphics{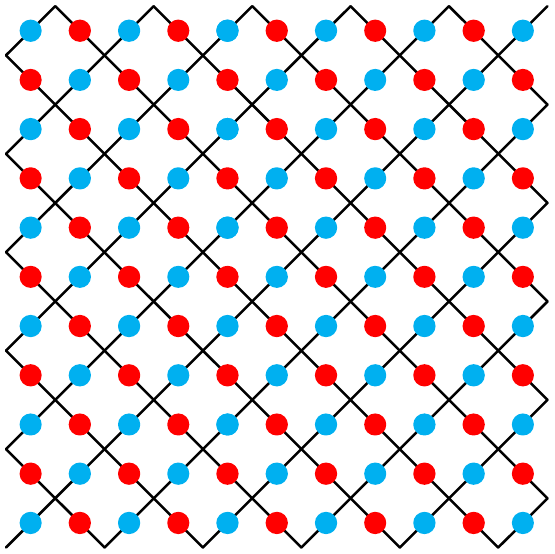}}
	\caption{Layout of the 121 PAF elements. The red and blue dots indicate the elements of two orthogonal polarisations.}
	\label{fig:elpos}
\end{figure}

The Apertif feed arrays each consist of 121 single polarised tapered slot (or Vivaldi) antenna elements \citep{5505527}, 61 in one polarisation and 60 in the orthogonal polarisation, see Fig.~\ref{fig:elpos}. The antenna elements are configured in a rectangular grid at a pitch of 10~cm. The feed array is installed in the (primary) focus cabin of the dishes, see Fig.~\ref{fig:photofeed}. For mechanical support, a grey polypropylene grid is attached to the front of the antenna elements. It has been verified that the impact of this grid on the electromagnetic and noise performance of the antenna array is negligible. For weather protection, a foam radome is installed in front of the antenna. With a focal length of $8.75$ m ($f/D=0.35$), the WSRT reflectors are relatively deep and have an opening angle of $\pm 71.1$ deg observed from the feed. A 1~m $\times$ 1~m sized PAF, the largest that fits in the focus cabin, provides a field of view of about $3 \times 3$ deg$^2$.

\begin{figure}
	\resizebox{\hsize}{!}{\includegraphics{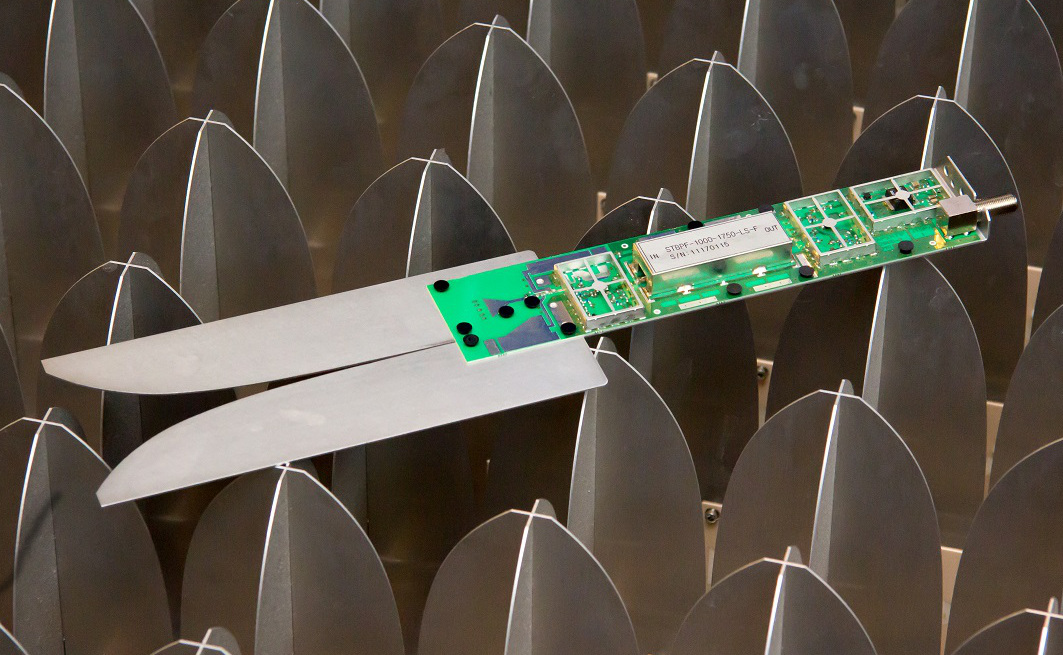}}
	\caption{Vivaldi antenna element with integrated low-noise amplifier.}
	\label{fig:antennaelement}
\end{figure}

Figure \ref{fig:antennaelement} shows one of the PAF elements with integrated Low Noise Amplifier (LNA). The entire front-end operates at ambient temperature. In the preliminary design phase, a cryogenically cooled system has been considered. Although a cryogenically cooled front-end was expected to lower the system temperature by 10 -- 20 K, it was infeasible due to its large impact on the system development, construction, and maintenance costs.

\begin{figure}
	\resizebox{\hsize}{!}{\includegraphics{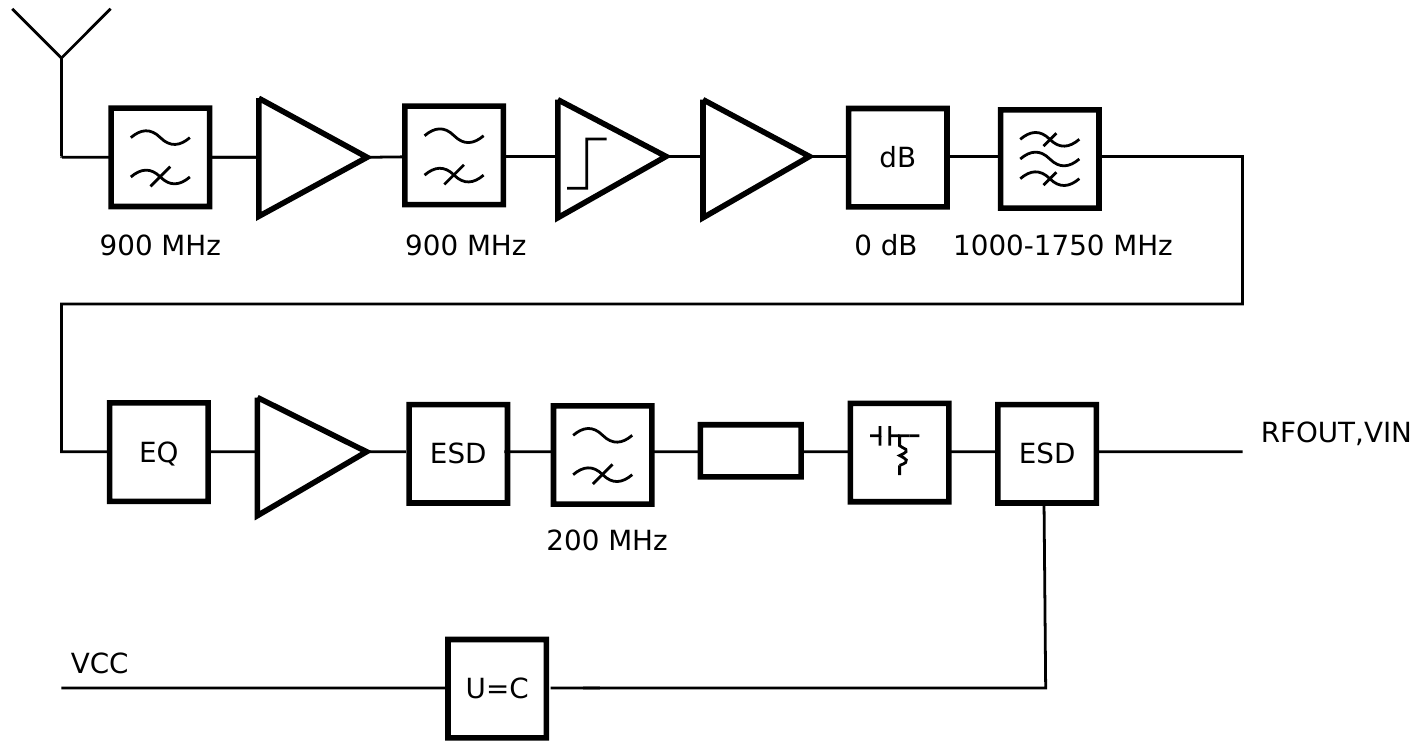}}
	\caption{Block diagram of the front-end board.}
	\label{fig:frontend}
\end{figure}

\subsection{Front-end}
\begin{figure}
	\resizebox{\hsize}{!}{\includegraphics{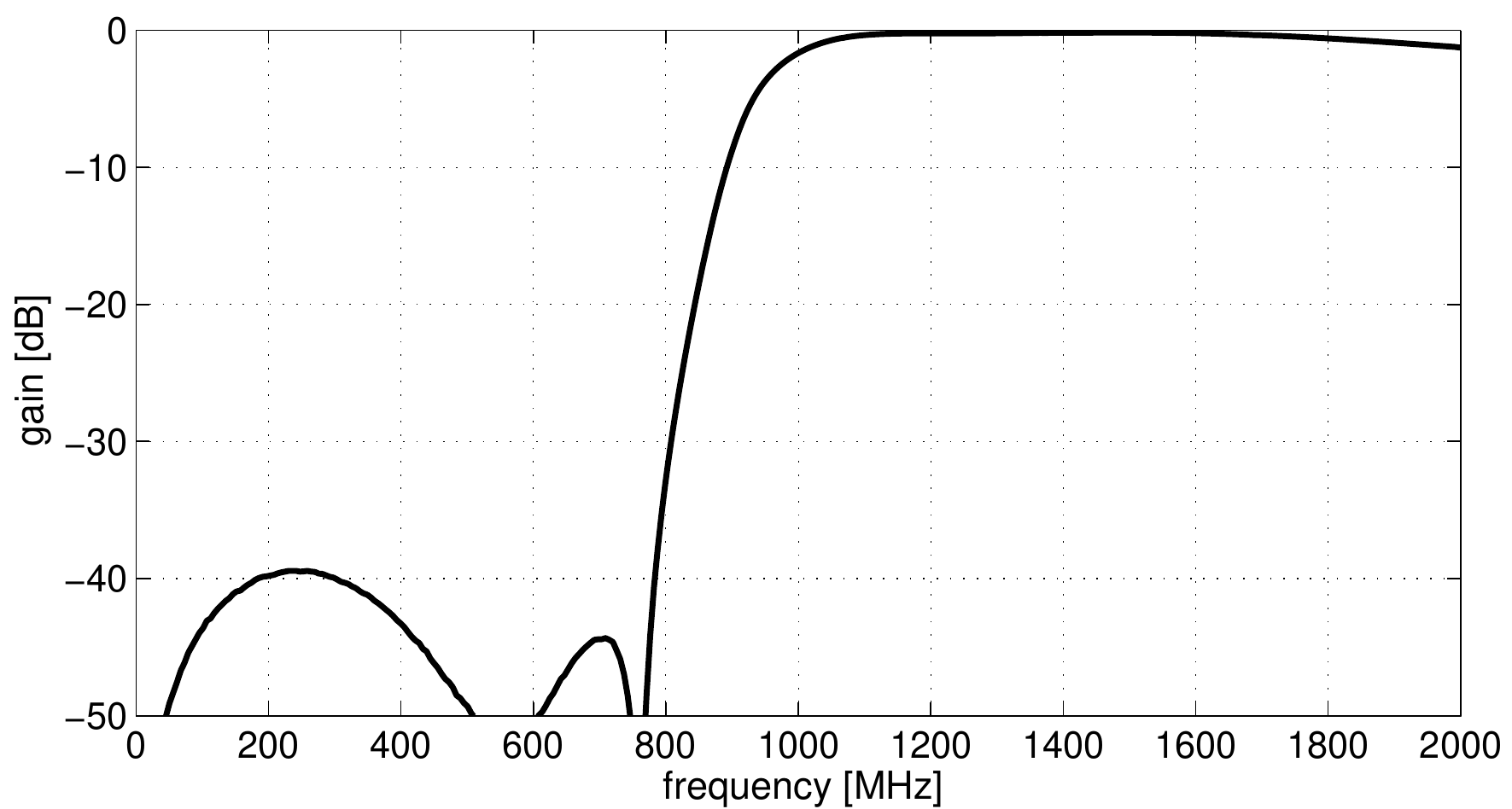}}
	\resizebox{\hsize}{!}{\includegraphics{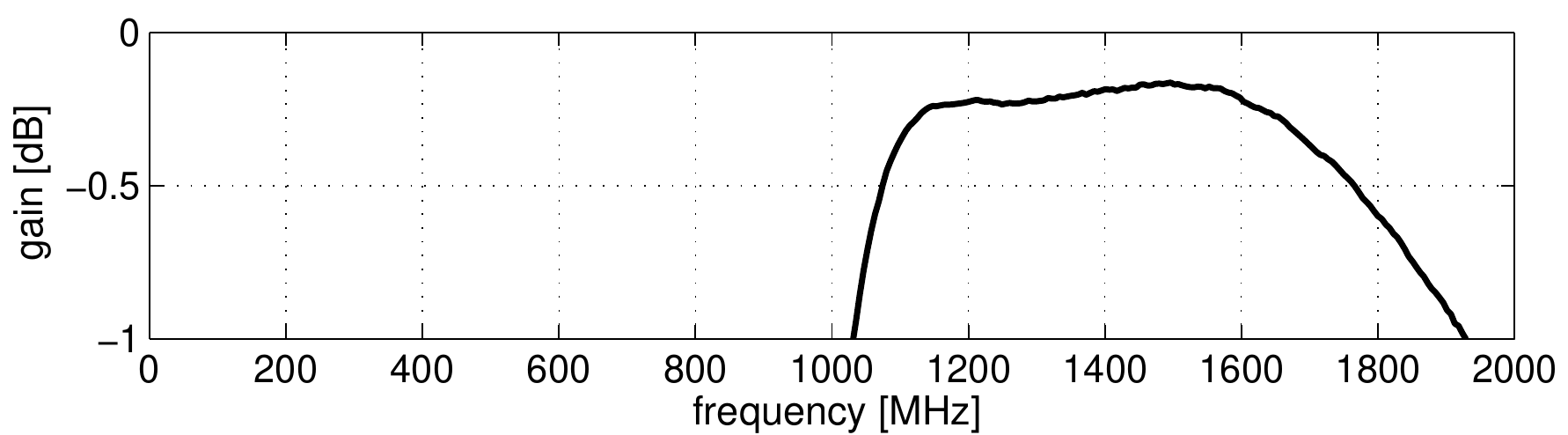}}
	\caption{Measured $S_{21}$ of the pre-LNA high-pass filter (top) and a close-up of the $S_{21}$ passband (bottom).}
	\label{fig:hpfilter}
\end{figure}

Fig.~\ref{fig:frontend} shows the block diagram of the front-end board. The board is tightly integrated with the antenna elements to minimise losses. As opposed to horn antennas, wideband array elements like a Vivaldi do not have a sharp cut-off at the lower end of their frequency band of operation. Consequently, most of the RFI below 1130 MHz will be present at the input terminals of the LNA's. To avoid the creation of intermodulation products in the Apertif band, a very low-loss high-pass filter is integrated in the LNA. The measured $S_{21}$ of the filter is shown in Fig.~\ref{fig:hpfilter}. It suppresses the DVB-T signals by more than 40~dB. The high pass filter uses a distributed architecture with three taps. The resonance frequencies of these taps are optimised to maximise the attenuation across the stop band below 1~GHz. The losses are minimised by using a low loss tangent board, and by finding a compromise between the conductive copper losses and the amount of attenuation. The $S_{21}$ in Fig.~\ref{fig:hpfilter} includes mismatch losses, which do not contribute to the noise budget. The net contribution of the filter to the system temperature is $\approx 10$~K (see Table~\ref{table:noise}). In addition, out of band RF filtering is installed between analogue gain stages through the entire signal chain. The second high pass filter is a discrete implementation of the same filter to further reduce the RFI.  The third band pass filter is a 9\textsuperscript{th} order discrete filter to filter out GSM-900 and GSM-1800 above and below the frequency band of interest. The first-stage amplifier is a Skyworks 67151-396LF, which is one of latest GaAs pHEMT generations, and has one of the lowest noise figures currently available using COTS devices. Its gain of 22~dB is used to overcome the largest part of the noise contribution of the other components in the RF system. The LNA receives its DC power supply over the same coaxial cable that is used to output the RF spectrum to the receiver. Protection against electrostatic discharges (ESD) is installed at the output of the front-end board. At the input and output discharge resistors are used as well to prevent static charge buildup.

\subsection{RF coaxial cables}
The 121 RF signals from the LNA's are transported to the telescope cabin by 40~m long RF coaxial cables. Because the cables cross both the elevation and the hour-angle axis, they are subject to repetitive bending. A number of candidates for the cables were selected and subjected to accelerated lifetime tests by repetitive bending at very low temperatures. The 75 Ohm COAX9 cable performed best and is now used in Apertif.

\subsection{Telescope cabins}
\begin{figure}
	\resizebox{\hsize}{!}{\includegraphics{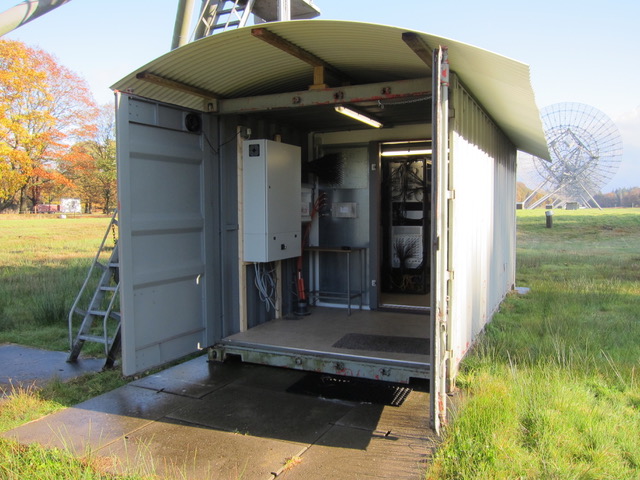}}
	\caption{Telescope cabin, an EMC shielded Faraday cage in a shipping container.}
	\label{fig:cabin}
\end{figure}

\begin{figure}
	\resizebox{\hsize}{!}{\includegraphics{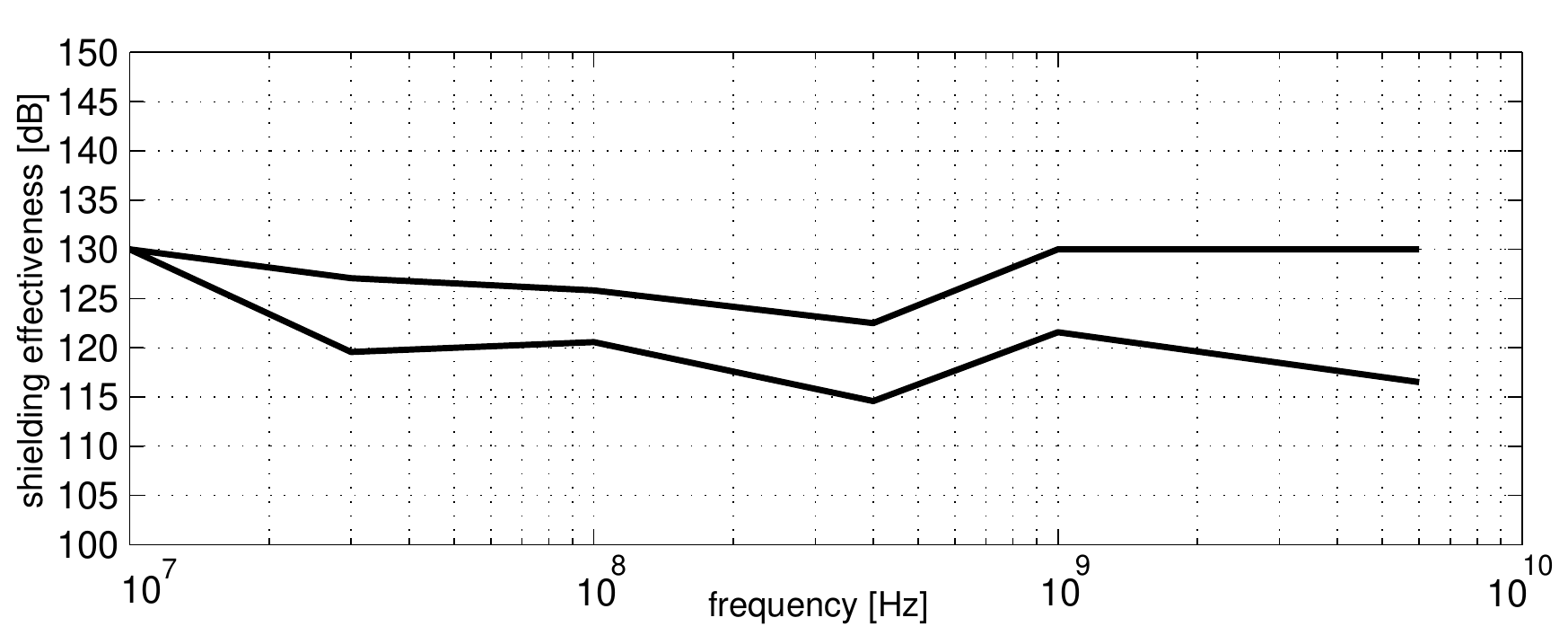}}
	\caption{Average measured shielding effectiveness according to EN 50147-1: March 1996 of the Apertif telescope cabins at two positions.}
	\label{fig:shielding}
\end{figure}

The output signal of the LNA's are connected to a telescope cabin (Fig.~\ref{fig:cabin}). The telescope cabin contains all dish-based systems (i.e. receivers and the PAF beam former). The cabin is a standard 20-ft shipping container with a thermally isolated Faraday cage inside. The cabins at the four movable dishes are placed on a frame and move with the telescope structure. Every Faraday cage is equipped with an air-conditioning system for cooling (and heating) up to 6.5~kW. The shielding effectiveness of the Faraday cages was measured and is shown in Fig.~\ref{fig:shielding}.

\subsection{Receiver}
\begin{figure}
	\resizebox{\hsize}{!}{\includegraphics{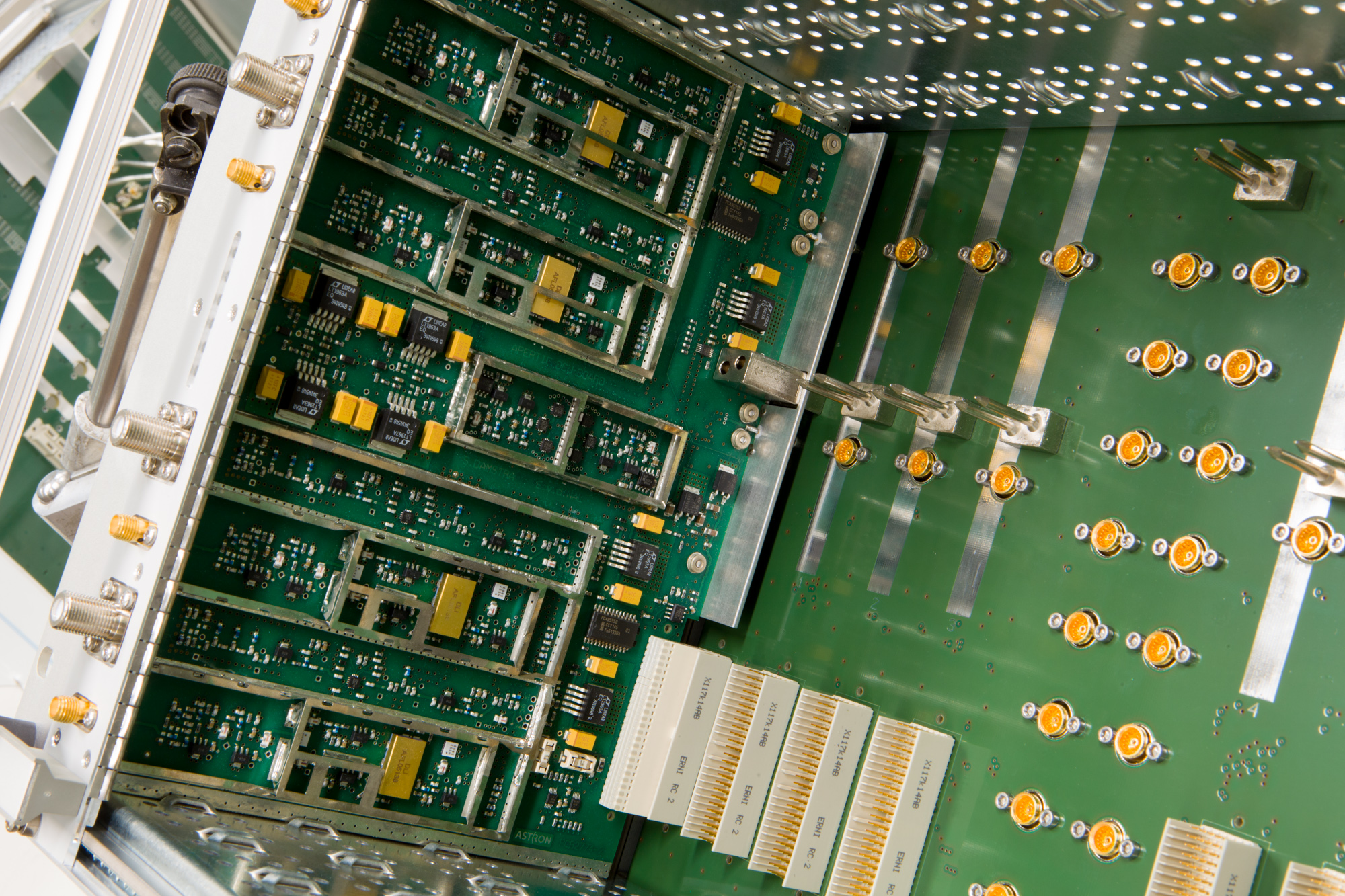}}
	\caption{The Apertif Down Convertor Unit, covers removed.}
	\label{fig:DCU}
\end{figure}

The Down Converter Unit (DCU) receiver of Apertif converts the input RF band to the IF band (Fig.~\ref{fig:DCU}). The DCU also supplies DC power to the LNA’s via the coaxial RF cables. All settings and monitoring information in the DCU sub-rack can be accessed through Ethernet. Four RF channels are combined on one printed circuit board. It also houses the last Local Oscillator (LO) distribution stages, power supply and control circuits, and the status monitoring of every channel. Eight DCU boards are combined to a DCU rack, which also houses the passive LO distribution. A separate rack is built for the filtered DC power supply. The DCU rack is shielded to suppress interference from EMI sources within the Faraday cage, like the digital back-end, power supplies, control computer and the air conditioning system.

\begin{figure}
	\resizebox{\hsize}{!}{\includegraphics{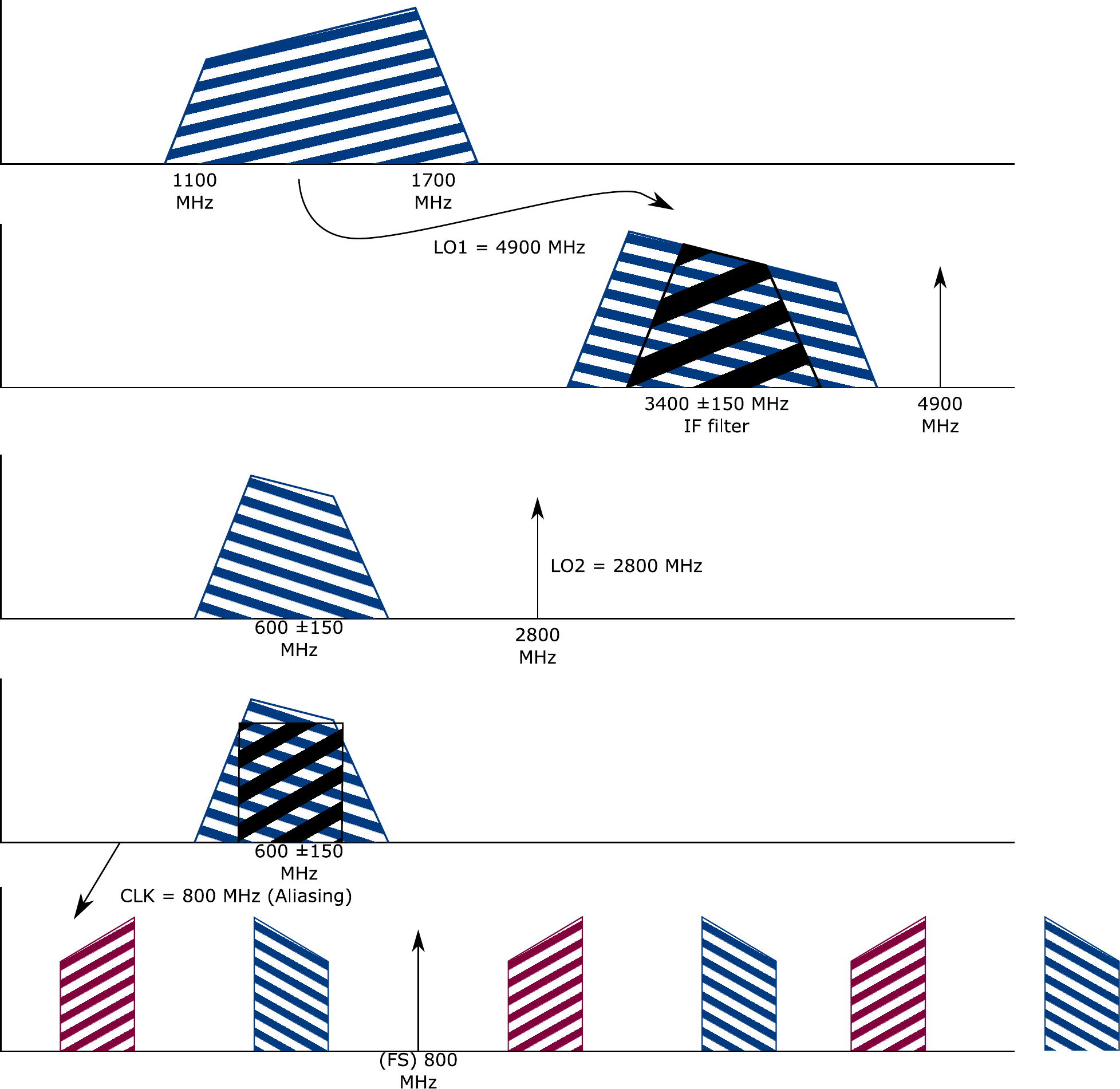}}
	\caption{Overview of the mixing schema used in the Apertif receivers.}
	\label{fig:mixingscheme}
\end{figure}

From an analysis of the RFI input spectrum and prerequisites like the ADC’s capabilities, it is concluded that a double conversion scheme is most appropriate. The down-conversion scheme converts the RF frequency from 1130 -- 1750~MHz to the second Nyquist zone of the ADC (450 -- 750~MHz). It was very hard to find a relatively clean frequency mixing scheme without mixing products falling back into the frequency band, given the input spectrum (Fig.~\ref{fig:wsrtspectrum}). Multiple options have been considered by modelling the RF spectrum throughout the receiver chain. The implemented mixing scheme (see Fig.~\ref{fig:mixingscheme}) gives the best compromise between a clean spectrum throughout the channel, performance requirements on the filters and feasible local oscillator frequencies using standard building blocks, such as PLLs, gain blocks, coax cables and standard PCB board technology.

Starting at the top of Fig.~\ref{fig:mixingscheme}, the RF input spectrum is up-converted using an adjustable LO1 frequency (4680 -- 5000~MHz). This allows an intermediate frequency of 3.4~GHz. A custom bandpass IF filter rejects the image frequency and selects the 300 MHz band, while a second down-conversion mixer converts from IF to the second Nyquist zone of the ADC. The IF filter is based on ceramic coupled line technology to ensure a repetitive performance for all $\approx$1500 channels and a stable response across frequency. The downconversion mixer uses an LO2 frequency of 2800~MHz, which is 3.5 times the ADC sampling clock frequency. If the clock or its harmonics would couple into signal it will be aliased back to DC or to 400~MHz. Finally an anti-aliasing filter selects the second Nyquist zone of the ADC.

\begin{figure}
	\resizebox{\hsize}{!}{\includegraphics{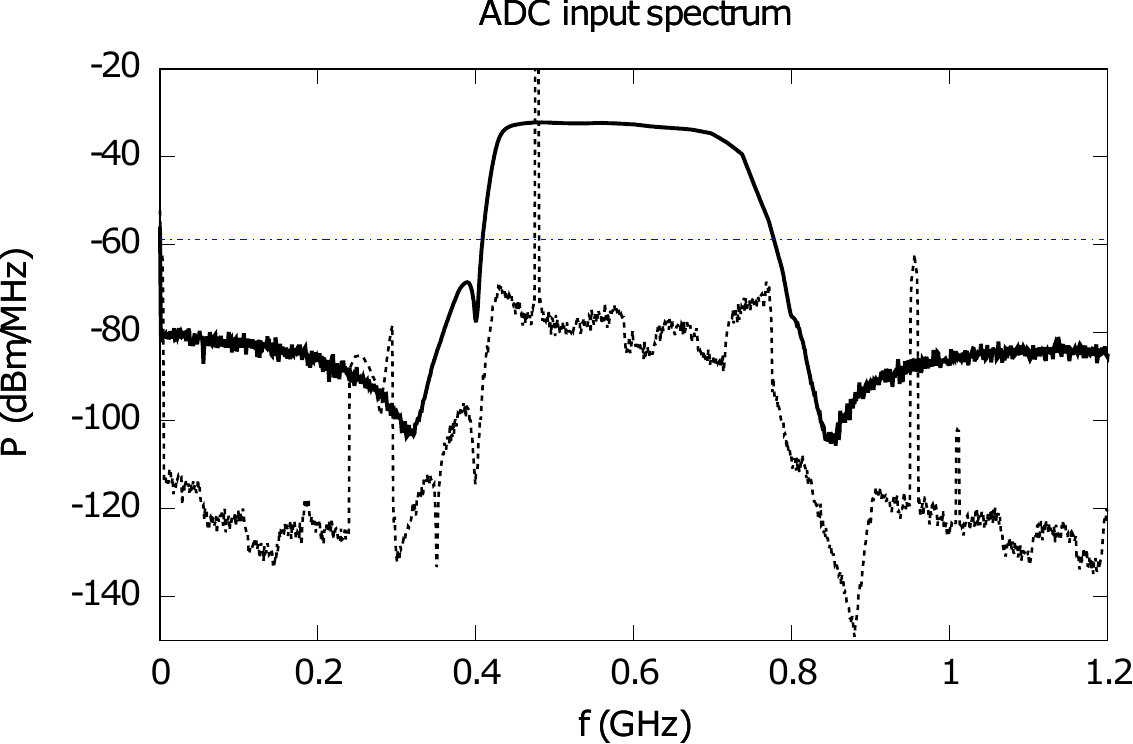}}
	\caption{Simulated power spectrum at the ADC input of the noise power (solid line) and the worst-case interference (dashed line), including its intermodulation products.}
	\label{fig:adcinputspectrum}
\end{figure}

Fig.~\ref{fig:adcinputspectrum} presents the simulated output spectrum of the RF system, when the dish is placed in the worst case position for RFI. The self-developed simulation code is loosely based on the harmonic balance method. The solid line represents thermal noise. The dashed line presents the power of the intermodulation products. The horizontal line around -60 is the maximum allowed level of the intermodulation products (i.e.\ 25~dB below the noise floor). It can be clearly seen that over the vast majority of bandwidth (450 – 750~MHz) the intermodulation products are well below the specification line. The line around 470 MHz is not and intermodulation product but is an in-band RFI signal from Iridium satellites.

\subsection{Frequency and time references}
\begin{figure}
	\resizebox{\hsize}{!}{\includegraphics{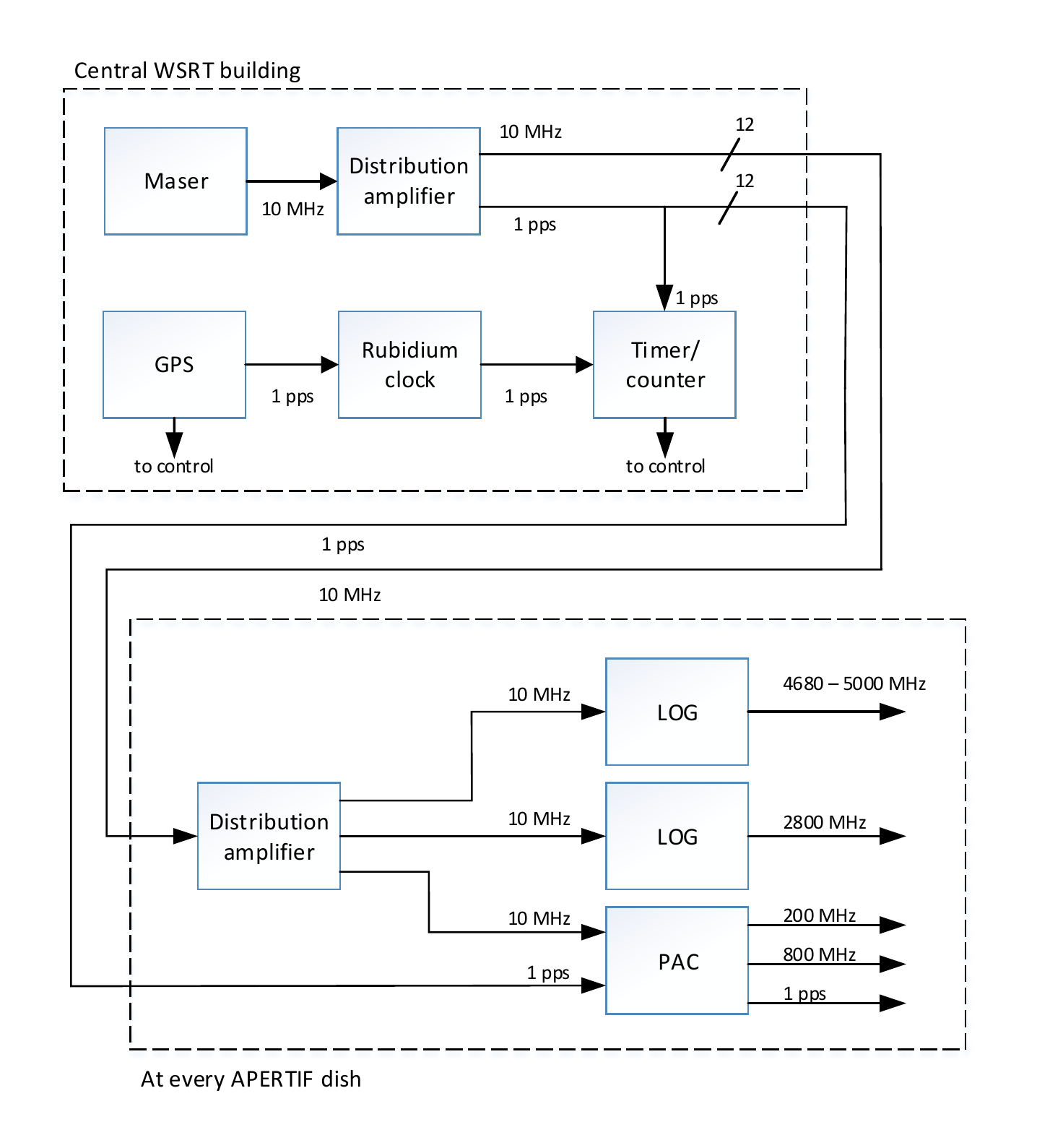}}
	\caption{Simplified diagram of the Apertif frequency and time reference distribution system.}
	\label{fig:TDS}
\end{figure}

The clock system in Apertif (Fig.~\ref{fig:TDS}) is used to synchronise all dishes with respect to each other, and to establish a link to the absolute time. All Apertif systems are locked to the 10~MHz clock from a hydrogen maser. A higher clock frequency would make the frequency selection grid too sparse, while lower clock frequencies would increase the phase noise contribution produced by the the Phased Locked Loop (PLL) circuits. An additional advantage of the 10~MHz reference frequency is the relative low loss at these frequencies of the coaxial cables used to distribute the clock signals. A distribution amplifier derives the 1~pps signal from the 10~MHz reference provided by the maser. The time difference between this 1~pps signal and a 1~pps signal from a GPS based clock is logged to establish an absolute time reference for the observations.

Commercially available distribution amplifiers are used to send the 10 MHz and 1 pps signals to the telescopes over existing buried coaxial cables. In the telescope cabin, two Local Oscillator Generators (LOGs) generate the two local oscillator signals from the 10~MHz clock reference. The LOG system is based on the Hittite HMC833 Phased Lock Loop and is built in an EMC shielded 19 inch sub-rack. For each local oscillator frequency range, a separate LOG system is built using different filters. The frequency can be configured remotely by an IP based Ethernet link. Furthermore, it has 4 isolated outputs with remotely controllable output powers. Thermal insulation of the PLL gives the unit its stability over time. To ensure phase repeatability between dishes after a reset, the output frequency is an integer multiple of the reference frequency.

The 200~MHz data processing clock that is used for the Uniboards in the dishes and the correlator, is locked to the same 10~MHz reference as the ADC sample clock. Therefore, the processing rate in all FPGAs cannot drift between FPGAs. Zero clock drift helps to keep the internal memory usage of input FIFO's for the data transport between the FPGA small.

\subsection{Real-time Digital Signal Processing} \label{sec:adu-dsp}
\begin{figure}
	\resizebox{\hsize}{!}{\includegraphics{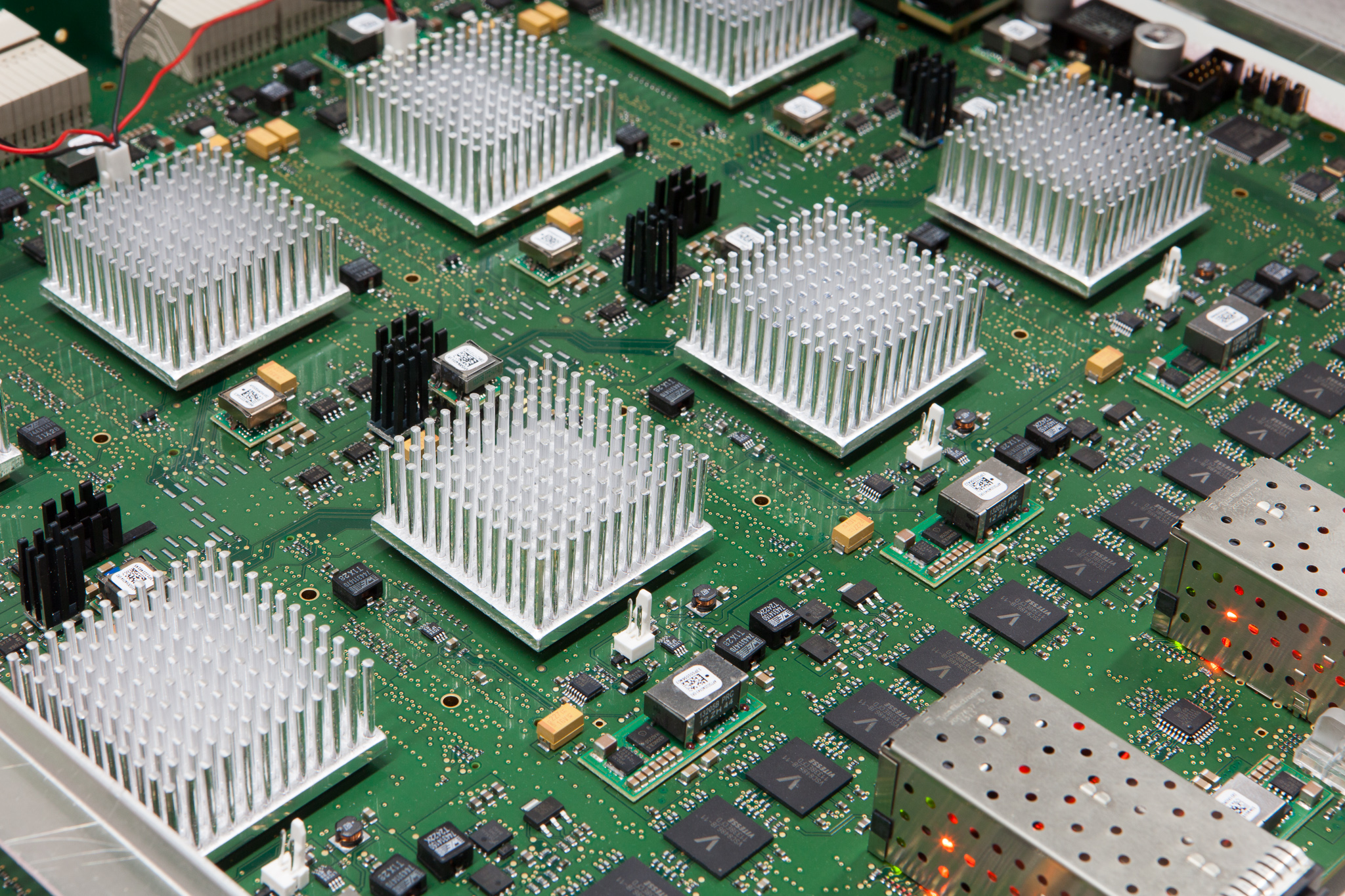}}
	\caption{UniBoard.}
	\label{fig:uniboard}
\end{figure}

The analogue baseband signals from the DCUs are input to the UniRack: A shielded subrack with 64 analog inputs that performs the analogue to digital conversion and the digital signal processing of the PAF signals, and sends the resulting 40 beams to the central correlator over optical fibres. The UniRack also adds timestamps to the digital data. Apertif uses the Uniboard \citep{uniboard} as the hardware platform for all digital signal processing, see Fig.~\ref{fig:uniboard}.

\begin{figure}
	\resizebox{\hsize}{!}{\includegraphics{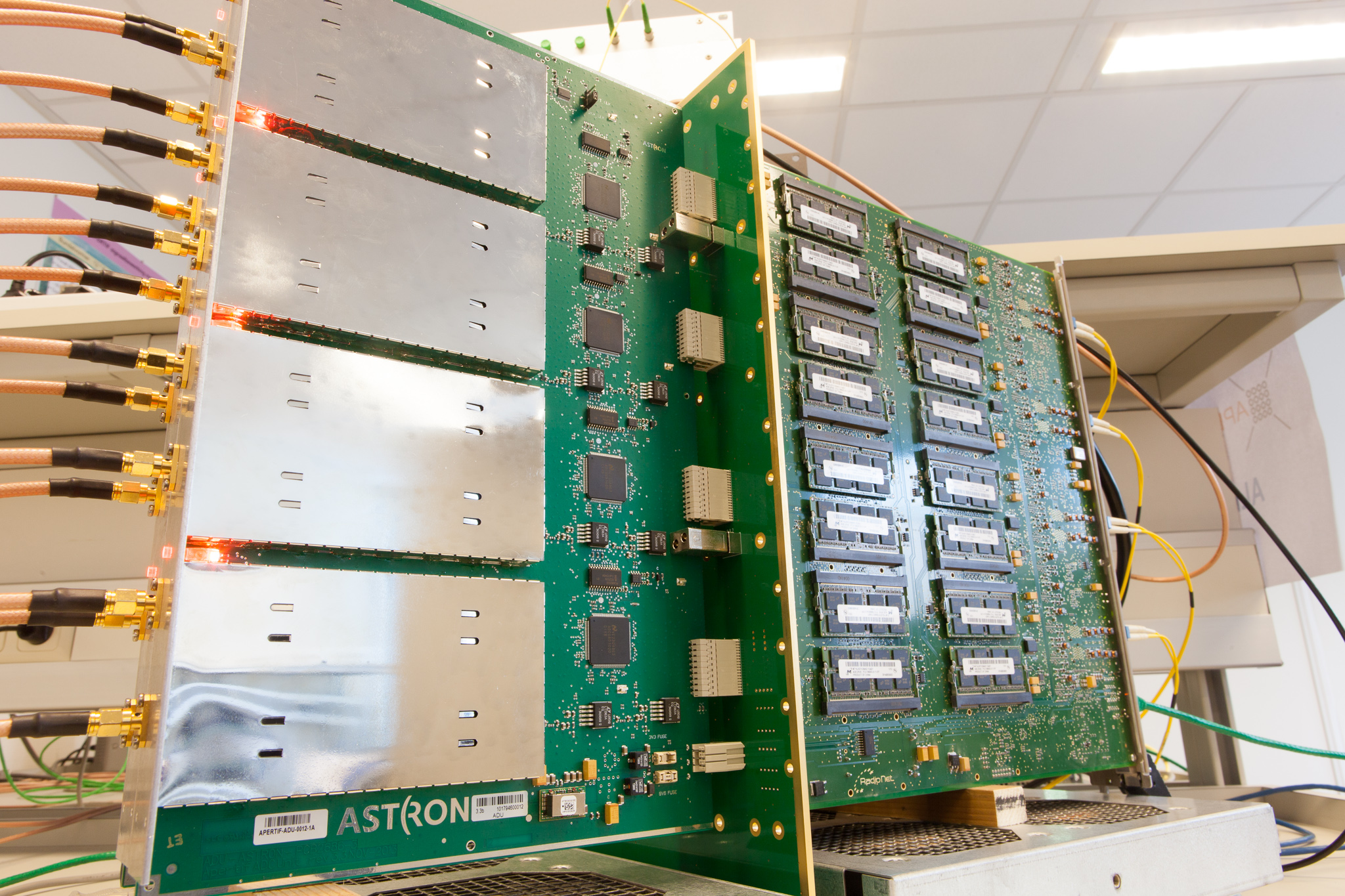}}
	\caption{The Analog to Digital conversion board (left), midplane (center), and UniBoard (right), cover removed.}
	\label{fig:ADU}
\end{figure}

\begin{figure}
	\resizebox{\hsize}{!}{\includegraphics{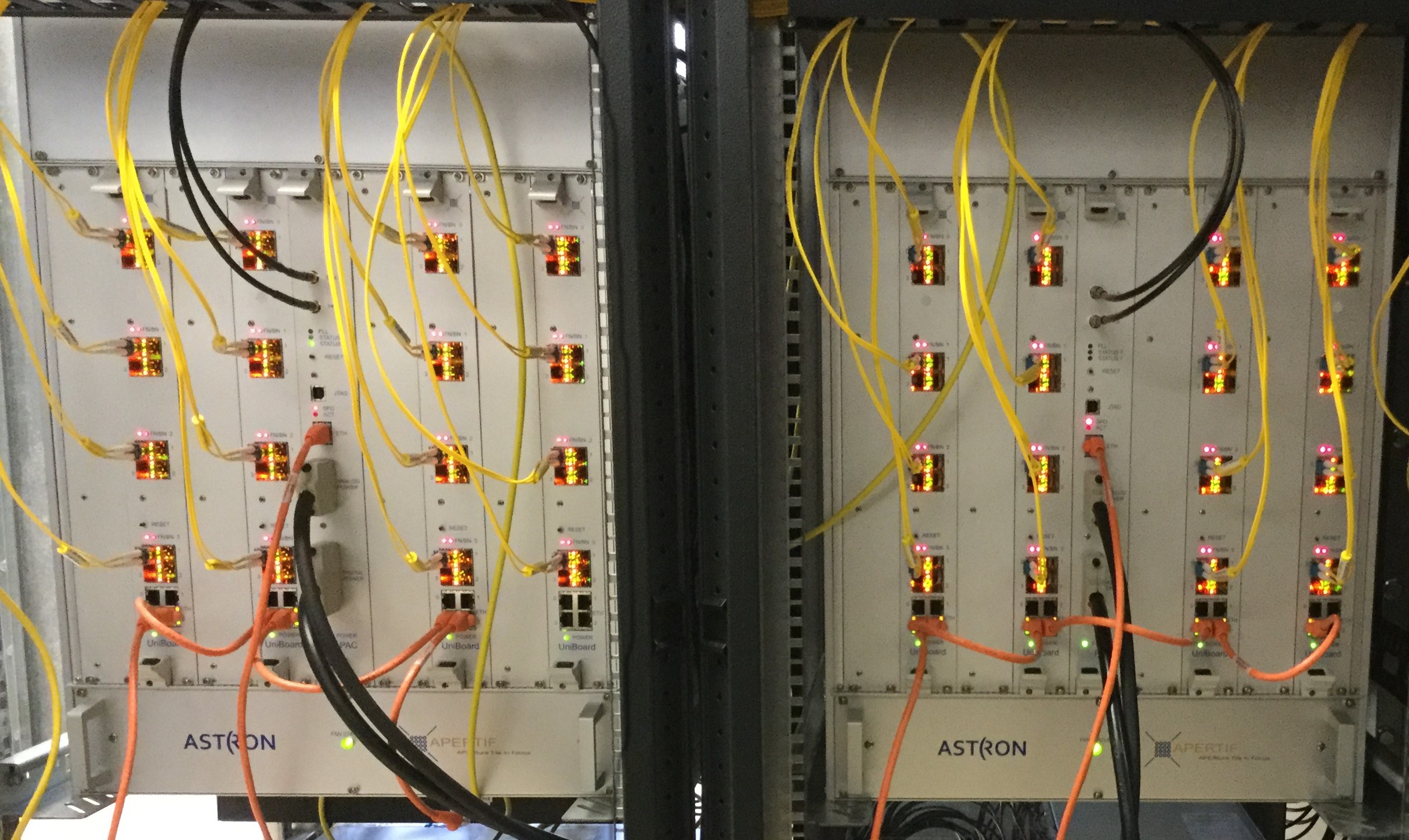}}
	\caption{Photo of two UniRacks in a Telescope Cabin (UniBoard side of the rack).}
	\label{fig:unirack}
\end{figure}

\begin{figure}
	\resizebox{\hsize}{!}{\includegraphics{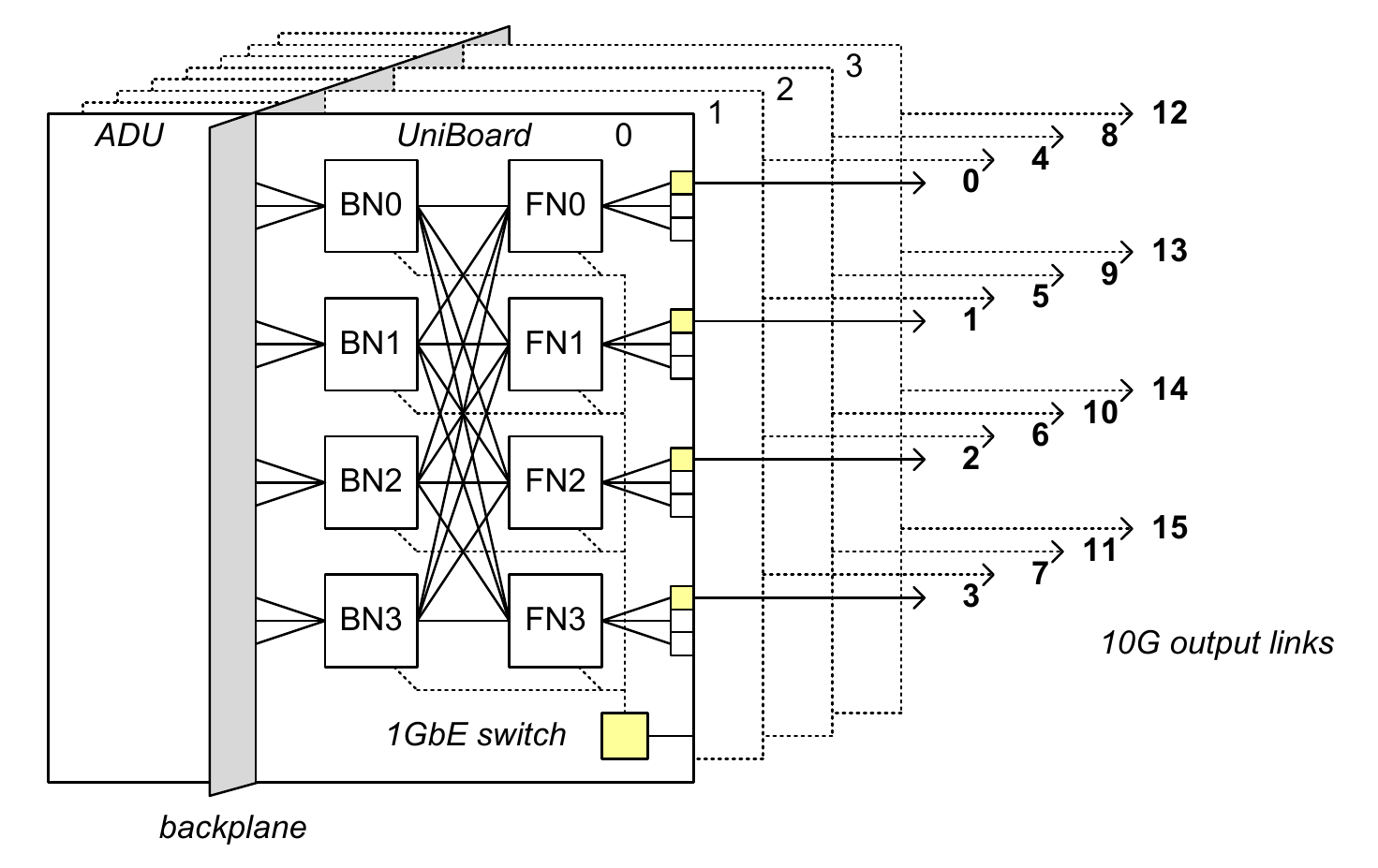}}
	\caption{Schematic overview of datapaths in a UniRack: ADU boards on the left side, a midplane to distribute the data and FPGA processing board with 10 Gbps outputs on the right side.}
	\label{fig:datapaths}
\end{figure}

One UniRack houses 8 Analog to Digital Unit (ADU) boards, 4 UniBoards, and a Power And Clock (PAC) board. A midplane distributes the data between the ADUs on one side of the rack to the UniBoards on the other side (see Figs.~\ref{fig:ADU} and \ref{fig:unirack}). The PAC board provides the power and reference clock signal to all other boards via the midplane. The 8-bit ADCs operate at 800~MSPS, sampling the baseband signal in the second Nyquist zone. A variable attenuator with a resolution of 1~dB allows tuning the input signal optimally to the dynamic range of the ADC. The mixing scheme (Fig.~\ref{fig:mixingscheme}) shifts the negative frequencies of the original RF band into the positive Nyquist zone of the ADC, so the band is flipped. Sampling in the second Nyquist zone flips the band again. Therefore the net result is that after sampling, the RF frequencies are represented at baseband in the same incrementing order. The UniRack itself, including the ADU boards, is shielded to reduce the coupling of spurious signals from the digital electronics into the analogue signal chain.

\begin{figure*}
	\centering
	\includegraphics[width=17cm]{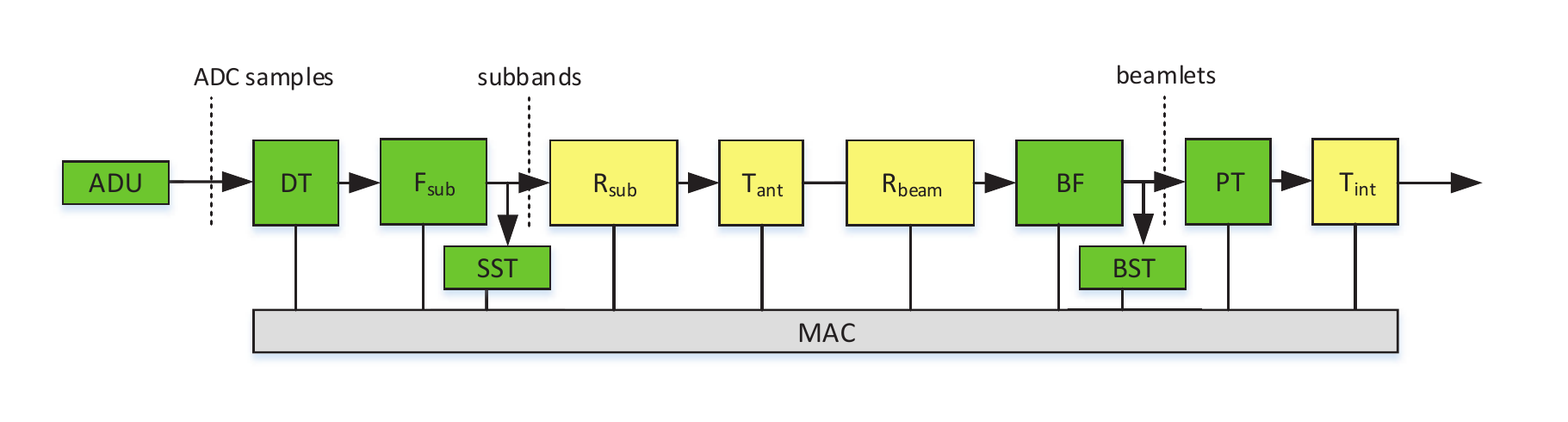}
	\caption{Overview of the phased array feed dish-level digital processing}
	\label{fig:bffirmware}
\end{figure*}

 Fig.~\ref{fig:bffirmware} provides an overview of the digital signal processing at the dish-level. First, the 400~MHz wide signal of every PAF element is split into 512 $\times$ 781\,250~Hz wide subbands (also known as coarse channels) by a critically sampled polyphase filter bank (F\textsubscript{sub}). Inevitably, aliasing occurs at the band edges of the critically sampled subbands, resulting in sharp dips in the bandpass function that are hard to compensate. The R\textsubscript{sub} block selects the 384 central subbands covering 300~MHz bandwidth, and discards the 50~MHz band edges at each side. The beam former for one subband requires the input from all PAF antenna elements, so therefore there needs to be a transpose T\textsubscript{ant} that groups the subbands from all antenna elements. For each of the 384 subbands the beam former (BF) forms 40 beamlets from the antenna inputs. A beamlet is a quantity that represents the information of one beam spanning one subband. The beam former is distributed over 64 BF units per UniRack. The R\textsubscript{beam} block holds and selects the subband such that the beam former can create the beamlets of the CB. In each beam former, the responses of all elements are multiplied with complex weights (i.e. magnitude and phase steering) and summed. The T\textsubscript{int} function is a corner turn that transposes and groups the data in time to provide the data in an optimal way for the correlator.  The Monitoring and Control (MAC) function takes care of the proper operation, the subband selection and the beam former weights.
 
 To monitor and to tune up the system, the power spectrum of every PAF element is obtained by calculating the autocorrelation statistics of the subbands. Similarly, the power spectrum of every compound beam is calculated. These power spectra are produced by the SST and BST blocks in Fig.~\ref{fig:bffirmware}. The integration interval for these spectra is 1.024~s. These monitoring facilities run in parallel to the regular data path and are always available.

The statistics function also has a special mode in which it calculates the subband crosscorrelation statistics between any two PAF elements \citep{Schoonderbeek}. The resulting covariance matrix of the PAF elements is used to determine the beam former weights as described in Sec.~\ref{sec:bf}. The measurement of a full $64\times 64$ cross correlation matrix takes about 2~minutes for all subbands. 

The two PAF polarisations are processed independently in two UniRacks. The beam forming for one polarisation of the PAF cannot be done on a single FPGA node for the full bandwidth, so the subband load has to be distributed across 16 FPGA nodes. Each UniBoard has 4 back node (BN) FPGAs and 4 front node (FN) FPGAs (see Fig.~\ref{fig:datapaths}). In a subrack the back node FPGAs implement the F\textsubscript{sub} and R\textsubscript{sub} of Fig.~\ref{fig:bffirmware}, and the front node FPGAs implement R\textsubscript{beam} and BF. The transpose T\textsubscript{ant} is implemented in two hops. The first hop is via the backplane that implements a full mesh interconnect between the corresponding BN on the 4 UniBoards. The second hop is via the mesh interconnect between BN and FN FPGAs on each Uniboard.

In total, the Apertif digital beam formers at the dishes contain 1536 ADCs on 192 ADU boards, and 768 FPGAs on 96 UniBoards. The total input data rate into the real time DSP is 9.8 Tbps. The central Apertif correlator (see section \ref{sec:correlator}) is implemented on 16 UniBoards with 128 FPGAs. Hence in total Apertif has 112 UniBoards and 896 FPGAs. Table \ref{tab:fpgaresources} shows what fraction of the Stratix IV EP4SGX230KF40C2 FPGA resources are used by the BN filterbank image, the FN beam former image at the dishes, and the central correlator image. The following resources are distinguished: The Adaptive Logic Module (ALM) that contains 2 Flip Flops + logic, Flip Flop (FF), 1 kByte block RAM (M9K), 16 kByte block RAM (M144K), and multipliers (DSP). The correlator uses more than about 85\% of the FPGA resources, which is relatively high. As a consequence, it was challenging to synthesise an image that can achieve the required 200~MHz clock.

\begin{table*}
	\caption{FPGA resource usage, as percentage of the maximum available number of resources.} % title of Table
	\label{tab:fpgaresources} % is used to refer this table in the text
	\centering % used for centering table
	\begin{tabular}{l r r r r r} % centered columns (4 columns)
		\hline\hline % inserts double horizontal lines
		  & ALM & FF & M9K & M144K &  DSP \\
		\hline
		Available max  & 91\,200 & 182\,400 & 1235 & 22 & 1288 \\
		\hline
        BN Filterbank & 88\% &  61\% &  61\% &   36\% &  68\% \\
        FN Beam former & 62\% &  43\% &  83\% &   68\% &  82\% \\
        Correlator    & 99\% &  83\% &  87\% &  100\% &  84\% \\
		\hline %inserts single line
	\end{tabular}
\end{table*}

\subsection{Geometrical delay compensation}
As the direction of incidence of a celestial signal on the telescope changes during an observation, its time of incidence varies with the dish location and the angle of incidence. Delay compensation is needed to compensate for the change in geometrical delay between dishes. The geometrical delay compensation has to be regularly updated to follow the rotation of the Earth. The delay compensation in Apertif consists of true integer sample delay tracking (DT in Fig.~\ref{fig:bffirmware}), followed by a residual phase tracking (PT). The delay tracking is known as the coarse delay tracking and the phase tracking is known as the fine delay tracking. The DT operates on the ADC samples at the input of the signal processor. The phase tracking operates on the beamlets per compound beam and thus stops the fringe for the centre of each CB. The maximum fringe rate of the Apertif telescope is 1.28 Hz. To avoid too much decorrelation the update rate for the phase tracking needs to be about a factor 100 faster than the maximum fringe rate, so a fine delay update interval of $T_{fine} = 7.8$ ms. This update rate is too high for the MAC interface and therefore the PT function uses piecewise linear interpolation, whereby the MAC only needs to provide the phase offset and phase slope coefficients for each linearised phase interval. The update interval for these linear PT coefficients is $< 10.3$ s. The fringe rate can also be interpreted as a differential Doppler shift between two dishes. The maximum fringe rate of 1.28~Hz is very small compared to the subband bandwidth. Therefore the delay compensation in Apertif does not need Doppler shift correction on the sampled RFBW band. The residual delay that remains after the delay tracking is 5~ns, or 4 ADC samples. To avoid significant decorrelation, the minimal subband period needs to be at least a factor 100 larger than the delay step, so $> 500$ ns and thus the subband bandwidth must be $< 2$ MHz. The subband bandwidth is about equal to the subband rate of 781,250~Hz, so the 512 subbands that are created by F\textsubscript{sub} are enough.

\subsection{PAF beam forming}\label{sec:bf}
As introduced in Sec.~\ref{sec:systemoverview}, the PAF beam former produces 40 dish-level compound beams by forming weighted superpositions of the element responses. These compound beams have the same role as the primary beams of dishes with a single receiver. The properties of the compound beams are defined by the scheme that is used to determine the weights. Apertif implemented a beam forming scheme that maximises the sensitivity \citep{6018280} of the compound beams. Although not yet implemented in Apertif, the flexibility of the PAF concept allows beam shape constraints to be added, for example to reduce the beam width variation with frequency \citep{5725169} or mitigate radio frequency interference (RFI) \citep{4705793}.

The Apertif dish-level beam former implements a bi-scalar beam forming approach, in which the two sets of orthogonally placed PAF elements are processed independently. Bi-scalar beam forming incurs only a marginal loss in sensitivity (\textasciitilde 4\% for Apertif) compared to optimal full polarimetric beam formers \citep{6236092,6050647,6206652} and saves a significant data handling and processing load. 

Apertif determines its weights by measuring array covariance matrices using the PAF subband correlator while observing a strong celestial point source (Cassiopeia A) and a relatively cold spot on the sky. Because the weights are determined with the actual system, all instrumental effects such as gain and phase differences between receiving channels, antenna mutual coupling, spill-over noise, and feed blockage, are implicitly taken into account. 

In the weight calculation, two normalisation steps are implemented. The first ensures that the beam former adds no gain, since that would conflict with the quantisation levels in the beam former firmware. The second ensures a smooth frequency response, since the above mentioned method to determine weights is done for each subband independently, which could lead to jumps from subband to subband. To ensure a smooth frequency response, the weights are normalised such that the receiver chain that has the largest weight amplitude is real by definition. In some cases, there is no unique receiver chain with a highest weight. In that case a particular receiver chain has the largest amplitude for one subband, while in a second subband another receiver chain has the largest amplitude. To prevent the normalisation from toggling between receiver chains over the frequency band, this selection is made based on the data for a reference subband.

\begin{figure}
	\resizebox{\hsize}{!}{\includegraphics{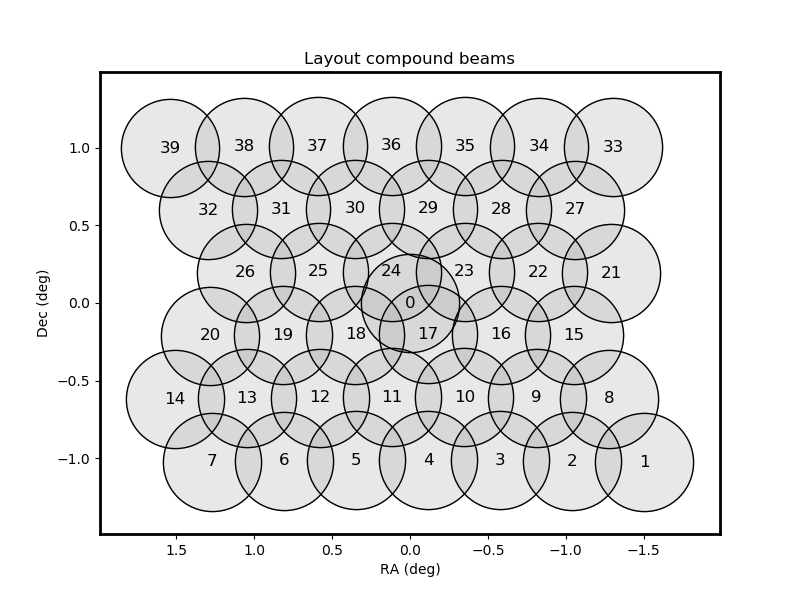}}
	\caption{Layout of locations of the 40 compound beams over the field of view (Hess et al., in prep.).}
	\label{fig:beampos}
\end{figure}

\begin{figure*}
    \resizebox{0.48\hsize}{!}{\includegraphics{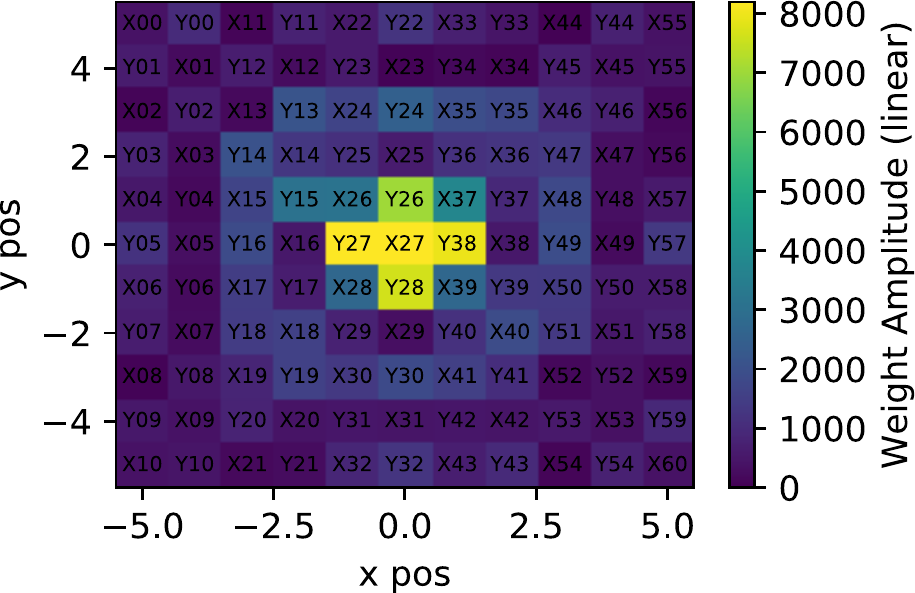}}   
	\resizebox{0.48\hsize}{!}{\includegraphics{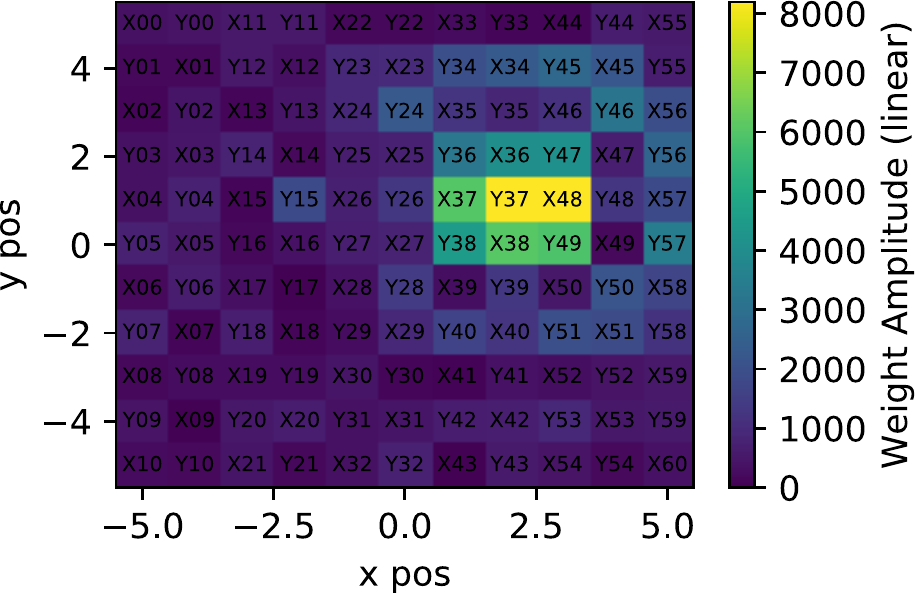}}
	\caption{Magnitude of the beam former weights of compound beams 0 (left) and 9 (right).}
	\label{fig:measweights}
\end{figure*}

For all observations, Apertif uses the beam layout shown in Fig.~\ref{fig:beampos}. Thirty nine beams are formed  which  are laid out in a hexagonal pattern. One additional beam (beam 0) is formed at the location of the optical axis of the antenna dish, which is convenient for test purposes. The weight magnitudes of two beams are shown in Fig.~\ref{fig:measweights}. It can be seen that in a PAF system only a few elements per beam have high weight. All other elements have weights that are at least 10 dB lower, but are still important to optimise the details of the compound beam. The on-axis beam (beam 0) is dominated by the response of the central X element and the surrounding Y elements. The off-axis beam (beam 9) is dominated by two elements. The position of the highly weighted elements with respect to the geometric focus point determines the pointing offset of the compound beam with respect to the central beam.

\section{Central Processing}\label{sec:centralprocessing}
The Apertif central processing combines the data from the 12 Apertif telescopes. The WSRT-Apertif system has two modes of operation: Beam formed modes that sum the response of the telescopes, and an imaging mode in which the telescopes are correlated into visibilities.

\subsection{Tied-array beam former}
\begin{figure*}[ht]
\centering
\includegraphics[width=\textwidth]{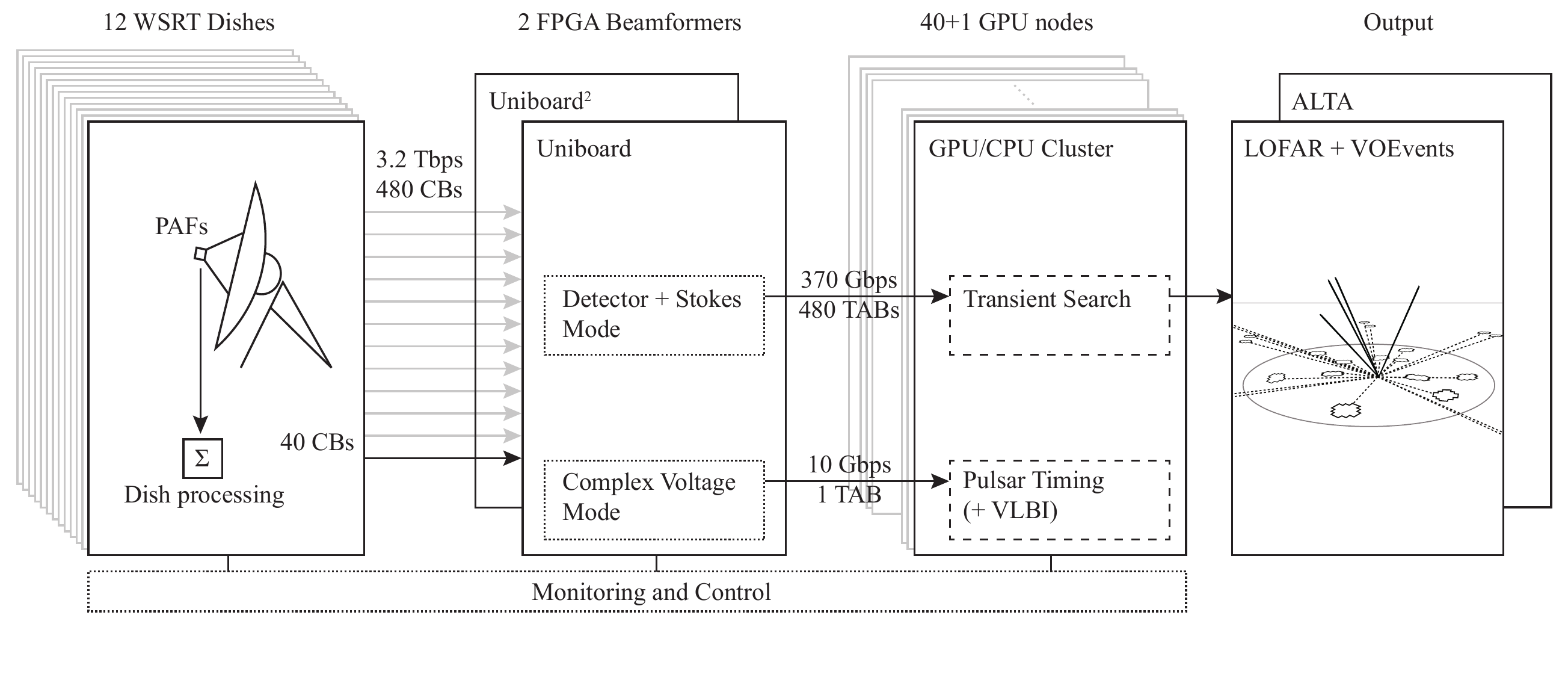}
\caption{Top-level diagram of ARTS, using the Apertif beam formed mode \citep{ARTS2}.}
\label{fig:toplevel}
\end{figure*}

The beam formed modes are primarily used for the Apertif Radio Transient System (ARTS), a supercomputing radio-telescope instrument that performs real-time Fast Radio Burst (FRB) detection and localisation on the WSRT, and can immediately alert LOFAR for incoming bursts. 
The system is extensively described in \citet{ARTS2}. As shown in Fig.~\ref{fig:toplevel}, 
the implementation of the beam former  is similar to that of the correlator, described in the next subsection. 
Compound beam data from the dishes are combined at the central location, using Uniboards and Uniboard$^2$s. 
On the Uniboards, the beam-forming modes are an alternative to the correlator mode described below. 
The system can run in a one of these two modes at a time.
The subsequent transient detections functions \citep{2016A&C....14....1S, sclocco_amber_2020} and post-processing software \citep[][]{oostrum_darc_2020}
are implemented on a powerful GPU cluster.

\subsection{Correlator} \label{sec:correlator}
\begin{figure}
	\resizebox{\hsize}{!}{\includegraphics{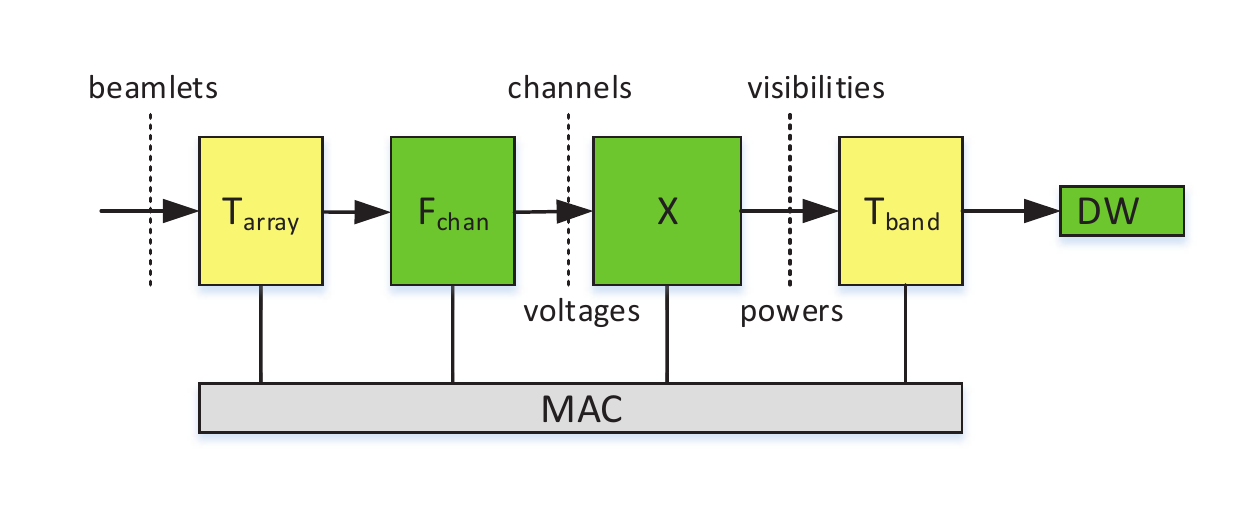}}
	\caption{The Apertif correlator subsystem.}
	\label{fig:corfirmware}
\end{figure}

For imaging operations, a full Stokes correlator is implemented. The correlator is based on the FX architecture. Fig.~\ref{fig:corfirmware} shows a functional overview. First a polyphase filterbank filters the 781\,250~Hz wide subbands into 64 channels (or \emph{fine channels}) of 12.2 kHz each. The correlator then performs the crosscorrelation between all pairs of the 24 telescope inputs (12 telescopes times 2 polarisations). The correlator calculates 276 crosscorrelations and 24 autocorrelations, so 300 visibilities, for all 24\,576 channels of all 40 compound beams. The cross products are integrated for 1.024 s. After each integration interval the visibilities are offloaded to the Data Writer. The MAC takes care of the proper operation of the correlator, which mainly concerns monitoring the input and configuring the output, because the correlator itself has no control.

\begin{figure}
	\resizebox{\hsize}{!}{\includegraphics{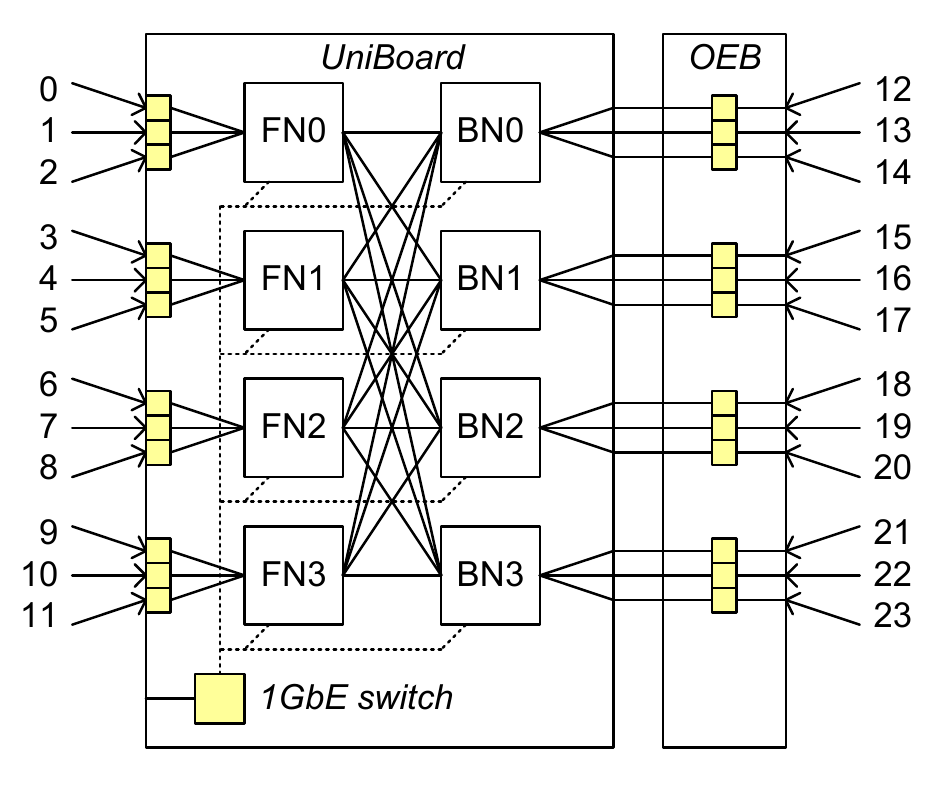}}
	\caption{Datapaths in the Apertif correlator.}
	\label{fig:cordatapaths}
\end{figure}

The Apertif correlator uses 16 UniBoard processing boards. To increase the number of optical inputs into the correlator, 16 Optical-Electrical Boards (OEB) are connected directly to the Uniboards (see Fig.~\ref{fig:cordatapaths}). In the Apertif correlator application, all 8 FPGAs on a UniBoard have the same function. Each FPGA has 3 10G links, so in total one UniBoard deals with 24 10GbE links from the dishes. This is just enough IO to connect to the 24 10GbE links that are processed by each board.  Each FPGA in the Apertif correlator correlates all 300 visibilities from the 24 telescope inputs. The visibilities, a data stream of 18.4 Gbps, are output via the 1GbE interfaces on the UniBoards and arrive via 4 10GbE interfaces at the data writer, see Fig.~\ref{fig:datapaths}. The transpose function (T\textsubscript{band}) is implemented through Ethernet switches. 
 
\subsection{Data Writer}
The main goal of the Data Writer is data reduction. The correlator delivers visibility data that is integrated for approximately one second, but the scientific requirements on time smearing allow for further integration of the data up to approximately 30~s, yielding a data reduction of a factor thirty. Secondly, the Data Writer applies phase corrections to the individual frequency channels within the subbands that were formed in the compound beam former at the dishes. This correction is required to avoid phase variations of up to $\pm 34$ degrees between the lowest and highest channels in a subband for the most off-centred compound beams. The phase slope across the band for off-center CBs arises because the delay tracking (DT) is the same for all CBs and optimal only for the central one. Lastly, the Data Writer writes the visibility data into separate Measurement Sets, one per compound beam. Each Measurement Set is augmented with all the metadata that is necessary to fully describe the data, hence ensuring that each data product is completely self-describing.
%Hardware and configuration

Apertif is equipped with two identical Data Writer systems that are used alternately to avoid that one node has to write and read at a high date rate at the same time. While one node is recording a running observation, the dataset of the previous observation is offloaded to a post-processing cluster. Each Data Writer node is equipped with an Intel Xeon E5-2630 v3 CPU, having eight physical cores with hyper-threading; 64 GiB of RAM; eight 4~TB disks configured in RAID-6, yielding a net storage of 24 TB; and four 10GbE network interfaces. Several adjustments were made to the configuration of the system to improve the handling of large amounts of incoming UDP packets.

%Input Reader
The Input Reader is the most performance critical part of the Data Writer software. It needs to handle the incoming UDP data stream of approximately 19 Gbps. The header of each data packet contains a time stamp. Since data packets may not arrive in the correct time order, the Input Reader must reorder them. Packets with a time stamp after the observation’s stop time are discarded immediately. The Data Writer uses a best-effort queue that consists of a number of so-called Integration Blocks. Each Integration Block holds data packets associated with a given integration interval, which is determined by the time stamp in the header. A new Integration Block is created in the best-effort queue when the time stamp in the packet belongs to the next integration interval. If the Output Writer cannot keep up, the best-effort queue may become full. In that case, no new Integration Block can be created, and the incoming data packet is dropped.

%Processing
In order to reduce the size of the data products as much as possible, the visibility data are further integrated in time by the Data Writer. In theory, the correlator could already do so, but the UniBoards do not have enough memory available for buffering. The Data Writer also takes care of aggregating the different flags that can be set to indicate problems with the data. Flagged data is excised when averaging. Therefore, the time centroid of each integrated frequency channel must be recorded as meta data in the Measurement Set, as well as the weight that needs to be applied to normalise the data.

%Fringe Stopping
The beam-former in the PAF only performs fringe stopping per frequency subband. However, when each subband is further split up into channels at the correlator, we need to correct for the phase slope that remains between the lowest and highest frequency channels in that subband. If we do not correct for this, then the residual phase error can become as large as $\pm 34$ degrees for the most off-centred compound beams.

%Output Writer
The Output Writer is responsible for writing the integrated and fringe stopped visibilities of each compound beam to a separate Measurement Set, together with all the metadata that is needed in order to make the Measurement Set fully self-describing. All data processing takes place on data that is stored in one of the Integration Blocks in the best-effort queue. When all data of a given Integration Block have been processed, they are written to disk. Up to four compound beams are handled in parallel by the Output Writer.

\section{Software control}\label{sec:mac}
\subsection{Monitor And Control system}\label{sec:mac2}
The Apertif Monitoring And Control (MAC) subsystem \citep{2016ADASS_mac} has four major responsibilities:
\begin{itemize}
    \item Distribution of information
    \item Solve timing and sequence related issues
    \item Control all the involved hardware that is distributed over many locations
    \item Provide monitoring information for state and incident tracking
\end{itemize}

\begin{figure}
	\resizebox{\hsize}{!}{\includegraphics{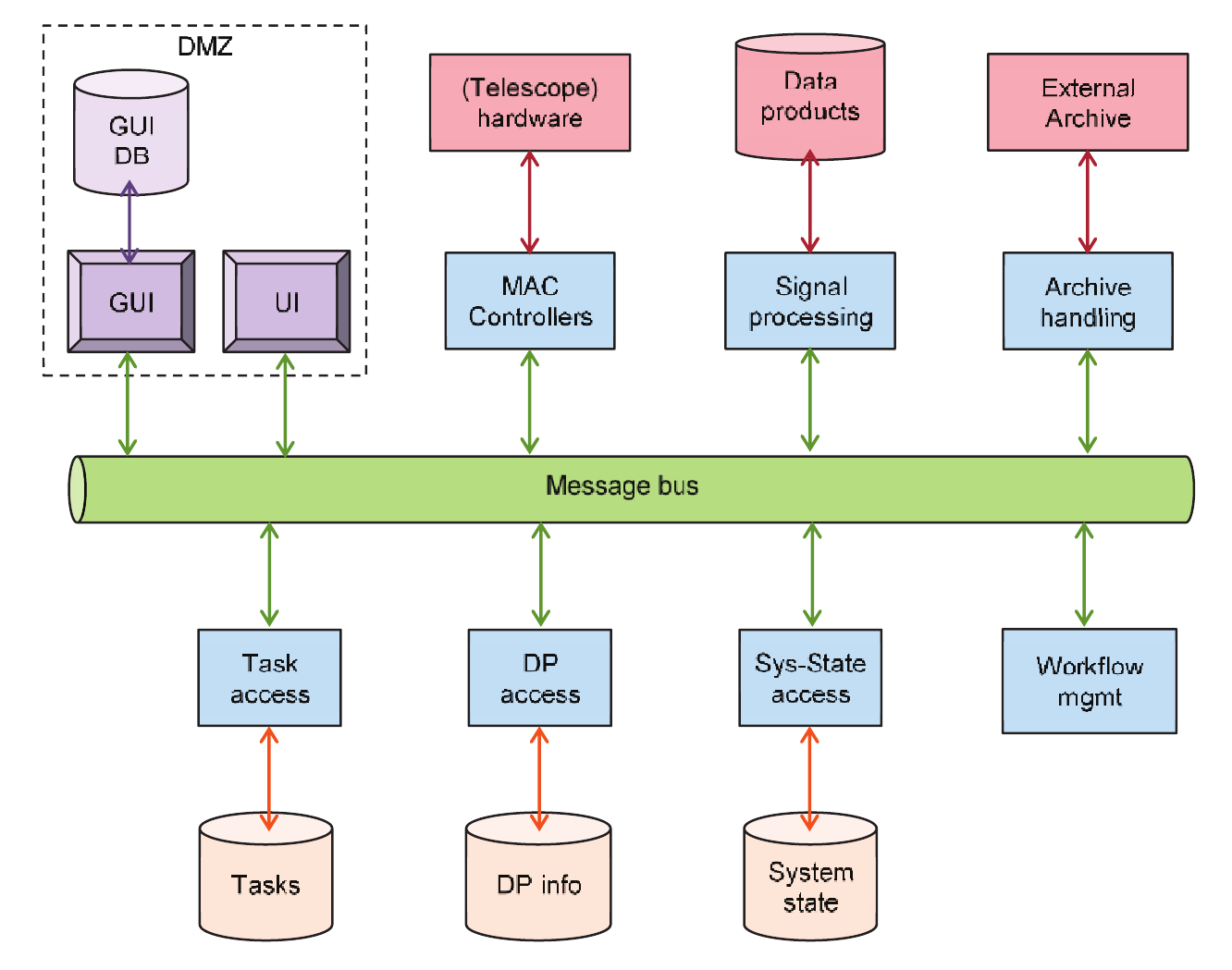}}
	\caption{High-level design of the MAC system of Apertif.}
	\label{fig:mac}
\end{figure}

The architecture of the MAC subsystem is based on loosely coupled components that communicate via a message bus. The message bus technology assures, among others, guaranteed information delivery, persistency where needed, distribution of information over multiple machines, and solves many other communication paradigms. 
Complementary to this technology, many components were developed to implement the required functionality of the MAC layer of Apertif. Since information exchange between components is easy and reliable, the scope of each component can be kept small thereby making each component less complicated. Fig.~\ref{fig:mac} shows the high-level design of the MAC system.

The basis for the distribution of information is the message bus technology. We chose to use the open source package QPID which is based on the widely supported messaging standard AMQP~1.0. This package offers many features like guaranteed delivery, routing, multi-machine support, controlled lifetime of messages, persistency, broadcasting, etc. On top of these features, a thin layer was developed for Apertif to simplify communication paradigms like publish-subscribe, remote procedure call, scatter and gather. Furthermore, we defined different message categories like command, event, notification, etc. each having its own properties that serve their way of information exchange.

\subsection{MAC components}
To create a reliable MAC system, we also defined the behaviour of any component that is part of the MAC system. This means that any form of communication between two or more components is well defined and understood. To go one step further we developed a component framework that implements this required communication behaviour. Guarded by timers and catching exceptions, the framework waits for messages on the message bus, executes the corresponding user-defined code and finally sends a response to the message bus. The framework offers many benefits: All communication paradigms are implemented only once which, to a large extent, decreases and simplifies maintenance, and the implementation of a component by a developer is reduced to the implementation of the functionality that is needed to handle the messages.

\begin{figure}
	\resizebox{\hsize}{!}{\includegraphics{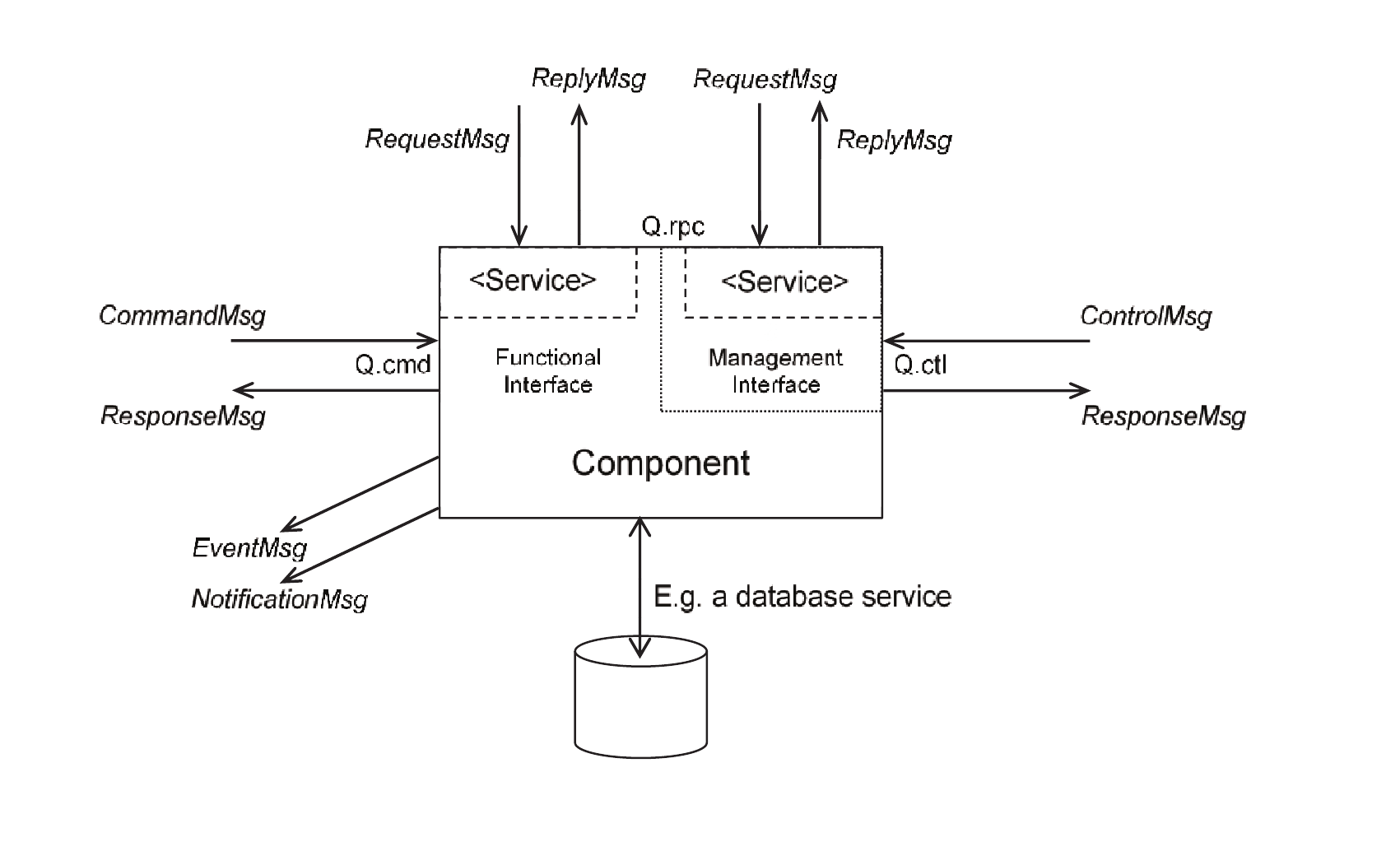}}
	\caption{Typical interfaces of a component.}
	\label{fig:interfaces}
\end{figure}

The diagram in Fig.~\ref{fig:interfaces} gives a schematic view on an Apertif component.

Currently the Apertif system contains three controllers for controlling all hardware: DirectionController, SignalController, and CorrelatorController. There is one instance of the DirectionController that is responsible for the pointing of all 12 dishes of Apertif. To control all the other hardware that is situated at each telescope, like DCUs, LOGs, PACs, ADUs, and UniBoards, each telescope has its own SignalController. The control of all these boards, more than 60 per telescope, is a complex task given the mutual dependencies of the boards. To prepare the telescope for an observation many settings have to be checked or set for each of the 121 data streams that will come from the PAF. Only when all settings are correct the telescope can participate in the observation. The configuration of all the boards can take up to approximately 6 seconds when all the boards need a reset, during this period no other commands may interfere with the complex sequence of initialisation. The component framework takes care of this, so there is no extra code needed from the developer to accomplish this. CorrelatorControl is responsible for configuring the 384 input streams of the telescopes in the right way and also for configuring the output streams of the correlator so that they connect to the machine where the DataWriter is started.

\subsection{Monitoring}
Monitoring the system can be done in two ways: actively and passively. Active monitoring can be done by sending a component a request message to report on its health, settings, or other information. The component framework will then execute the related user code to collect this information and send the answer to the requester. To collect the information of the whole system all components need to be polled.

Another way of monitoring the system is to let each component by itself report the state changes of the part of the system it is controlling. To support this, the controller framework implements a table that can contain commands that will be executed at a given interval. Depending on the implementation of the command by the developer, the command can deliver all current values as a notification message or deliver only the differences since the last time it was executed as an event message. In that case, the monitoring component receives all notifications and events, and registers the information in a database. It enables easy implementation of collecting the trends of values. The monitoring component can send out alarm messages when values or combination of values run outside their limits. In Apertif we implemented the passive mechanism but the information can still be collected in an active way also.

\subsection{ARTAMIS}\label{sec:artamis}

To collect and visualize monitor data for the Apertif operator and engineers, we employed Siemens' WinCC-OA Scada framework. WinCC-OA provides functionality such as a real-time event database, archiving, alarming, GUI development in a single toolkit. Being a single package, installation and implementation of the required functionality is easy and straightforward. To be able to store the collected monitor data in the database, we had to create a software component that collects data from the QPID messagebus (see Sec.~\ref{sec:mac2}) and translates those messages into the proper database points of the WinCC-OA database. Once in the database, WinCC-OA provided tooling can be used for processing, alarming and visualisation of that data.

\begin{figure}
	\resizebox{\hsize}{!}{\includegraphics{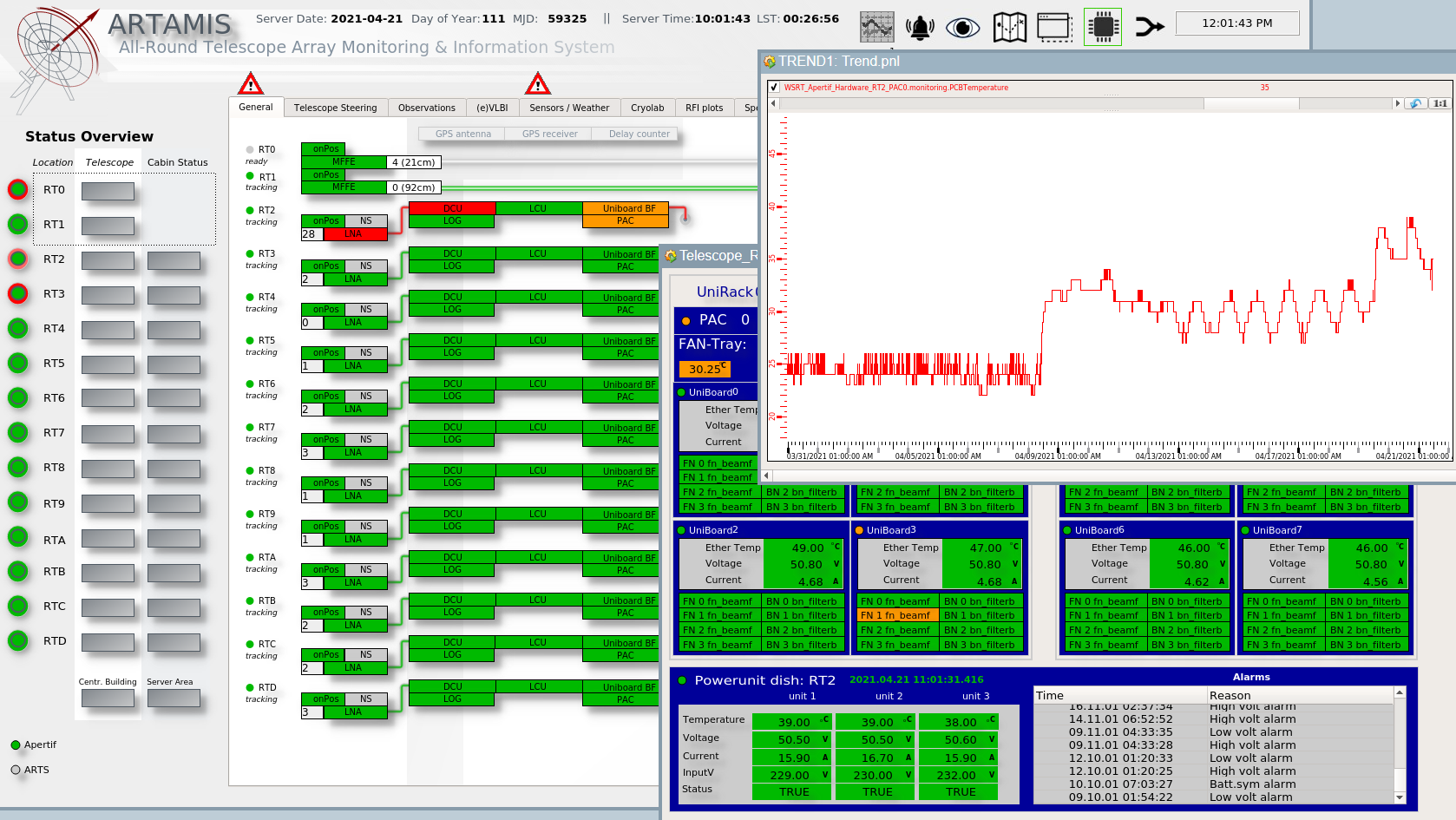}}
	\caption{Screenshot of the ARTAMIS monitoring tool of Apertif.}
	\label{fig:artamis}
\end{figure}

The specific tool that we developed for Apertif has been named ARTAMIS, which is an acronym for "All-Round Telescope Array Monitoring and Information System". 
ARTAMIS can be logged into via a dedicated desktop tool, or via a web based interface, which makes it possible to actively monitor the system from any location: the operator control room in the ASTRON main building, the WSRT-based control room, the operator's home, or any other connected location. This flexibility has proven to be essential in the recent COVID-19 period, where access to ASTRON's building has been severely limited by government regulations.
ARTAMIS provides multiple views on the system: hierarchical views based on hardware components at different scales (Array -> Telescope -> Beamformer -> Uniboards), up to the level of individual electronics boards. Furthermore, we provide a datastream-centric view showing the network topology and science data streams through the network from reception to storage. These views are complementary and which one most useful depends on the user's intent. In Figure \ref{fig:artamis} we show an example screenshot of the graphical user interface.
Apart from the hardware view, ARTAMIS provides views of the ATDB-based schedule (see section \ref{sec:atdb}), of the Archive (see section \ref{sec:alta}), but also of the current state of non-APERTIF telescopes. With ARTAMIS, we thus built a single tool for the monitoring of the entire WSRT system.

\subsection{Apertif Task Database}\label{sec:atdb}
The Apertif Task Database (ATDB) forms the core of Apertif's operational software. Based on the survey schedule provided by astronomers, ATDB autonomously manages the full operational flow. From sending the observation parameters to the telescope system, up to data ingest into the Apertif Long Term Archive (ALTA), and removal of the raw data from the data writers. Despite being autonomous, ATDB also provides a web GUI for fine grained control by operators to troubleshoot, and to make changes in the data flow during the process. Astronomers can use ATDB to qualify observations as good or bad, after which a signal is sent to ALTA to either remove the data from the archive or transfer it to our cold storage facility at SURF.

ATDB is a web application that consists of a central database, surrounded by a pattern of micro services that all perform individual tasks based on the status of an observation in the central database. A stabilising feature of ATDB is that all the required logic is spread over small independent services that run on a myriad of different machines, but all communicate with the central database. The services are built and run in such a way that they restart whenever they fail for whatever reason, so that the mechanism can be seen as self correcting. Services can also be taken offline and changed individually without affecting the rest of the system. The central ATDB back-end can therefore remain simple and stable, limiting both maintenance and the risk of introducing bugs. In the 2.5 years of continuous operation ATDB has only failed once, due to hardware failure of an overheating computer.

\section{Apertif Long Term Archive (ALTA)}\label{sec:alta}
\begin{figure}
	\resizebox{\hsize}{!}{\includegraphics{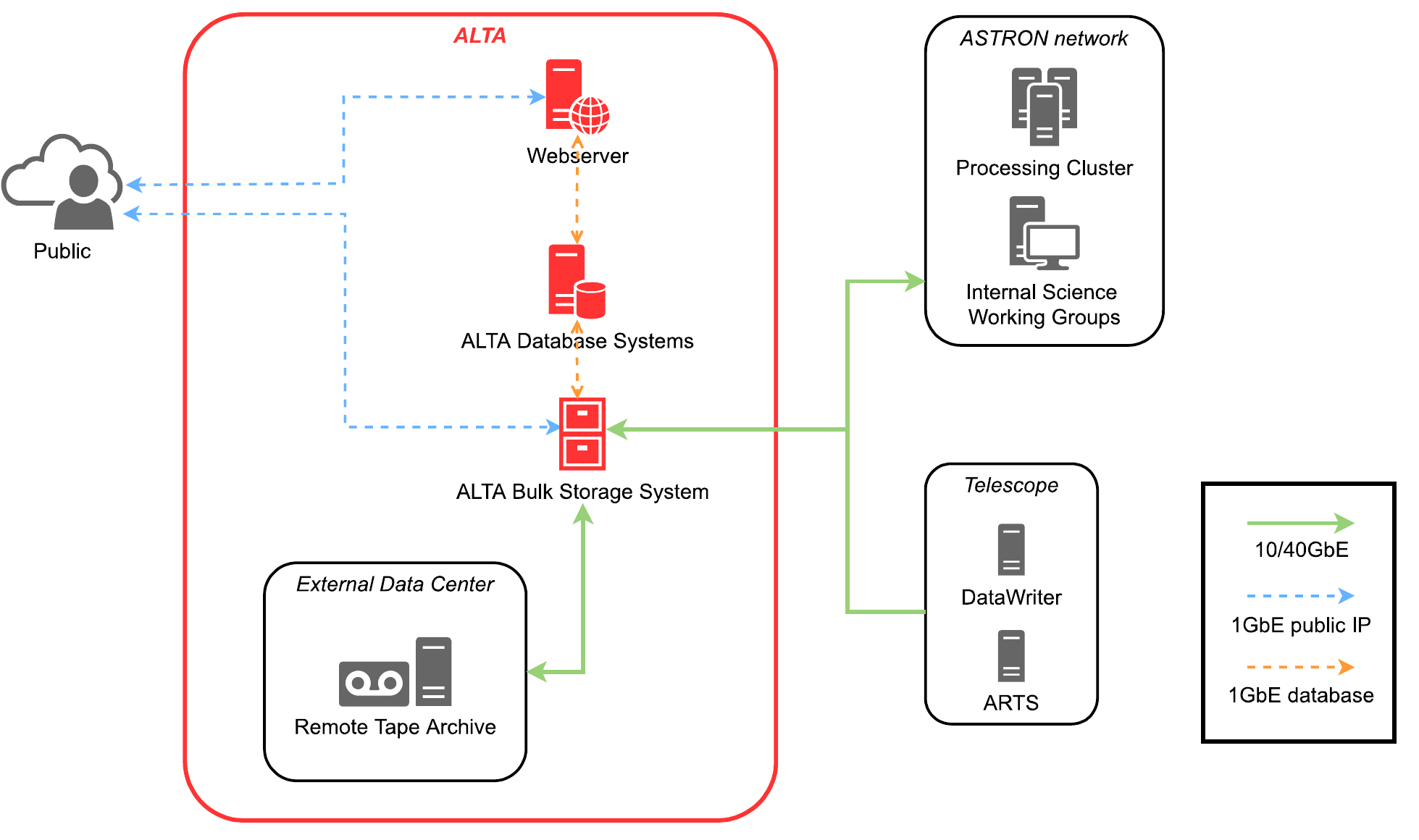}}
	\caption{The Apertif Long-Term Archive (ALTA) system.}
	\label{fig:alta}
\end{figure}

The Apertif Long-Term Archive (ALTA) \citep{2019ADASS_alta} provides storage and dissemination capabilities for the scientific data generated by the Apertif system. It supports telescope operations by providing continuous data intake capabilities. The scientists are supported with data management capabilities including overviews of observed and processed datasets, data selection and retrieval, quality annotations, and data ingest facilities for derived data products. Released data-products are published through ALTA to the public domain. A web-based user and application interface provides the required functionality for the targeted user groups to interact with ALTA. Fig.~\ref{fig:alta} provides an overview of the ALTA system, its main components, and the main external systems it is connected to.
 
ALTA has been designed to collect and manage millions of data products with a total volume of twenty petabyte over the operational lifetime of the Apertif instrument. Internal and external network connections for data transfers are based on 10 and 40 Gbps Ethernet technology. ALTA encompasses storage resources in two locations: ASTRON hosts online storage that provides 1.8 petabyte of capacity to support the operational telescope workflow and the analysis workflows for the Survey teams. The tape-based long-term storage of ‘cold’ data is hosted by SURF, the Dutch national HPC and e-Science support centre. The nearline/tape back-end supports an annual growth of up to four petabytes.

The basis for the data management components of ALTA is provided by the iRODS\footnote{https://irods.org/} framework. This framework allows the deployment of storage resources in so-called zones. Zones typically represent independently managed storage resources and can be mapped to resources hosted by different organisations. For ALTA, a single Zone has been created that spans iRODS resources at both the ASTRON (four data servers) and the SURF location (one server connected to the tape backend) as these are managed centrally as a single storage environment. Integrated functionality is implemented in Python, and executed as scheduled services in the iRODS environment. These services validate data-products during the ingest process, extract the metadata that is incorporated in the database, generate inspection plots for data quality assessment, and handle the migration between disk and tape storage based on a set of policies. The system allows for future implementation of user oriented services such as on-demand image-cut-out and –stitching. 

The ALTA data model is based on the LOFAR data model and designed such that it can be easily mapped to VO-compliant data structures. For incorporation of full data provenance, inspiration has been drawn from the W3C provenance which has been adapted to fulfil the constraints of the ALTA data model. In particular, the ALTA provenance model is strongly process-oriented, enabling the inspection of the full chain of processing steps and intermediate data-products, starting from the observations. A central Postgresql relational database stores ALTA metadata using a schema that closely maps the structure of the ALTA data model and is fully exposed via a REST interface through the web user interface.

The GUI and API components of ALTA consist of a javascript-based interactive user front-end and a RESTful back-end. The backend builds on Django\footnote{https://www.djangoproject.com/} and the Django REST framework\footnote{https://www.django-rest-framework.org/} and handles the interaction with the database and data management components, both for the front-end and as an API for automation purposes. Data and information that is not yet released to the public domain requires an account with appropriate authorisation.

ALTA data collections are made publicly available through a release process where observations and processing pipeline runs are marked and subsequently exposed to anonymous users of the ALTA web interface. The associated data products are made available for retrieval via the WEBDAV protocol. Not all data is available for immediate access, in particular those products that are stored on tape. This applies mainly to the lower level data products such as timeseries data, uncalibrated and calibrated visibilities and the calibration tables generated by the calibration pipeline. Users are directed to a helpdesk service if they wish to retrieve data products not accessible immediately.  Released observations, processing runs, and associated data products are registered in the ASTRON virtual observatory server, running GAVO DaCHS\footnote{https://dachs-doc.readthedocs.io} software, from where they become findable and accessible via standard VO aware client tools. The first scientific data release supported by ALTA has been published in November~ 2020\footnote{http://hdl.handle.net/21.12136/B014022C-978B-40F6-96C6-1A3B1F4A3DB0}.

\section{Operations}\label{sec:operations}
The WSRT-Apertif telescope is operated as a survey instrument conducting surveys both in imaging and time-domain (tied array) mode. The operations are supported by both ASTRON observers and astronomers from the science teams. The operations are designed and run to support this dedicated survey observing mode. The cadence of observing is two to three weeks of dedicated survey observations. At the end of an observing period, system maintenance is performed. Following that, a beamweights measurement is performed. These beamweights are then used for the full period of observing that follows. Further information about how operations were designed and are implemented on a daily basis can be found below.

\subsection{Designing Apertif operations}
\label{designing_operations}
The focus of Apertif on conducting large surveys in a systematic way presented an opportunity to design an operational workflow that would suit this context. Like many new instruments and telescopes, the transition between commissioning and survey operations had fuzzy boundaries which meant that we needed to make use of commissioning experience to the greatest extent possible to prepare for full survey operations. During this period, a strong emphasis was placed on automation and streamlined workflows in order to guarantee the maximum efficiency possible with limited resources. We sought to minimise the locations in the data flow which relied on humans making decisions with the goal of limiting any potential bottlenecks, which in some cases marked a deviation from initial plans for the Apertif data flow. Doing this required looking critically at each stage of the observation and determining what steps could be made via automation to reduce these risks. 

\begin{figure}
    \centering
    \includegraphics[keepaspectratio,width=0.9\linewidth]{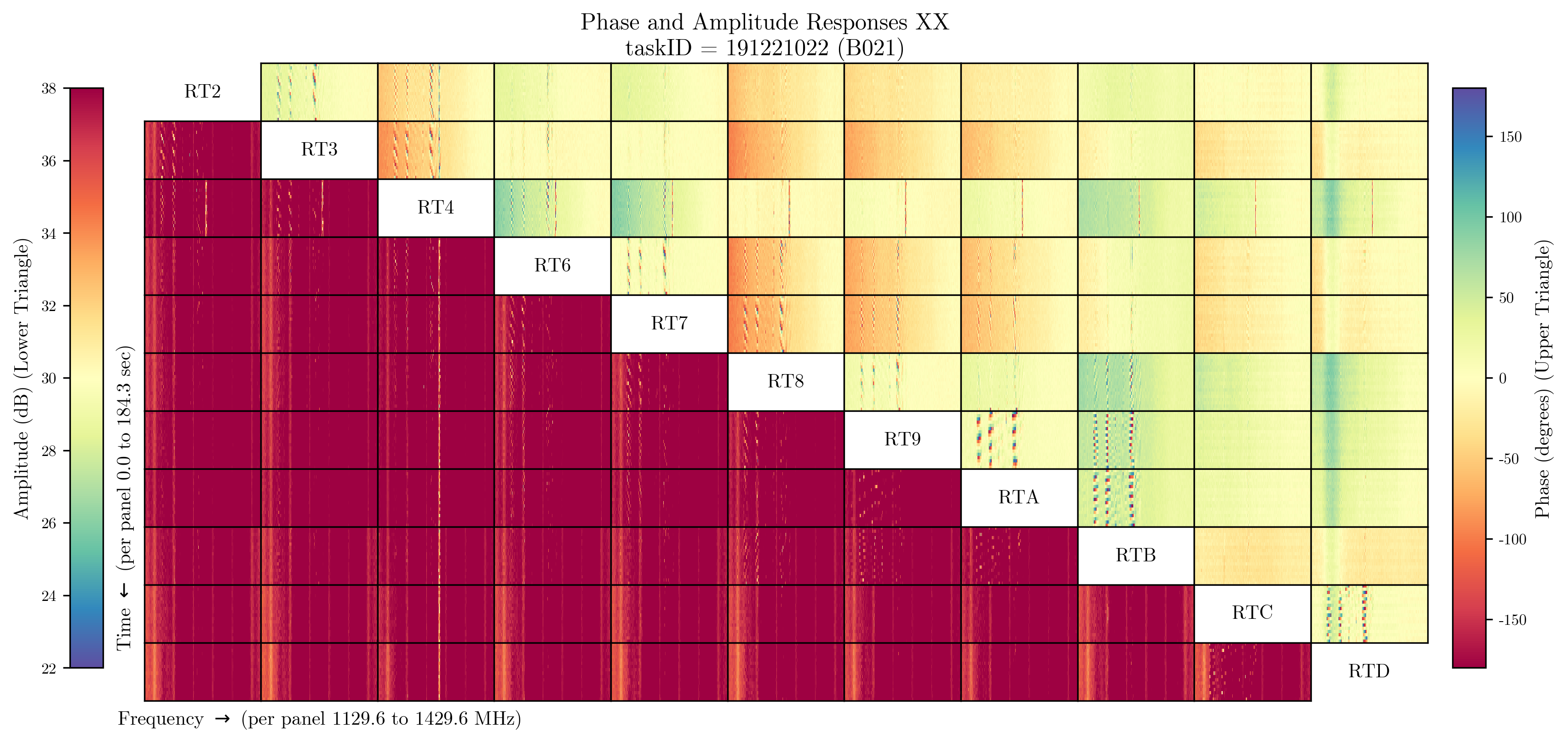}
    \quad
    \includegraphics[keepaspectratio,width=0.9\linewidth]{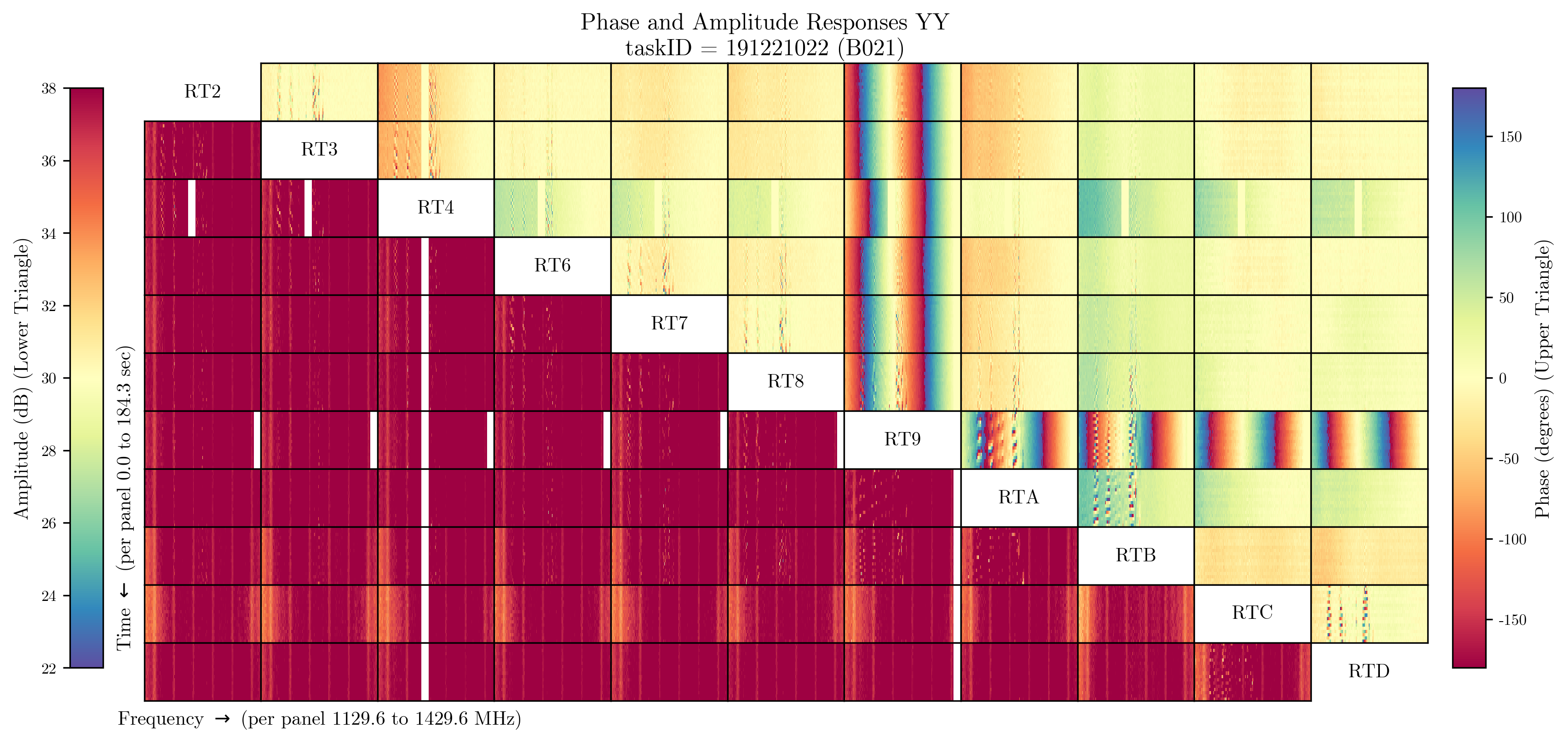}
    \caption{Waterfall plots (time vs. frequency) of amplitude (lower left) and phase (upper right). Each baseline is shown for a single beam; for survey fields the central beam is shown and for calibrator observations the beam containing the calibrator. A plot is made for both XX (upper image) and YY (lower image). These plots provide an initial check on data quality, especially the system tuning. RT9 has poor tuning in the YY polarization; this is due to the missing data in the frequency range used for tuning.}
    \label{fig:waterfall_baseline}
\end{figure}

\begin{figure}
    \centering
    \includegraphics[keepaspectratio,width=0.9\linewidth]{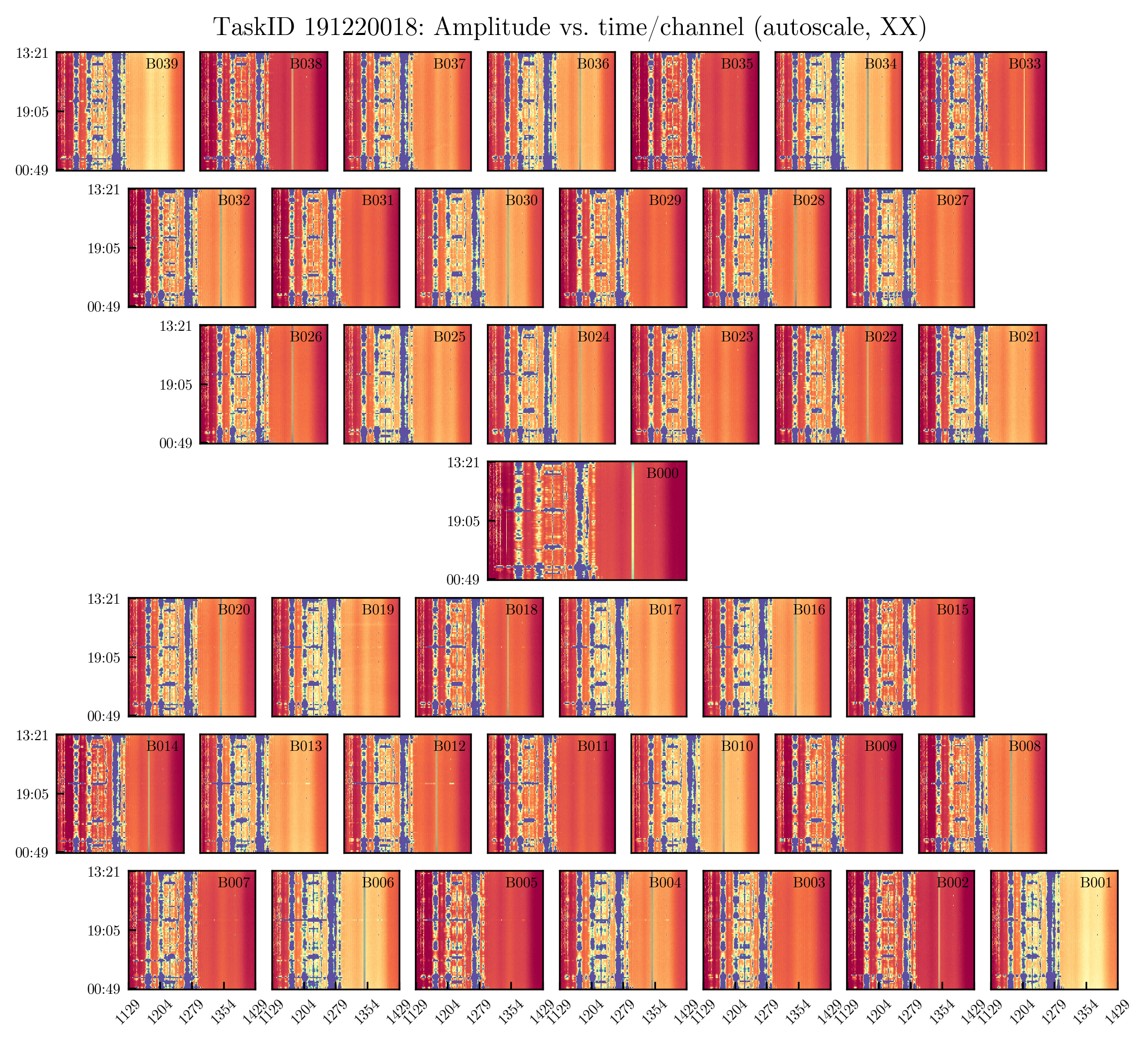}
    \caption{Waterfall plots of amplitude for every beam of an observation. All baselines are averaged together for the XX polarisation, and these plots are created on a fixed colour scale. These inspection plots are particularly useful for monitoring the RFI environment.}
    \label{fig:waterfall_beams}
\end{figure}

Decisions that we made during this time, motivated by the above, include:
\begin{itemize}
\item Unified machine-readable system specification parameters for both imaging and ARTS: we identified the key parameters needed from each team and united these into one machine-readable format, which serves as the input into the scheduling intepreter tool (ATDBSpec, described below). This involved iteration during the commissioning phase to ensure we captured all requirements but has remained unchanged since the start of full survey operations. Schedules can be uploaded by science teams and are automatically synced to the telescope system from this uploaded location. This machine-readable format was developed so that any future interpreter could potentially replace the one currently in production (ATDBSpec).
\item Semi-automated translation of schedule files to system schedule: the ATDBSpec package was built in Python to sit between the machine-readable schedule files and the telescope system. ATDBSpec interfaces directly with ATDB, meaning that once it is run on a given schedule file, the observations are scheduled in the system and no further action is needed. We did not automate the process of picking up schedule files, as current procedures involve telescope operators making the final decision about whether the system is ready to observe before any observations are scheduled. ATDBSpec was later expanded to calculate schedules for drift scans, test observations and pointing calibration in addition to handling the translation of survey schedules. ATDBSpec runs entirely on the command line, developed with the goal of minimising the need to interact with a visual interface as much as possible (ATDB itself provides this interface if necessary) and with the intention of enabling future automated processes to interface with it if needed. 
\item Automated system setup: the Apertif system can be readily set up in any observing mode using a Python script with minimal input parameters. This Python script wraps around existing system setup scripts and commands, bringing various aspects of setup together in one location with unified argument specification, and performs a system check at the end of the setup process to verify if it worked as expected. This Python script evolved from a shell script that was developed during commissioning, and adapted to make modifying input variables as straightforward as possible for operations. During commissioning, it was found that it was very important to ensure the right sequence of steps were followed for the various different modes of operations (e.g. imaging vs. transient surveys, or ``cold" vs. ``warm" start of firmware which resets or preserves system status), and as such this script has been vital to setting things up with the right parameters.
\item Semi-automated calibration processes: we adapted Jupyter notebooks that had been used for the commissioning period (for delay and phase tuning) to be as automated as possible. A single Jupyter notebook is used to control the entire process, sending the necessary observations to the system based on which antennas are available and processing the results with visual diagnostic plots. The notebook can then be used by an operator to make a judgement call on the calibration process. By using the Jupyter notebook approach hosted on the internal server but accessed remotely via VPN, it is possible for the system to be set up from any location. Once the outcomes of calibration are approved, all relevant system files are copied to the right location and become immediately active. This process works reasonably well, but given more time and resources we would have aimed to additionally automate the validation of the calibration process and retire the notebook approach. 
\item Automatic generation of raw data diagnostics: in order to provide a means of easily assessing data quality in the stage between observation and processing, a series of diagnostic plots were designed. These plots slice through the data across different relevant dimensions of time, frequency, amplitude and phase in order to make trends (in particular, bad data) readily visible. These plots are automatically generated once the data is successfully uploaded to ALTA, and are directly linked in the online archive to the relevant TID. A total of 18 raw data diagnostics (covering both XX/YY polarisation) are generated for each imaging observation, including the waterfall plots shown as examples in Figures \ref{fig:waterfall_baseline} and \ref{fig:waterfall_beams}.  
\item Automated diagnostic plots sent to Slack for immediate diagnostic by the first line operators: a crontab was set up to monitor the incoming observations, and automatically generate diagnostic plots if the observation is detected to be of a calibrator. For imaging, a plot of waterfall time-frequency for amplitude/phase for each cross-correlated antenna pair for the beam containing the calibrator is generated for XX/YY polarisations, and sent to the relevant operations channel ({\#}apertif-verification) in the ASTRON Slack space. Figure \ref{fig:waterfall_baseline} is an example of these particular diagnostic plots. In the case of time-domain observations, a crontab was set up on the ARTS cluster to monitor the production of local diagnostic plots from observations. This service picks up the generation of new plots and sends these to the same channel in Slack, allowing time-domain data to also be verified for quality. These approaches provide a quick and easy-to-access way to verify the results of calibration tuning and monitor calibrator or observation quality status from any location.
\item Automated monitoring and notifications about disk space: a crontab was set up to keep an eye on the disk space on both the datawriters and the imaging processing server, with notifications above certain limits triggering warnings sent to Slack ({\#}apertif-happili) at regular intervals. Sending these warnings to an easily-accessible central location helped immensely with ensuring timely reactions to disk space problems and increasing transparency of the current disk status. 
\item Automated triggering of the imaging pipeline: we extended the automated workflow for Apertif from operations into processing via the Autocal package. This sits between ALTA and the imaging pipeline Apercal, and automatically picks up new observations which it passes on to the pipeline. In its current form, Autocal automatically determines which observations are targets, and seeks then to match the 40 flux and polarisation calibrator observations that are generally found either side of the target. Complexity in Autocal arises from the need to sometimes schedule multiple targets in a row, the fact that not all observations are successful (meaning it needs to determine archive status for each calibrator beam and adjust accordingly), exceptions to normal scheduling and errors in scheduling either due to the system or operator input. Autocal generally performs equal to or better than a human in automatically starting up the pipeline, and operates 24/7 with systematic reproducible behaviour, thus ensuring a high efficiency for Apertif imaging processing. Upon triggering the Apercal pipeline, it sends a notification message to the relevant Slack channel ({\#}apercal-notifications) and also reports back when the processing is complete. Autocal has also been extended to trigger the quality assessment part of Apertif imaging, and it both prepares and executes the ingest of processed data products back to ALTA. More detail about both Autocal and Apercal can be found in the Apercal paper (Adebahr et al., in prep).
\end{itemize}

Our efforts to implement as much automation as possible for Apertif survey operations were driven by the desire to ensure that human effort was invested effectively. This has enabled operations to generally proceed smoothly, and most issues that have surfaced tend to be due to real system problems or human error rather than issues with automated approaches. The removal of manual decision-making processes as much as possible in critical workflow paths was a key part of ensuring sustainable Apertif operations within resource constraints. For example, data is automatically ingested to ALTA without waiting for a human decision on quality, ensuring the datawriters are kept clear and deferring quality assessment till a later stage. On reflection, it is clear that there are more avenues in which these processes of automation could be improved and streamlined, and we hope to see continued advances in automation as a key part of observatory workflows both for Apertif and for telescopes worldwide. 

\subsection{Daily operations}

%The two weeks observing blocks are interleaved with maintenance and calibration runs.
The operational tools described above are
a crucial part of the daily operations
of observing with Apertif.
The operations routine encompasses
the following main aspects:
\begin{itemize}
    \item Daily monitoring: Both the state of the system and the quality of the observations are monitored on a daily basis, by both operators and astronomers. ARTAMIS (Sec \ref{sec:artamis}) provides information about the instantaneous state of the system, over the full system chain,
    from the pointing position of the dishes to the
    state and health of the electronics and hardware and up to the data occupancy of the various on-line clusters (ALTA, ARTS cluster, data writer).  Monitoring available data space and taking action to ensure ample space for continued observing are a core part of daily operations. The diagnostic plots described above provide a first view of the quality of the data being recorded and are regularly inspected to ensure ongoing observations meet the minimum scientific requirements. Overall, the monitoring of the system and data quality is used to inform the maintenance plan.
    \item Maintenance: Every two to three weeks, survey observations are paused for system maintenance. In extraordinary circumstances, maintenance may interrupt the planned survey observations if required to return the Apertif system to an operable state that meets the minimum scientific requirements for observing. Maintenance includes both the hardware and software components of Apertif.
    Hardware maintenance mainly focuses on the health of the WSRT dishes (namely steering) and fixing the signal path of single antenna elements in a PAF. 
    As Apertif is in active operational mode, software maintenance is
    focused on necessary bug fixes and upgrades of operating systems, with occasional deployments of small new functionalities that improve the operational flow.
    \item System setup: Following the completion of maintenance, the system must be set-up to resume survey observations. The automated setup script is used to fully reset the system with a ``cold" start, along with setting the central frequency of the 300 MHz  
    band to be used in the coming observations\footnote{At the start of operations, we supported observing both the 1130-1430 MHz and 1220-1520 MHz bands. Currently, all survey observations are carried out in the 1220-1520 MHz band.}. The beamweights are then measured using Cassiopeia A, as described in Sec.~\ref{sec:bf}\footnote{Note that while Cassiopeia A is a circumpolar source, the equatorial mounts on WSRT limits the times during which it can be observed, providing a strong constraint on the operational schedule.}. After the measurement, the beamweights are calculated and sensitivity plots of all compound beams are produced, which are used to validate the beamweights before they are applied to the system. After the application of the beamweights, the delay and phase tuning is undertaken using the Jupyter notebook described above. 
    The accuracy of this step in the calibration chain is crucial to ensure high signal-to-noise for tied array observations and a smooth continuum for spectral line studies.
    \item Observing: After the PAF and array calibration are completed, the system is setup in the correct observing mode\footnote{The delay and phase tuning are undertaken in imaging mode, so for transient surveys the system mode needs to be changed.}. Scientists provide an observing file, and the operator on duty uses ATDBSpec to send the observations to this system. At this point, the daily monitoring resumes, and the cycle of operations continues. 
\end{itemize}

\section{System performance}\label{sec:performance}
Here we summarise  the performance of the Apertif system. More details are given in the documentation of the first data release of the Apertif imaging surveys\footnote{http://hdl.handle.net/21.12136/B014022C-978B-40F6-96C6-1A3B1F4A3DB0}  and in a number of papers (\citealt{Morganti20}, \citealt{2020A&A...635A..61O}, Adams et al.\ in prep, Hess et al.\ in prep, van Leeuwen et al.\ in prep.).

\subsection{Sensitivity}
\subsubsection{System noise temperature}
The design of the RF system aims to minimise the receiver noise temperature. Table~\ref{table:noise} gives a breakdown of the various contributions. Starting at the antenna, the received EM wave passes through a radome, constructed from polystyrene plates. Aluminium is used for the antenna element because of its low conductivity losses, low weight and cost effectiveness. Entering the LNA board, the transition from antenna to microstrip line has been designed to be as efficient as possible. The LNA is assembled on a RO4003 substrate, which is low loss and has stable RF parameters over time. The noise contribution of the LNA includes the effects of mutual coupling of the antenna elements.

At the DCU, the mixers are the main noise contributors due to their passive topology. No other mixers were available that satisfy the high linearity, the isolation, and the large bandwidth required. The last component contributing to the noise figure is the ADC. It is used in the second Nyquist zone, having 7.4 effective number of bits and a signal to noise ratio of 47 dB. This in total gives the ADC a noise figure of 45 dB. Sufficient gain has been placed in front of the ADC to overcome the huge noise figure. In addition to the components of the receiver noise listed in  Table~\ref{table:noise}, other factors, such as spillover, atmospheric emission, and emission from the cosmic microwave background and the Galaxy, contribute to the overall system noise.  The sum of all contributions results in a system temperature of about 70~K.

\begin{table}
	\caption{Overview of the receiver noise contributors, all referred to the input.} % title of Table
	\label{table:noise} % is used to refer this table in the text
	\centering % used for centering table
	\begin{tabular}{l c} % centered columns (4 columns)
		\hline\hline % inserts double horizontal lines
		  Component & $T_{\rm rec}$ contribution [K]\\
		\hline
		Radome, antenna element & 6.0 \\
		RFI filter & 10.0 \\
		LNA & 32.9 \\
		Coaxial cable & 0.2 \\
		DCU & 3.8 \\
		Coaxial cable & 0.0 \\
		ADU & 2.6 \\
		\hline
		Total & 55.5 \\
		\hline %inserts single line
	\end{tabular}
\end{table}

\subsubsection{Aperture efficiency}

\begin{figure*}
	\centering
	\includegraphics[width=17cm]{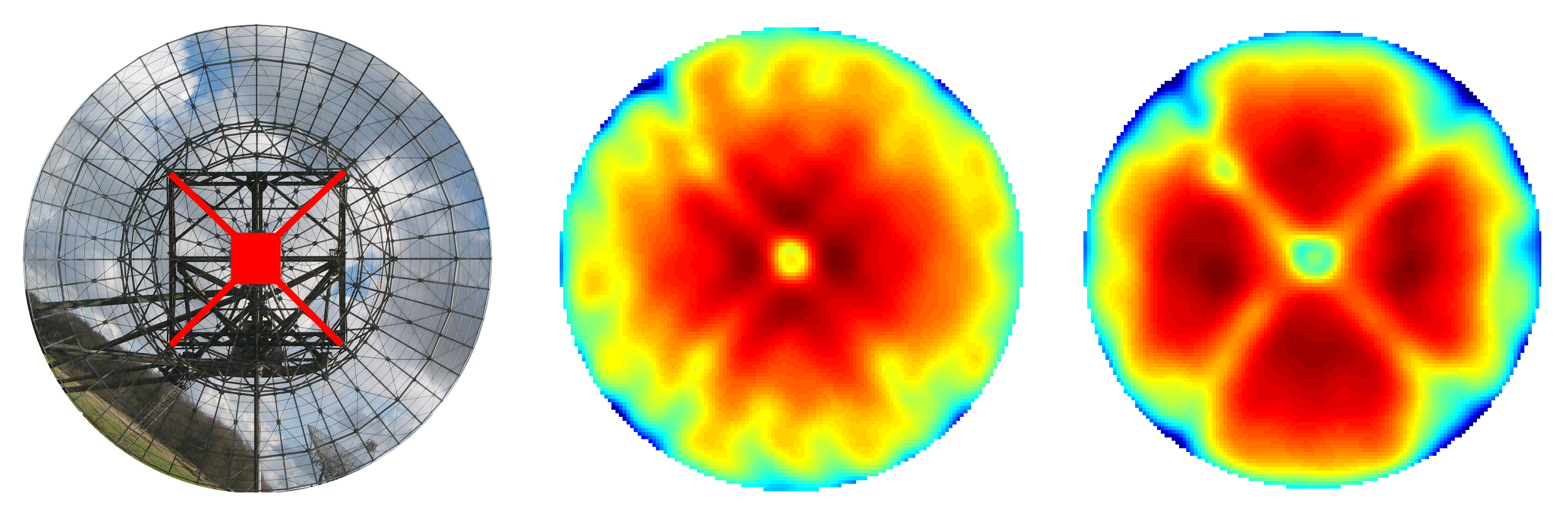}
	\caption{Front view of the reflector (left) and magnitude of the aperture field of a horn feed (centre) and PAF (right) on a logarithmic color scale \citep{PAST2010}.}
	\label{fig:Holog}
\end{figure*}

Due to its flexibility, the Apertif PAF  achieves a higher illumination efficiency than the  horn feed used at the WSRT before the upgrade with Apertif.  By optimising the weighting coefficients of the PAF, the illumination of the reflector can be made more uniform, while keeping the spillover contribution low \citep{PAST2010}. The illumination pattern of the  horn feed used previously and of the on-axis Apertif PAF beam have been determined by Fourier transforming the 2-dimensional complex (magnitude and phase) beam patterns (Fig.\ \ref{fig:Holog}). The X-shaped pattern in the illumination is due to blockage by the four struts of the dish and by the focus cabin. It is clear that the PAF illuminates the reflector more uniformly than the horn feed, while the edge illumination is similarly low to avoid spillover. Measurements show that an aperture efficiency around 75\% is achieved.

\begin{figure}
	\resizebox{\hsize}{!}{\includegraphics{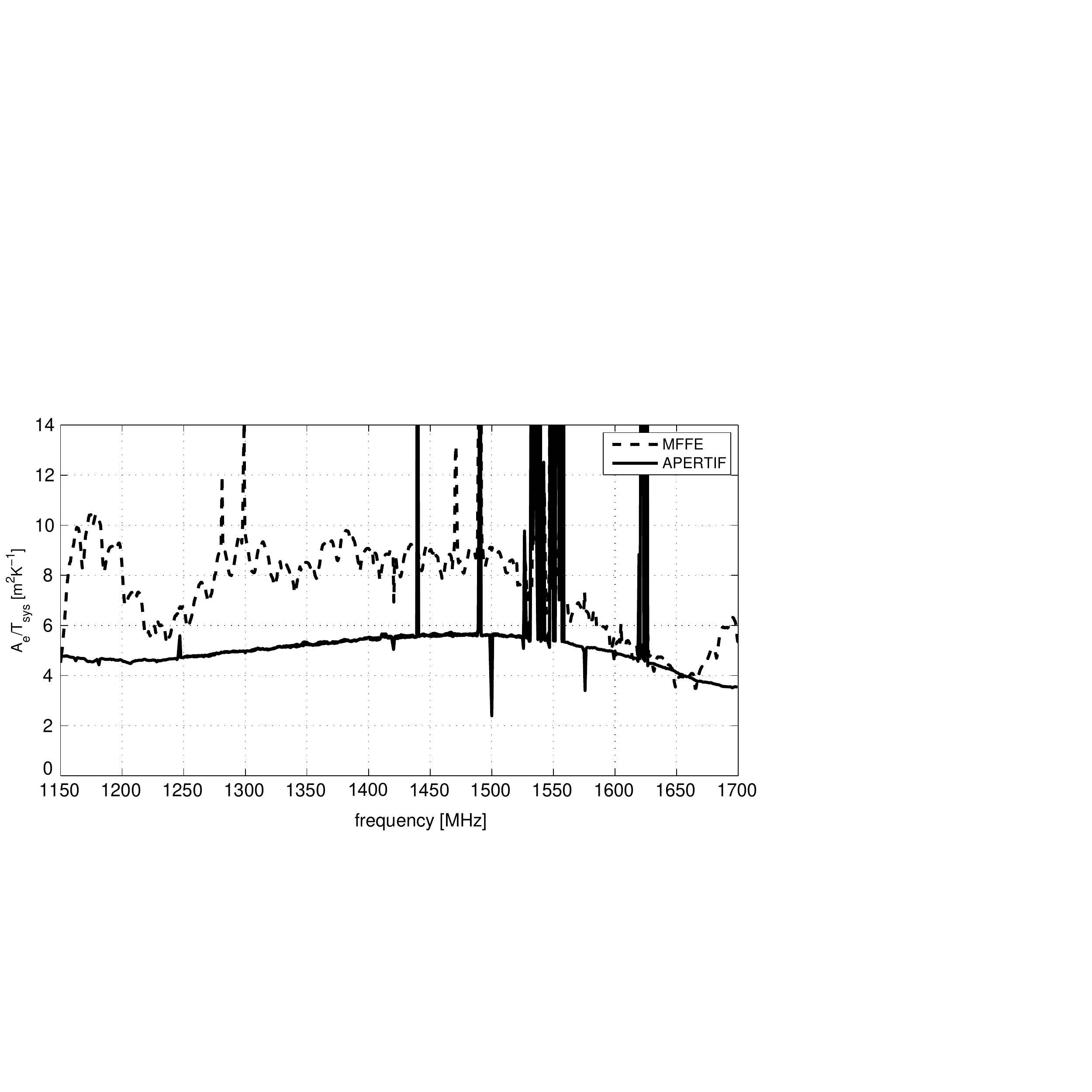}}
	\caption{Measured $A_{\rm eff}/T_{\rm sys}$ as function of frequency of the central compound beam of Apertif and the pre-Apertif WSRT frontend (MFFE).}
	\label{fig:aetsys}
\end{figure}

\begin{figure}
    \centering
    \includegraphics[angle=0,width=9cm]{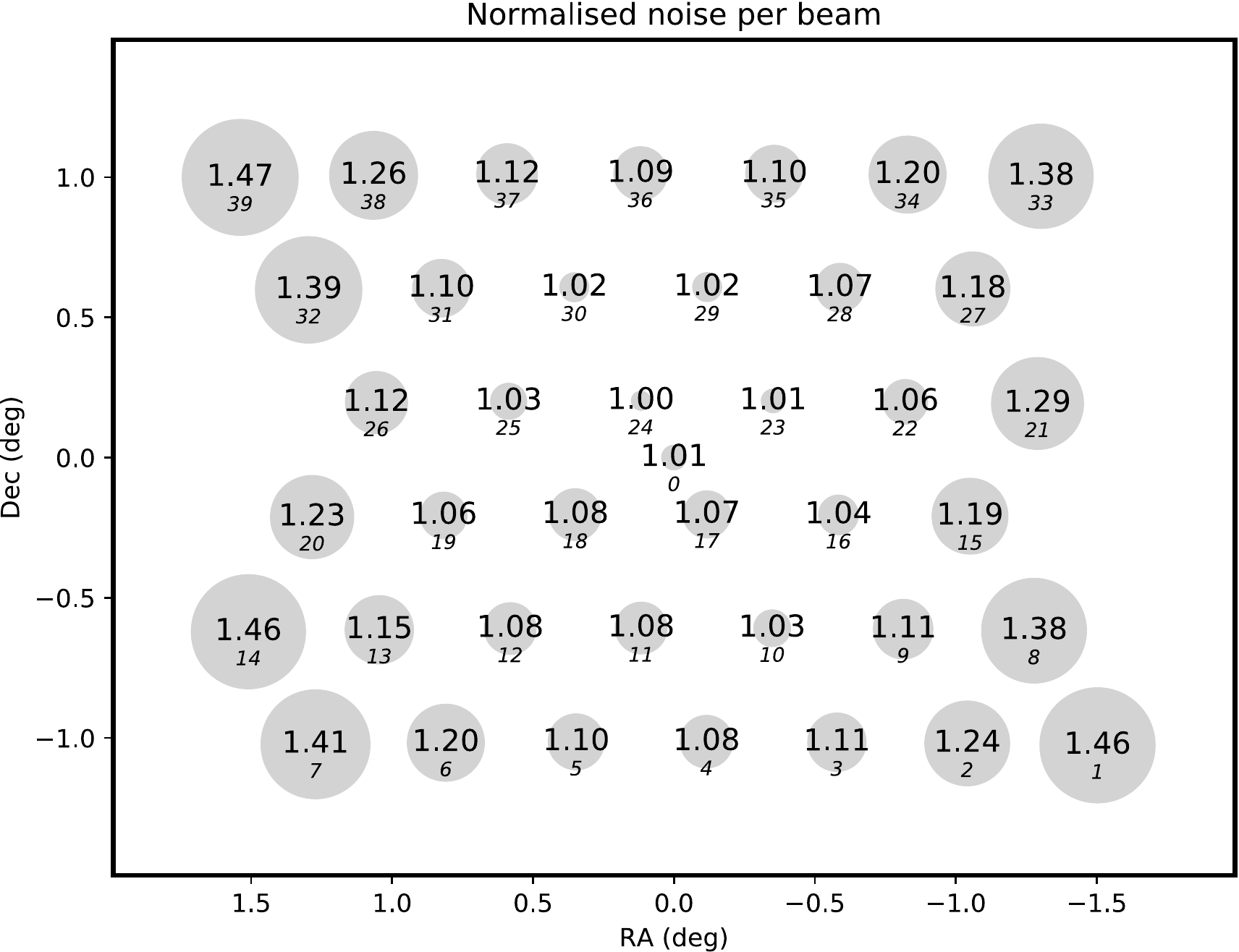}
    \caption{Relative noise of the  compound beams over the field of view. The size of the circles scales with the relative noise levels. Beam numbers are slanted. The noise values are normalised to that of the beam with the lowest noise.}
    \label{fig:noiseMap}
\end{figure}

%------------------ Beams
\begin{figure*}
    \centering
    \includegraphics[angle=0,width=9cm]{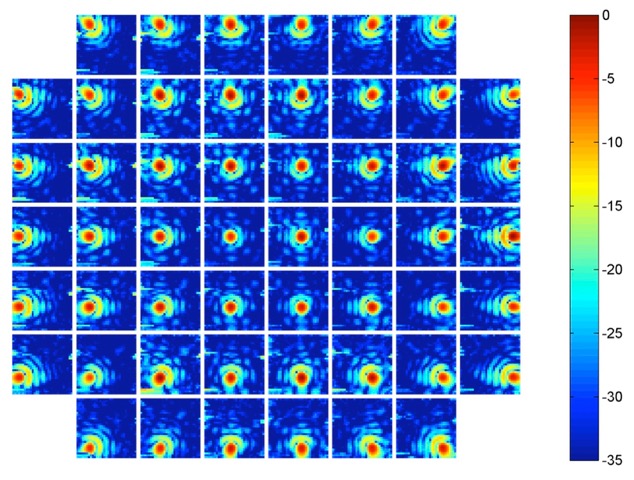}
    \includegraphics[angle=0,width=9cm]{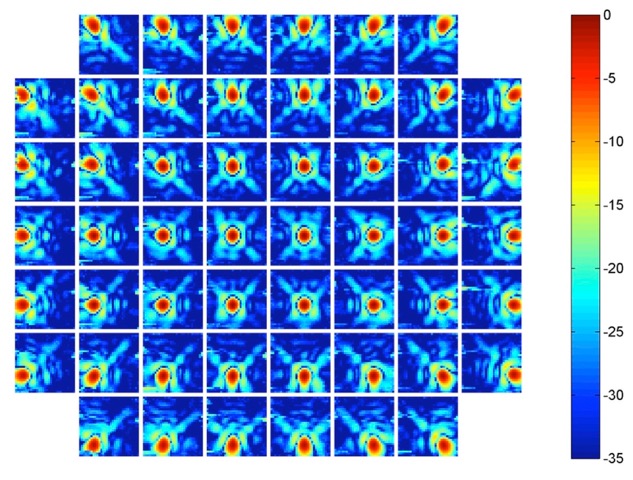}
    \caption{Left: Response patterns on the sky of the individual Vivaldi elements of the PAF. Each panel covers the same $3^{\circ} \times 3^{\circ}$ on the sky and shows the primary beam of a given element. Right: Response of compound beams formed in the same directions of the element beams. Note the significant} improvement in the beam shapes compared to the element beams. The same colour scales have been used in the two figures. This makes it easy to see that the sidelobe level in the optimised compound beam is much lower than that of the element beams.
    \label{fig:Beams}
\end{figure*}
%--------------------------------------

\subsection{SEFD and Image noise}
\label{sec:sensitivity}

Fig.~\ref{fig:aetsys} shows the measured $A_{\rm eff}/T_{\rm sys}$ as function of frequency of the central compound beam of Apertif and compares it with that of the old WSRT. This shows that the  $A_{\rm eff}/T_{\rm sys}$ of Apertif is about 35\% lower, somewhat depending on frequency. 
Although the aperture efficiency of Apertif is 50\% better than that of the pre-Apertif WSRT, its system temperature is about a factor 2.3 higher. This is mainly due to the Apertif  LNA's not being cooled, and that a filter had to be inserted before the LNA to suppress inter-modulation products due to strong out-of-band RFI leading to an increase in $T_{\rm sys}$ of 10 K. Figure \ref{fig:aetsys} also shows that for Apertif the behaviour of  $A_{\rm eff}/T_{\rm sys}$ is a smooth function of frequency, while for the pre-Apertif WSRT, it varies strongly with a fairly regular pattern with bumps spaced 17 MHz. These bumps are caused by standing waves in the dish which plagued the pre-Apertif WSRT and which are absent in Apertif (see Sec.\ \ref{sec:waves}).

The measured SEFDs of the different compound beams depend, as expected, on where they are within the overall field of view of Apertif, although the variation is very modest within the central 2$^\circ \times 2^\circ$ of the field of view (Fig.\ \ref{fig:noiseMap}). The measured SEFD of the central beam is around 560 Jy, which is $\sim$7\% higher than nominal. The other beams within the central 2$^\circ \times 2^\circ$ have an SEFD at most 10\% higher.  Outside this central region, the SEFD steeply increases with distance from the central beam, with the SEFD of the beams most distant of the field centre being around 850 Jy. This is due to the aperture efficiency of these beams being lower because only a limited number of PAF  elements can be used to do the beam forming. 

In practise, the noise in Apertif images is about 20\% higher than expected from the theoretical performance. Apart from the SEFD being slightly higher than nominal, in every observation some data have to be flagged due to RFI, leading to higher image noise. For the first Apertif data release, the observing band used was 1130 -- 1430 MHz. However, Apertif survey data products were  made by only using the upper 120 MHz of the observing band due to the RFI conditions in the lower part of the full 300 MHz Apertif band, mainly due to global navigation satellites. The best uniformly weighted continuum images have a noise level of about 30 $\mu$Jy beam$^{-1}$ while that of the line images is about 1.4 mJy beam$^{-1}$ for a velocity resolution of 7-8 km s$^{-1}$ and using Briggs weighting with robust = 0. Observations taken after January 2021 use a different frequency setup (1230 -- 1530 MHz) resulting in a larger fraction of the observing band being free of RFI. The noise in these later continuum images is  20 -- 25 $\mu$Jy beam$^{-1}$.

\subsection{Beam pattern and field of view}

The  layout of the 40 compound beams is shown in Fig.~\ref{fig:beampos}. As described by Hess et al.\ (in prep.), this layout has been chosen to obtain reasonable uniformity of the noise within the field of view of a single observation after the images of the individual compound beams have been mosaicked together while optimising the size of the field of view. The noise variations are below 20\%  in the central $2^\circ \times 2^\circ$ area. Thirty nine compound beams are put on a hexagonal grid with a separation of 28 arcminutes in Right Ascension  and 24 arcminutes in Declination. In addition, one compound beam is put on the optical axis of the telescope, which is convenient for test purposes. In a single observation, an area of roughly  3\fdg5 $\times$ 3$^\circ$ is imaged. For survey observations, which use overlapping observations spaced 2.8$^\circ$ in Right Ascension and 2.3$^\circ$ in Declination giving an effective field of view of 5.25 deg$^2$.
An example of a mosaic from a single Apertif observation is shown in Fig.~\ref{fig:ContinuumPointing}.

\subsection{Survey speed}

The survey speed of a radio telescope is defined as the inverse of the time it takes to image a large area on the sky  down to a certain noise limit. It therefore scales as SEFD$^{-2}\cdot \Delta\nu \cdot\Omega$ with $\Delta\nu$ the total bandwidth and $\Omega$ the field of view. This implies that the survey speed of Apertif (SEFD = 560 Jy, 39 beams, $\Delta\nu = 300$ MHz) is about 22 times  that of the pre-Apertif WSRT (SEFD = 310 Jy, 1 beam, $\Delta\nu = 160$ MHz).  In practice, however, a significant part of the observing band of Apertif is unusable due to strong RFI below 1300 MHz. Due to this loss of bandwidth due to RFI, the relative survey speed is about 10 for data taken before January 2021 and  15 for observations after that  date.  So despite the higher SEFD of Apertif and the impact of RFI, the gain in survey speed is  substantial.

%------------------ Beams
\begin{figure}
    \centering
    \includegraphics[angle=0,width=8cm]{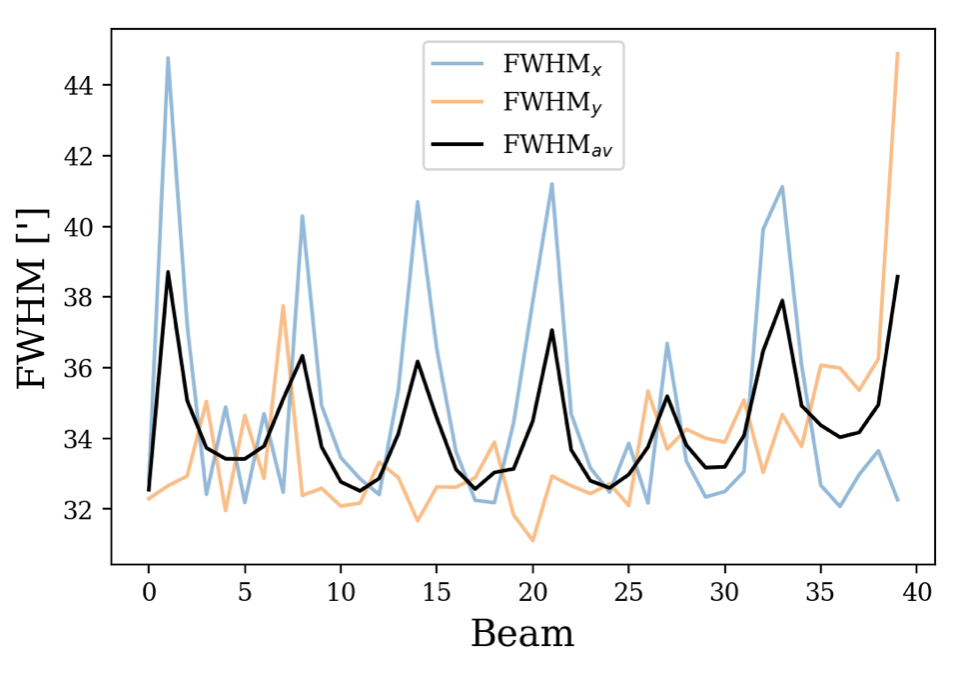}
    \caption{FWHM as a function of beam number for a frequency of 1363 MHz. The black line shows the average FWHM when fitting a 2-D Gaussian function to the beam maps. The blue line shows the FWHM of the 1-D Gaussian along the major axis and the orange line shows the FWHM along the minor axis. The shape of the compound beams is not perfectly Gaussian, but a 2-D Gaussian function is a good approximation for the beam shapes within a few percent.  }
    \label{fig:BeamShapes}
\end{figure}
%--------------------------------------

%------------------ Spectra
\begin{figure*}
    \centering
    \includegraphics[angle=0,width=8cm]{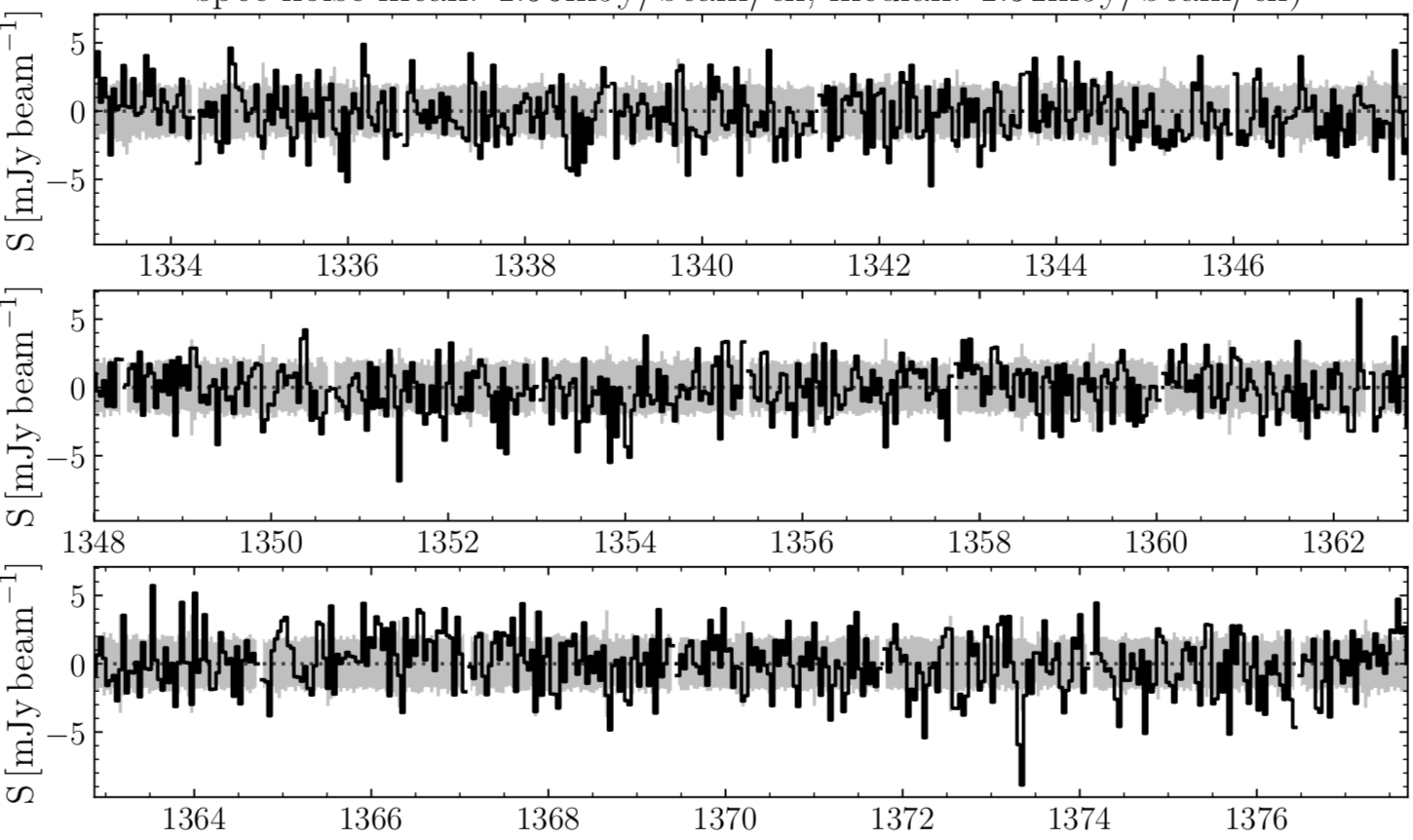}
    \includegraphics[angle=0,width=8cm]{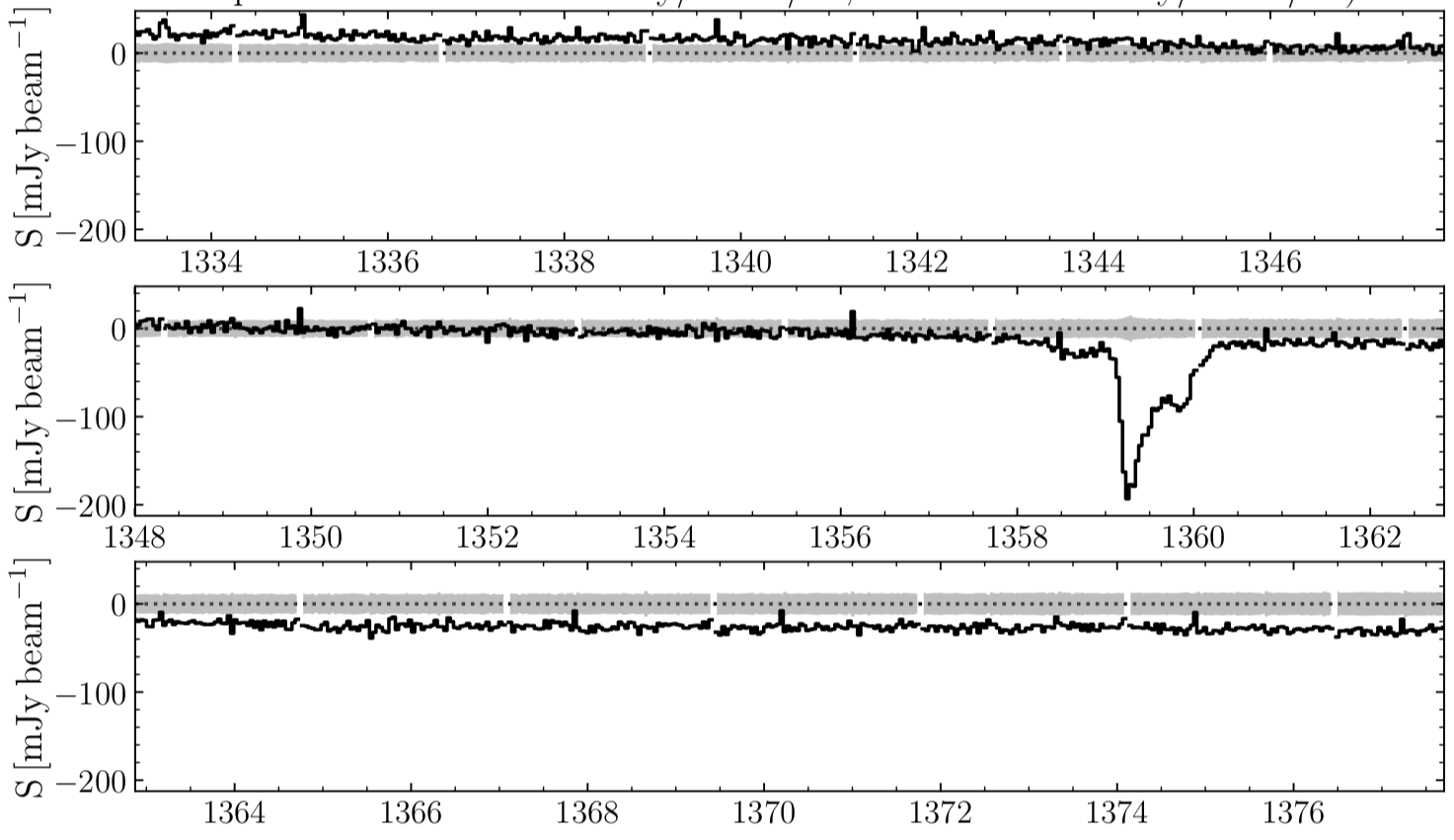}
    \caption{Two continuum subtracted spectra extracted from  data cubes produced by the imaging pipeline (see Adams et al.\ in prep for details) against radio continuum sources of very different flux density.  The source on the left has a flux density of $\sim$20 mJy while the flux density of the source on the right is $\sim$2900 mJy. The horizontal axis is  frequency covered by the corresponding cube. In both spectra  redshifted \HI\ absorption is detected. Given the high flux density of the source on the right, the flatness of the spectra illustrates the good quality of the bandpass calibration  that can be achieved even for bright radio sources. The offsets in the spectra relative to the zero intensity line are due to limitations in the continuum subtraction in the present version of the imaging  pipeline.  }
    \label{fig:SpectrumExample}
\end{figure*}
%--------------------------------------

\subsection{Beam shapes}

\subsubsection{Element beams vs compound beams}

The left panel of Fig.~\ref{fig:Beams} shows the beam patterns   measured on the sky, in a single polarisation, for all single elements of the PAF. Each panel covers the same $3^{\circ} \times 3^{\circ}$ on the sky and shows how the primary beam of a given element is positioned on the sky. It is clear that each element is sensitive to a slightly different area, but together they cover a region of about 8 deg$^2$.

One can clearly see the large optical distortions in the reception patterns of the single elements, in particular for those that are close to the edge of the PAF. However, because the PAF fully samples the radiation field in the focal plane, the signals from all elements can be combined in amplitude and phase to counteract these optical distortions and to optimise  the shape of the combined beam. 

The panel on the right in Fig.~\ref{fig:Beams} shows the result of this  optimisation for compound beams formed in the direction of the element beams. Note that all images in  this figure have the same colour scale. The right panel clearly shows that much of the optical distortions have been removed and that the shape of the beam patterns varies much less over the field of view. Some distortions remain, in particular towards the edges and corners of the field of view. To correct for these distortion one would need  elements located outside the physical PAF array of Apertif. 

\subsection{Measuring compound beams}
\label{sec:compoundBeams}
Beam weights are measured at the start of every imaging observing period (see Sec.~\ref{sec:bf}) and are typically used for 2--3 weeks in a row. 
The compound beam shapes are empirically derived in two separate ways: drift scans across a bright source, and by comparing continuum fluxes of sources in Apertif continuum images with those from the  NVSS using a Gaussian process modelling technique.

With the first method, beam maps are produced from drift scans performed periodically on Cygnus~A (occasionally Cassiopeia~A is used). During the drift scan measurement, the PAF is at a fixed position on the sky and Cygnus~A drifts through the field of view. The change in the auto correlation amplitudes during the drift scan measurements is used to reconstruct the response of the compound beam response (D\'enes et al., in prep.). This is then repeated 31 times to cover sufficient area to allow measurement of all 40 Apertif beams. 

In the second approach, a Gaussian process regression technique is used to reconstruct the primary beam response (Kutkin et al.\ in prep.). Continuum fluxes of sources in Apertif images for a given compound beam are compared to NVSS flux values. The ratio of the Apertif flux over the NVSS flux across the field of view corresponds to the primary beam response.

Both approaches find consistent shapes for the compound beams. Fig.~\ref{fig:BeamShapes} shows the sizes of the beams as measured using the drift scan method. The asymmetry of the beams located at the edge of the field is clearly seen in this plot.

The two approaches have different strengths and weaknesses. The drift scan methodology provides frequency and polarisation information about the compound beam shape, in addition to measuring the shape per WSRT dish. However, this approach can suffer from also detecting other sources than the main calibrator source, especially at the edges of the primary beam. It is  also time-consuming observationally, leading to measurements that are performed infrequently. The Gaussian process regression approach utilises existing data, can be refined to provide shapes over a narrow range in time, and compensates any offset in flux scale between Apertif continuum images and NVSS. However, this method provides no information on the frequency or polarisation dependence of the beam, and the measurement is an average across all dishes.

For the first Apertif data release, the primary beam images based on the Gaussian process regression method, with frequency scaling information from the drift scans, were released. This choice is motivated in Adams et al.\ (in prep), along with a comprehensive look at the impact on the continuum fluxes.

\subsection{Polarisation}

The image noise for the  Stokes Q- and U-cubes with a frequency resolution of 0.78125 MHz  (Adebahr et al.\ in prep.) is $\sim120\ \upmu$Jy/beam per channel, leading to a final image noise after RM-Synthesis \citep{2005A&A...441.1217B}  for data with a total 150\,MHz bandwidth of $\sim17 \upmu$Jy/beam in the polarised intensity images. This is slightly lower than the expected noise from the Stokes I continuum, which would normally be $\sqrt{2}$ higher. Since source densities in polarisation are generally lower and emission is fainter, the resulting images are less affected by sidelobe confusion and image artefacts.

Leakage from total power emission can significantly confuse emission in Stokes Q, U and V. To determine the instrumental leakage we used two different strategies. For linear feeds, instrumental leakage in Stokes Q is mostly directly correlated  with Stokes I, while leakage into Stokes U is correlated with Stokes V. Since several sources within an Apertif beam are physically linearly polarised, a separation of instrumental leakage and physical polarisation is difficult. To access the Stokes Q-leakage we used the drift scans taken for determining the shape of the compound beams. Here the difference in the measurements of XX and YY corresponds directly to Stokes Q. Leakages of the order of 0.001 are found at the beam centres and 0.02 at a distance of 0.5\,deg from the beam centres. For the edge beams, the leakage can be a bit higher, up to 0.1 for beams 01 and 39. For the analysis of the Stokes U-leakage, Stokes V images were analysed. Sources appearing in Stokes V are very rare and therefore any emission in this Stokes parameter can be assumed to be leakage. By analysing the fractional polarisation ${\rm FP}_V=V/I$ for a large number of sources in  a large number of images, we have determined the leakage dependence. Values of ${\rm FP}_V<0.001$ were found for the pointing centres, increasing linearly to values of $FP_V<0.02$ at distances of 0.5\,deg, which are very similar to the values for Stokes Q.

\subsection{uv-coverage and resolution}
\label{sec:uvcoverage}

For many years, the WSRT has been participating in VLBI observations, for example as part of the European VLBI Network, and the aim of the Westerbork Observatory is to continue to do so. Many of these VLBI observations are performed at frequencies not available to Apertif. Therefore, Apertif was installed in only twelve of the fourteen  WSRT dishes, with the two remaining antennas keeping the existing multi-frequency frontends so that the  WSRT can continue to perform VLBI observations  at  frequencies other than L band. After extensive study of the various options of which two antennas not to include in Apertif, it was decided to exclude the two most western dishes (RT0 and RT1). Although this leads to a loss of spatial resolution of about 10\%,  this choice creates the fewest gaps in the $uv$-coverage, resulting in dirty PSFs with the lowest average sidelobe levels. In addition, for images tapered to 30-arcsec resolution (which are often used for spectral line work), no gap is present in the $uv$-coverage and the imaging performance is very similar to that of the pre-Apertif WSRT.

It was also decided to maintain regular coverage of the very inner uv plane, as was  common practise with the old WSRT (the so-called maxi-short configuration). This means that the shortest four baselines are 36, 54, 72 and 90 m. This ensures that smooth structures with an extent of up to $\sim$10 arcminutes can be well imaged with Apertif. Four of  the twelve Apertif dishes are mounted on rails and in principle the array can be reconfigured. However, in contrast to the pre-Apertif WSRT, this option is not used for Apertif. The longest baseline of the array is 2412 m. This gives a spatial resolution in uniformly weighted images of about $12^{\prime\prime}$ in right ascension and, due to the fact that the WSRT is an east-west array, $12^{\prime\prime}/\sin\delta$ in declination at declination $\delta$.

%------------------ Standing Waves
\begin{figure}
    \centering
    \includegraphics[angle=0,width=8cm]{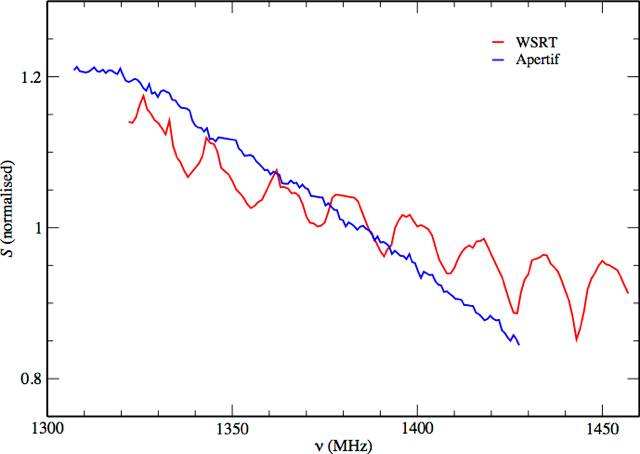}
    \caption{Illumination efficiency as function of frequency as measured for the pre-Apertif WSRT frontend (MFFE) and for  the Apertif  PAF. The MFFE data clearly show the periodic modulation due to standing waves, while the Apertif PAF shows no sign of such effects. }
    \label{fig:StandingWaves}
\end{figure}
%--------------------------------------
\subsection{Standing waves}
\label{sec:waves}

Before Apertif was installed on the WSRT, images made with  the telescope were strongly affected by standing waves due to reflections between the vertex of the reflector and the frontend in the telescope focus \citep{Popping2008}. Effectively, such standing waves affect the illumination of the dish as function of frequency. In the case of the WSRT, the 8.75~m distance between the vertex and the primary focus results in variations of the illumination pattern with a period of 17 MHz. The main impact of this on the images is that it modulates the observed spectrum of off-axis sources. Instead of being smooth, the spectrum of such sources show periodic variations (Fig.~\ref{fig:StandingWaves}). In addition, these variations are polarisation dependent. This presents a huge complication for the calibration and analysis of wide-band  observations since it is very difficult to take these spectral  modulations into account in the calibration and data reduction, leading to very significant imaging errors for off-axis sources. 
Measurements show that these negative effects are absent in  Apertif data (Fig.~\ref{fig:aetsys}). Fig.~\ref{fig:StandingWaves} shows the impact of this on the spectrum of a strong off-axis source as observed with the pre-Apertif WSRT and with Apertif.
The pre-Apertif spectrum clearly shows the periodic modulation  induced by standing waves in the dish which are completely absent in the Apertif data.
Electromagnetic simulations confirm that the absence of standing waves in Apertif is due to a much lower reflection  of radiation  (a factor 1000) by the frontend back into the reflector because the Apertif PAF absorbs more energy than the WSRT horn system. In addition, the PAF acts as a diffusor for the small amount of radiation reflected back into the dish because it is reflected over a much larger solid angle.

\subsection{Bandpass}

The observing band of Apertif is the combination of 384 subbands of 0.78125 MHz each with every sub-band having 64 channels, each channel 12.2 kHz wide. Because the subbands are critically sampled as described in Sec.\ \ref{sec:adu-dsp}, aliasing results in a sharp drop in overall response for channels at the subband edges, namely channels 0, 1 and 63 of every subband. At the moment this issue is solved by flagging these three channels with low signal at the edges of the subbands. 

The stability of the bandpass is key for a proper calibration of the data. The changes to the amplitude of the bandpass during an observation are  a fraction of percent. This has been derived from long (12~h), dedicated observations of strong calibrators, and from the comparison of the bandpass solutions derived from the standard calibration observations at the beginning and at the end of full synthesis observations.  
The well-behaved calibrated bandpasses can be seen in the two examples shown in  Fig.~\ref{fig:SpectrumExample} which shows two spectra taken at the location of radio sources, after subtraction of the continuum flux and showing detections of \HI\ absorption. The two sources have very different  flux density,  $\sim 20$ mJy for the source in the left panel, while the one on the right is a much stronger source of $\sim$2900 mJy.
The flatness of the  spectra  indicate good spectral stability, in particular in the case of the very strong radio source on the right. 

\section{Conclusions}
We have presented  an overview of the Apertif system, the phased-array upgrade of the WSRT: its science goals, system design, and key characteristics. The WSRT with Apertif is one of the first telescopes worldwide that successfully uses phased array feeds for all its science operations, and is living up to the expectations. It is demonstrated that the WSRT-Apertif data can be calibrated and imaged using automated pipelines, that deliver good quality science data.

\begin{acknowledgements}
This work makes use of data from the Apertif system installed at the Westerbork Synthesis Radio Telescope owned by ASTRON. ASTRON, the Netherlands Institute for Radio Astronomy, is an institute of the Dutch Research Council NWO (De Nederlandse Organisatie voor Wetenschappelijk Onderzoek).

This work is part of the research programme "Investeringen NWO-groot" with project numbers 175.010.2005.015 and 175.010.2009.012, which are (partly) financed by NWO.

UniBoard was a Joint Research Activity in the EC-funded RadioNet FP7 programme, under contract no. 227290, and was led by the Joint Institute for VLBI ERIC (JIVE). Optimisations of the board for production were funded by Samenwerkingsverband Noord-Nederland (SNN). The integration of multiple UniBoards into one mechanical construction was realized through the ExBox project, funded by NWO.
MV acknowledges support by NWO through Vici grant 639.043.511. 
EAKA is supported by the WISE research programme, which is financed by NWO.
The research leading to these results received funding from the European Research Council
under the European Union's Seventh Framework Programme (FP/2007-2013), 
through 
ERC Grant Agreement No. 291531 (‘HIStoryNU’, PI: JMvdH),
ERC Advanced Grant RADIOLIFE-320745 (PI: RM) and
ERC Grant Agreement No. 617199 ('ALERT', PI: JvL).
JvL further acknowledges funding from Vici research programme `ARGO' with project number 639.043.815, financed by NWO.
EP acknowledges funding from an NWO Veni Fellowship.
SMS  was supported by the National Aeronautics and Space Administration (NASA) under grant number NNX17AL74G issued through the NNH16ZDA001N Astrophysics Data Analysis Program (ADAP).
DV acknowledges support from the Netherlands eScience Center (NLeSC) under grant ASDI.15.406
\end{acknowledgements}

\bibliography{Apertifbib}
\bibliographystyle{aa}

\end{document}